\documentclass[floats,floatfix,showpacs,amssymb,prd,onecolumn,superscriptaddress,nofootinbib]{revtex4-2}

\usepackage{amsfonts,amsmath,amssymb,ascmac,bm,tensor}
\usepackage{fnpct} 
\usepackage{comment}
\usepackage{ifpdf}
\usepackage{slashed}
\usepackage{esint}
\usepackage{color}
\usepackage[mathscr]{eucal}
\usepackage[utf8]{inputenc}
\usepackage{physics}
\usepackage{cancel}
\usepackage{soul}
\usepackage{simpler-wick}
\usepackage{mathrsfs}

\ifpdf
  \usepackage{graphicx}     
  \usepackage[bookmarksopen,colorlinks=true,linkcolor=bblue,citecolor=bblue,urlcolor=ppink]{hyperref}
\else     
\fi

\usepackage{tikz}
\usepackage[compat=1.1.0]{tikz-feynman}

\hypersetup{
           breaklinks=true,   
           colorlinks=true,   
           linkcolor=bblue,
           citecolor=bblue,
           urlcolor=ppink
        }

\makeatletter
\newsavebox{\@brx}
\newcommand{\llangle}[1][]{\savebox{\@brx}{\(\m@th{#1\langle}\)}%
  \mathopen{\copy\@brx\kern-0.5\wd\@brx\usebox{\@brx}}}
\newcommand{\rrangle}[1][]{\savebox{\@brx}{\(\m@th{#1\rangle}\)}%
  \mathclose{\copy\@brx\kern-0.5\wd\@brx\usebox{\@brx}}}
\makeatother

\definecolor{red}{rgb}{1,0,0}
\definecolor{darkred}{rgb}{0.6,0,0}
\definecolor{darkgreen}{rgb}{0.992447,0.623778,0.034597}
\definecolor{ppink}{rgb}{1,0.4,0.4} 
\definecolor{bblue}{rgb}{0.284602,0.317763,0.963947}
\definecolor{purple}{rgb}{0.5 ,0, 0.7}

\newcommand{\fo}{{(1)}}
\renewcommand{\so}{{(2)}}
\newcommand{\tho}{{(3)}}
\newcommand{\foo}{{(4)}}
\newcommand{\no}{{(n)}}

\newcommand{\tre}{\text{tr} }
\newcommand{\vx}{\text{vx}}

\newcommand{\sm}{\text{sm}}

\newcommand{\dbphi}{{\dot{\bar \phi}}}

\newcommand{\inte}{\text{int}}

\newcommand{\Pl}{\text{Pl} }
\newcommand{\ee}{\text{e}}
\newcommand{\tad}{\text{tad}}

\newcommand{\uv}{\text{UV}}
\newcommand{\ir}{\text{IR}}
\renewcommand{\min}{\text{min}}

\newcommand{\bfk}{\mathbf{k}}
\newcommand{\bfp}{\mathbf{p}}
\newcommand{\bfq}{\mathbf{q}}
\newcommand{\bfr}{\mathbf{r}}

\newcommand{\bfx}{\mathbf{x}}
\newcommand{\bfy}{\mathbf{y}}

\newcommand{\tauf}{{\tau_f}}
\newcommand{\taui}{{\tau_i}}
\newcommand{\tauis}{{\tau^*_i}}

\newcommand{\inh}{\text{inh}}
\newcommand{\homo}{\text{hom}}

\renewcommand{\Re}{\text{Re}}
\renewcommand{\Im}{\text{Im}}

\newcommand{\svev}[1]{\langle #1 \rangle}
\newcommand{\ssvev}[1]{\llangle #1 \rrangle}

\makeatletter
\newcommand\footnoteref[1]{\protected@xdef\@thefnmark{\ref{#1}}\@footnotemark}
\makeatother

\allowdisplaybreaks[1]

\begin{document}


\title{
Cancellation of one-loop time dependence in superhorizon curvature perturbations from all scales
}

\author{Keisuke Inomata}
\affiliation{William H. Miller III Department of Physics and Astronomy, Johns Hopkins University, 3400 N. Charles Street, Baltimore, Maryland, 21218, USA}

\begin{abstract} 
\noindent
We show the conservation of the superhorizon curvature perturbations at one-loop level in spatially-flat gauge, including contributions from loop wavenumbers on all scales in the decoupling limit, $\epsilon (\equiv -\dot H/H^2) \to 0$.
In contrast to previous works, we do not assume a hierarchy $k \gg q$ between the wavenumber of the loop integral, $k$, and that of the power spectrum, $q$, and we explicitly include the regime $k \lesssim q$. 
Taking into account the nonlinear relation between the inflaton fluctuation and the curvature perturbation with the $\delta N$ formalism, we show that the apparent time dependence of the one-loop curvature power spectrum cancels once all contributions, including boundary terms, are combined consistently. 
\end{abstract}

\date{\today}
\maketitle

\tableofcontents

\section{Introduction} 

Cosmological perturbations are produced in the early universe and set the initial conditions for the cosmic microwave background (CMB) anisotropies and the large-scale structure (LSS)~\cite{Mukhanov:1981xt,Mukhanov:1982nu,Starobinsky:1982ee,Guth:1982ec,Panagiotakopoulos:1982rn,Bardeen:1983qw}.
The details of inflation are imprinted in these perturbations, since quantum fluctuations during inflation seed them.
Curvature perturbations are often used to characterize the amplitude of cosmological perturbations because they are thought to become constant (or conserved) once they exit the horizon in single-field inflation models.
The conservation of superhorizon curvature perturbations is a key feature in cosmology because it allows us to relate inflationary models to current observations. 
In this paper, we discuss the conservation of curvature perturbations beyond the linear perturbation level in single-field inflation models.

The nonlinear effects for the curvature perturbations can be important when there is a large enhancement of the perturbations on small scales, which is often studied in the context of primordial black holes (PBHs) and stochastic gravitational waves (GWs). 
In the presence of the perturbation enhancement, a sizable amount of PBHs can be produced and they can explain dark matter~\cite{Chapline:1975ojl,Ivanov:1994pa,Yokoyama:1995ex,GarciaBellido:1996qt,Afshordi:2003zb,Frampton:2010sw,Belotsky:2014kca,Carr:2016drx,Inomata:2017okj,Espinosa:2017sgp} and/or BHs detected by the LIGO-Virgo-KAGRA collaborations~\cite{Bird:2016dcv,Clesse:2016vqa,Sasaki:2016jop,Eroshenko:2016hmn,Sasaki:2018dmp,Carr:2020gox,Green:2020jor,Escriva:2022duf}. 
Large-amplitude curvature perturbations can also induce strong GWs at second order in perturbation. 
Such induced GWs could explain the stochastic GWs recently detected by the PTA experiments~\cite{NANOGrav:2023gor,NANOGrav:2023hvm,EPTA:2023fyk,EPTA:2023xxk,Reardon:2023gzh,Xu:2023wog}.

The nonlinear effects during inflation are often discussed in the in-in (Schwinger-Keldysh) formalism~\cite{Weinberg:2005vy,Senatore:2009cf}.
In this formalism, the nonlinear effects on the power spectrum can be calculated as loop corrections to the power spectrum. 
Over the past few years, there have been debates on whether superhorizon-limit curvature perturbations are conserved or not at one-loop level~\cite{Kristiano:2022maq,Riotto:2023hoz,Choudhury:2023vuj,Kristiano:2023scm,Riotto:2023gpm,Firouzjahi:2023aum,Motohashi:2023syh,Firouzjahi:2023ahg,Franciolini:2023agm,Tasinato:2023ukp,Cheng:2023ikq,Maity:2023qzw,Firouzjahi:2023bkt,Davies:2023hhn,Iacconi:2023ggt,Saburov:2024und,Ballesteros:2024zdp,Kristiano:2024vst,Kristiano:2024ngc,Sheikhahmadi:2024peu,Frolovsky:2025qre,Fumagalli:2023zzl,Tada:2023rgp,Inomata:2024lud,Kawaguchi:2024rsv,Fumagalli:2024jzz,Inomata:2025bqw,Fang:2025vhi,Inomata:2025pqa,Fang:2025kgf,Braglia:2025cee,Braglia:2025qrb,Kristiano:2025ajj,Iacconi:2026uzo,Ema:2026dop,Tanaka:2026zew}.
The original claim by Ref.~\cite{Kristiano:2022maq} was that superhorizon-limit curvature perturbations can be modified by small-scale perturbations enhanced during a transient ultra-slow-roll (USR) period at one-loop level.
This claim seems inconsistent with the existing references~\cite{Lyth:2004gb,Langlois:2005qp,Pimentel:2012tw}.
It has also been refuted by Refs.~\cite{Fumagalli:2023zzl,Tada:2023rgp,Inomata:2024lud,Kawaguchi:2024rsv,Fumagalli:2024jzz,Inomata:2025bqw,Fang:2025vhi,Inomata:2025pqa,Fang:2025kgf,Iacconi:2026uzo,Ema:2026dop,Tanaka:2026zew}.
In particular, in Ref.~\cite{Inomata:2025pqa}, we pointed out that the counterterms were overlooked in the original claim. 
Once we properly take into account the counterterms with an appropriate UV regularization, we can see that the small-scale perturbations do not affect the superhorizon-limit curvature power spectrum~\cite{Pimentel:2012tw,Fumagalli:2024jzz,Inomata:2025pqa,Braglia:2025cee,Braglia:2025qrb,Ema:2026dop}.
We also refer to Refs.~\cite{Ballesteros:2025nhz,Braglia:2026fle,Fang:2026off} for the recent developments of renormalization of the loop correlators.

In our previous works~\cite{Inomata:2024lud,Inomata:2025bqw}, we took spatially-flat gauge, where the spatial metric is flat, and showed that the superhorizon curvature perturbations are protected against the one-loop corrections from much smaller-scale perturbations. 
Assuming a large hierarchy $k \gg q$ with $k$ and $q$ being the wavenumbers of the loop integrals and the power spectrum, respectively, we saw that the calculation of the one-loop power spectrum in spatially-flat gauge is much simpler than that in comoving gauge, where $\delta \phi = 0$ with $\delta \phi$ the inflaton fluctuation.
This advantage is mainly thanks to the decoupling between the inflaton fluctuations and the metric perturbations during inflation in spatially-flat gauge~\cite{Baumann:2011su,Pajer:2016ieg}.

The main goal of this paper is to extend our proof of the curvature conservation in Refs.~\cite{Inomata:2024lud,Inomata:2025bqw} by removing the assumption of the large hierarchy $ k\gg q$. 
We here note that Ref.~\cite{Maldacena:2002vr} has shown that the nonlinear contributions from $k \lesssim q$ are negligible on superhorizon scales in comoving gauge. 
However, given a huge gap between the loop calculation in comoving gauge and spatially-flat gauge, it is still meaningful to independently prove the curvature conservation at the one-loop level in spatially-flat gauge without assuming the hierarchy.
Also, we remark that spatially-flat gauge is often taken in the lattice simulations during inflation, where the nonlinear couplings between the inflaton fluctuations on close scales are studied~\cite{Caravano:2021pgc,Caravano:2024tlp,Caravano:2024moy,Caravano:2026hca}.
Even in this sense, deepening the understanding of the nonlinear behavior in spatially-flat gauge is worthwhile.

To show the curvature conservation with the contributions from $k \lesssim q$ in general situations, we extend our previous setups~\cite{Inomata:2024lud,Inomata:2025bqw}. 
The major extensions are as follows:

1) We take into account the nonlinear relation between the curvature perturbation $\zeta$ and the inflaton fluctuation $\delta \phi$ in spatially-flat gauge. 
In our previous works, we focused on the situation where the linear relation between them is a good approximation for simplicity.
In this work, we will see that the loop contributions from $k \lesssim q$ generally cancel through the nonlinear relation between them. 

2) We consider the transition: the initial period $\to$ slow-roll (SR) period $\to$ the late period. 
We assume that the tree-level curvature perturbations on $k \lesssim q$ remain constant after the initial period.
In contrast to our previous works, we do not assume $V_\tho = 0$ ($V_\tho$: third derivative of the inflaton potential) for any of the above periods, including the SR period.
We show that the power spectrum of superhorizon curvature perturbations is constant at one-loop level during the late period.
We will explain the details of the transition in the next section.

This paper is organized as follows.
In Sec.~\ref{sec:setup}, we explain our setup by summarizing the basic equations that we use throughout this work. 
In Sec.~\ref{sec:one_loop_ps}, we summarize the curvature power spectrum by taking into account the nonlinear relation between the curvature perturbations and the inflaton fluctuations in spatially-flat gauge. 
Then, we disentangle the loop contribution in the two-point function of the inflaton fluctuations in Sec.~\ref{sec:in_in_loop}.
In Sec.~\ref{sec:cancellation}, we see the cancellation of the time dependence of the one-loop curvature power spectrum on superhorizon scales.
We conclude this paper in Sec.~\ref{sec:conclusion}.

Throughout this work, we use $\bfq$, $\bfq'$ for the wavenumber of the power spectrum and $\bfk$ and $\bfk'$ for the loop wavenumber. 
We often simply call the power spectrum wavenumber the $q$ mode in this paper.

\section{Setup}
\label{sec:setup}

In this paper, we basically take the same setup and notations as in our previous paper~\cite{Inomata:2025bqw}, except for the extensions mentioned in the Introduction.
We define the slow-roll parameter $\epsilon \equiv -\dot H/H^2$ with $H$ being the Hubble parameter and the dot denoting the physical time derivative.
We take spatially-flat gauge in the de-Sitter limit ($\epsilon \to 0$), where the metric perturbations are suppressed by $\epsilon$, compared to the potential derivative contributions.
This limit is often called the decoupling limit and allows us to neglect the metric perturbations~\cite{Baumann:2011su,Pajer:2016ieg}. 
Within this limit, we consider the Lagrangian for a canonical inflaton field $\phi$:
\begin{align}
  S = \int \dd \tau \dd^3 x a^4 \mathscr{L}, \  \mathscr{L} = -\frac{1}{2} \partial^\mu \phi \partial_\mu \phi - V_b(\phi),
\end{align}
where $V_b$ is the bare potential and decomposed into $V_b(\phi) = V(\phi) + V_c(\phi)$ with $V$ and $V_c$ being the background and counter potential, respectively.
We have introduced the counterterm contributions through the form of the counter potential so that it respects the diffeomorphism invariance of the theory. 
Then, we fix the background by substituting $\phi(\tau,\bfx) = \bar \phi(\tau) + \delta \phi(\tau,\bfx)$ into this and obtain the following equation of motion for the background:
\begin{align}
  \bar \phi'' + 2 \mathcal H \bar \phi' + a^2 V_\fo(\bar \phi) = 0,
  \label{eq:b_eom}
\end{align}
where the prime denotes the conformal time derivative and $V_\no(\bar\phi) \equiv \dd^n V(\bar \phi)/\dd \bar \phi^n$. 
In this work, we allow a nonzero tadpole ($\expval{\delta \phi(\tau,\bfx)} \neq 0$), which corresponds to a nonzero backreaction to the background $\bar \phi(\tau)$ from the perturbations.

Let us here clarify the $\epsilon$-dependence of $V_\tho$ when $V_\tho \neq 0$ in the decoupling limit with $\epsilon \to 0$. 
The decoupling limit corresponds to the limit $\epsilon \to 0$ with $H^2/(\epsilon M_\mathrm{Pl}^2)$, $\eta$, and $\xi_{n \geq 3}$ fixed, where $\eta = \dot \epsilon/(H \epsilon) = \xi_2$ and $\xi_n = \dd \ln \xi_{n-1}/\dd N$ with $\dd N = H \dd t$~\cite{Baumann:2011su,Pajer:2016ieg}.
$H^2/(\epsilon M_\mathrm{Pl}^2)$ is fixed to preserve the amplitude of the curvature perturbations.
Fixing $\xi_{n\geq 2}$ corresponds to fixing $\dd \ln V_{(n\geq 1)}(\bar \phi(t))/\dd N (= H^{-1} \dot {\bar \phi} \, \dd \ln V_{(n\geq 1)}(\bar \phi)/\dd \bar \phi )$ when $V_\tho \neq 0$, where $t$ is the physical time. 
From this, we can obtain the $\epsilon$-dependence of the potential derivatives as $V_\so \propto \epsilon^{0}$ and $V_\tho \propto \epsilon^{-1/2}$ when $V_\tho \neq 0$ in the decoupling limit.
See also Ref.~\cite{Inomata:2022yte} for the same relations in the inflaton potential with oscillatory features.
Note again, in this limit, metric perturbations can be neglected compared to the contributions from the potential derivative terms~\cite{Pajer:2016ieg}.

The Hamiltonian density for $\delta \phi$ can be decomposed into the free part $\mathscr{H}_0$ and the interaction part $\mathscr{H}_\inte$:
\begin{align}
  \label{eq:hamil}
  \mathscr{H} &= \frac{1}{2a^2} \left[(\delta \phi')^2 + (\partial_i \delta \phi)^2 \right]  + \sum_{n=2}\frac{1}{n!} V_{b,(n)}(\bar \phi) \delta\phi^n + V_{c,\fo}(\bar \phi) \delta\phi \nonumber \\
  &= \mathscr{H}_0 + \sum_{n}\mathscr{H}_{\inte,n},
\end{align}
where the free Hamiltonian density is 
\begin{align}
  \mathscr{H}_0 \equiv \frac{1}{2a^2} \left[(\delta \phi')^2 + (\partial_i \delta \phi)^2 \right] + \frac{1}{2} V_\so(\bar \phi) \delta \phi^2,
\end{align}
and the interaction Hamiltonian density is
\begin{align}
  \mathscr{H}_{\inte,n} \equiv \begin{cases}
  \cfrac{1}{n!} V_{c,(n)}(\bar\phi) \delta\phi^n & \quad (n=1,2) \\[2ex]
  \cfrac{1}{n!} V_{b,(n)}(\bar \phi) \delta\phi^n = \cfrac{1}{n!} \left[V_{(n)}(\bar \phi) + V_{c,(n)}(\bar \phi)\right]\delta\phi^n & \quad (n \geq 3) \\  
    \end{cases}.
    \label{eq:inte_hamil}
\end{align}
We express the interaction Hamiltonian as 
\begin{align}
  H_{\inte,n} = \int \dd^3 x \, a^4 \mathscr H_{\inte,n}, \ H_\inte = \sum_n H_{\inte,n}.
\end{align}

We expand the interaction picture $\delta \phi$ as
\begin{align}
  \delta \phi^I(\bfx,\tau) &= \int \frac{\dd^3 k}{(2\pi)^3} \ee^{i \bfk \cdot \bfx} \delta \phi^I_{\bfk}(\tau)\nonumber \\
  &= \int \frac{\dd^3 k}{(2\pi)^3} \ee^{i \bfk \cdot \bfx} \left[ u_k(\tau) a(\bfk) +  u^{*}_k(\tau) a^{\dagger}(-\bfk) \right],
\end{align}
where $[a(\bfk), a(\bfk')] = [a^\dagger(\bfk), a^\dagger(\bfk')] = 0$ and $[a(\bfk), a^\dagger(-\bfk')] = (2\pi)^3 \delta(\bfk + \bfk')$, and $u_k$ follows the equation of motion for the free part:
\begin{align}
  \left[\frac{\dd^2}{\dd \tau^2} + 2 \mathcal H \frac{\dd}{\dd \tau} + k^2 + a^2(\tau) V_\so(\tau) \right] u_k(\tau) = 0,
  \label{eq:u_eom}
\end{align}
where $V_\no(\tau) \equiv V_\no(\bar\phi(\tau))$.
$u_k$ also satisfies the Wronskian condition:
\begin{align}
  u_k(\tau){u_k^*}'(\tau) - u_k'(\tau)u_k^*(\tau) = \frac{i}{a^2(\tau)}.
\end{align}
Parameterizing $u_k(\tau) = |u_k(\tau)|\ee^{-i\theta(k,\tau)}$, we can rewrite this condition as
\begin{align}
  \label{eq:u_k_wrons_theta}
  |u_k(\tau)|^2 \theta'(k,\tau) = \frac{1}{2a^2(\tau)}.
\end{align}
Similar to our previous work~\cite{Inomata:2025bqw}, we impose the comoving IR and the physical UV cutoff for $u_k$ as 
\begin{align}
  u_k(\tau) \propto \Theta(k-k_\ir) \Theta\left(k_\uv \frac{a(\tau)}{a_*} -k \right),
  \label{eq:uk_cutoff}
\end{align}
where $\Theta$ is the Heaviside step function, $k_\uv$ is the physical UV cutoff scale at some time $\tau_*$, and $a_* = a(\tau_*)$.
Accordingly, we set the initial condition of $u_k$ at the time when $k = k_\uv a(\tau_\uv)/a_*$ to be the adiabatic (Bunch-Davies vacuum) solution:
\begin{align}
  u_k(\tau_\uv) = -\frac{H k\tau_\uv}{\sqrt{2k^3}}\ee^{i\frac{\left(2\nu + 1\right)\pi}{4}} \sqrt{\frac{-\pi k\tau_\uv}{2}} H_\nu^\fo(-k\tau_\uv), \ u_k'(\tau_\uv) = \frac{\dd u_k(\tau_\uv)}{\dd \tau_\uv},
\end{align}
where $H_\nu^\fo$ is the Hankel function of the first kind and $\nu = \sqrt{9/4-V_\so(\tau_\uv)/H^2}$.
To obtain this initial condition, we have assumed that the evolution timescale of $V_\so$ is much longer than the oscillation timescale of the UV mode $a_*/(a(\tau_\uv) k_\uv)$, which is true when $k_\uv$ is sufficiently large.
In this work, we use the in-in formalism to calculate the expectation value of some product $Q(\tau)$ of field operators as~\cite{Weinberg:2005vy} 
\begin{align}
  \expval{Q(\tau)} = \vev{ \left( T \ee^{-i \int^\tau_{-\infty} \dd \tau' H_\inte(\tau')} \right)^\dagger Q^I(\tau) \left( T \ee^{-i \int^\tau_{-\infty} \dd \tau' H_\inte(\tau')} \right)},
  \label{eq:in_in_def}
\end{align}
where $T$ is the time-ordering operator and $Q(\tau)$ can be $\delta \phi(\bfx)$, $\delta \phi_{\bfq} \delta \phi_{\bfq'}$, or $\delta \phi_\bfq \delta \phi_{\bfq'-\bfk} \delta \phi_{\bfk}$ in this paper.

Throughout this paper, we decompose the inflation era into the initial period ($\tau < \tau_i$), the SR period ($\taui \leq \tau < \tauf$), and the late period ($\tau \geq \tauf$).\footnote{
  In our previous works~\cite{Inomata:2024lud,Inomata:2025bqw,Inomata:2025pqa}, we considered the transition, SR $\to$ a transient non-SR period $\to$ SR, and assumed that the $q$-mode exits the horizon during the first SR period.
  The setup in this paper covers this transition because the initial period can be a SR period, and the late period can include a transient non-SR period followed by a SR period.
}
During the initial period, the $q$ mode exits the horizon. We define the end of the initial period, $\tau_i$, as the time when the tree-level curvature perturbations on $k \lesssim q$ become (approximately) constant as $\zeta_{\tre,\bfk} \propto u_k/\dbphi = \text{const}$. 
This requires $|q\taui| \ll 1$.
During the SR and the late period, the tree-level curvature perturbations on $k \lesssim q$ are always constant.
Note that $V_\tho$ can be nonzero during any period.
We will clarify the necessary conditions for the SR period in Eqs.~(\ref{eq:im_cond}) and (\ref{eq:im_cond_strict}).
The initial and the late period can have a transient non-SR period, such as the USR period~\cite{Kinney:1997ne,Inoue:2001zt,Kinney:2005vj,Martin:2012pe} or a parametric resonance period with oscillatory features~\cite{Inomata:2022yte}, as long as the tree-level curvature perturbations on $k \lesssim q$ are constant in $\tau \geq \tau_i$.
This means that the $q$ mode may exit the horizon during a transient non-SR period.
The main goal of this paper is to show that the curvature power spectrum on the scale of $q$ is constant during the late period even at one-loop level.

\section{One-loop power spectrum of curvature perturbations}
\label{sec:one_loop_ps}

In this section, we obtain the expression of the one-loop power spectrum of curvature perturbations by taking into account the nonlinear relation between the curvature perturbation $\zeta$ and the inflaton fluctuation $\delta \phi$ in spatially-flat gauge.

\subsection{Relation between curvature perturbation and inflaton fluctuation}

We use the $\delta N$ formalism~\cite{Starobinsky:1985ibc,Sasaki:1995aw,Sasaki:1998ug} to connect $\zeta$ and $\delta \phi$ during the late period ($\tau \geq \tauf$).
The $\delta N$ formalism requires the coarse-graining (smoothing) of small-scale perturbations to focus on the superhorizon perturbations, based on the separate universe picture, which claims that a sufficiently large region (much larger than the horizon) can be locally approximated as homogeneous and isotropic and therefore evolves as the FLRW universe~\cite{Lyth:2004gb}.
In the $\delta N$ formalism, we consider the e-folds from the flat time slice to the uniform density time slice and focus on how $\delta \phi$ at the initial (flat) time slice changes the e-folds.
Then, the e-fold difference ($\delta N$) induced by the $\delta \phi$ at the initial time corresponds to the curvature perturbation. 
Since we are interested in whether the curvature perturbations are constant on superhorizon scales, we take the two slices infinitesimally close and define the curvature perturbations at each time.
Specifically, we denote the $\phi$-dependence of the e-folds at the initial slice by $N(\phi)$, and focus on $\delta N = N(\bar \phi + \delta \phi) - N(\bar \phi)$.
Then, we can connect $\zeta$ to $\delta \phi$ on superhorizon at $\tau (\geq \tauf)$ as 
\begin{align} 
  \label{eq:zeta_dN}
  \zeta(\bfx,\tau) = \delta N(\bfx,\tau) = N_{\phi}(\tau) \delta \phi_L(\bfx,\tau) + \frac{1}{2}N_{\phi\phi}(\tau) \delta \phi_L^2(\bfx,\tau) + \frac{1}{6} N_{\phi\phi\phi} \delta \phi_L^3(\bfx,\tau) + \mathcal O(\delta \phi_L^4),
\end{align}
where we have changed the argument from $\bar \phi$ to $\tau$ with the use of their one-to-one correspondence and we have defined $N_{\phi}(\tau) \equiv \dd N(\bar\phi(\tau))/\dd \bar\phi(\tau)$, $N_{\phi\phi}(\tau) \equiv \dd^2 N(\bar\phi(\tau))/\dd \bar\phi^2(\tau)$, and $N_{\phi\phi\phi}(\tau) \equiv \dd^3 N(\bar\phi(\tau))/\dd \bar\phi^3(\tau)$. 
$\delta \phi_L$ is the coarse-grained field defined as 
\begin{align}
  \delta \phi_L(\bfx) \equiv \int \frac{\dd^3 k}{(2\pi)^3} \ee^{i \bfk \cdot \bfx} \Theta(1/L-k) \delta \phi_\bfk,
  \label{eq:d_phi_l}
\end{align}
where $L$ is the smoothing scale and we define the smoothing wavenumber cutoff as $k_\sm = 1/L$.
We impose $q \ll k_\sm \ll -1/\tau_i$ so that we can fully take into account the nonlinear contributions around the scale of $q$. 
Besides, we also impose that $\zeta_{\tre,\bfk}$ is constant on $k < k_\sm$ in $\tau \geq \taui$.
One might worry that, if we consider a transient USR period during the late period, the $\delta N$ formalism could also depend on $\delta \dot{\phi}$, which might modify the above $\delta N$ expression. However, we note that $\delta \dot \phi_\bfk/(H \delta \phi_{\bfk}) = \mathcal O((k/aH)^2)$ is satisfied on superhorizon scales during a SR, and $\delta \dot \phi$ at the beginning of the transient USR period (during the late period) determines the non-adiabaticity during that period. 
This means that, if we consider such a long SR period that $\mathcal O((k/aH)^2)$ is very small on $k < k_\text{sm}$ at $\tau_f$, we can safely neglect the non-adiabaticity and can use the standard $\delta N$ formalism with $N(\bar \phi)$, which is secured by our assumption $\zeta_{\tre,\bfk} = \text{const.}$ on $k < k_\sm$ in $\tau \geq \tau_i$.
As we will discuss in Sec.~\ref{subsec:bound_term}, the homogeneous (space-independent) smoothing cutoff $k_\sm$ is actually inconsistent with the separate universe picture. 
This leads to the non-conservation of the superhorizon curvature perturbations due to a boundary term at $k_\sm$. 
However, if we modify the smoothing procedure to be consistent with the separate universe picture, the remaining boundary term is canceled and the superhorizon curvature perturbations are conserved. 
For simplicity, we proceed with the loop calculation using the homogeneous smoothing cutoff until Sec.~\ref{subsec:bound_term}, where we come back to this issue.
We also note that, throughout the main text of this paper, we introduce the smoothing scale fixed in comoving scales as $1/L = k_\sm$, similarly to Ref.~\cite{Iacconi:2026uzo}.
On the other hand, in Appendix~\ref{app:smoothing}, we discuss the formulation with the smoothing scale fixed in physical scales. 
With the cutoff fixed in physical scales, we can see the conservation of curvature without modifying the cutoff (see Appendix~\ref{app:smoothing}).

Neglecting the higher-order contributions $\mathcal O(\delta \phi^4)$, we can reexpress Eq.~(\ref{eq:zeta_dN}) as 
\begin{align}
  \zeta(\bfx) &= -\frac{H}{\dbphi}\delta \phi_L(\bfx) + \frac{\eta}{4} \frac{H^2}{\dot{\bar \phi}^2} \delta \phi_L^2(\bfx) - \frac{\eta^2-\dot\eta/H}{12} \frac{H^3}{\dot{\bar \phi}^3} \delta \phi_L^3(\bfx),
  \label{eq:zeta_dN2}
\end{align}
where we have omitted the time arguments and used $N_\phi = - H/\dbphi$, $N_{\phi\phi} = (\dd N_\phi/\dd t)/\dbphi$, $N_{\phi\phi\phi} = (\dd N_{\phi\phi}/\dd t)/\dbphi$, and $\eta \equiv \dot \epsilon/(H\epsilon) = 2\ddot {\bar \phi}/(H\dot{\bar \phi}) + 2\epsilon \simeq 2\ddot {\bar \phi}/(H\dot{\bar \phi})$ in the decoupling limit $\epsilon \to 0$ with $\eta$ fixed.

\subsection{Curvature tadpole}

Before discussing the power spectrum, let us first discuss the tadpole of curvature perturbation at the one-loop level. 
Taking the ensemble average of both sides of Eq.~(\ref{eq:zeta_dN2}), we obtain the curvature tadpole,
\begin{align}
  \expval{\zeta} &= -H\frac{\expval{\delta \phi}}{\dot {\bar\phi}} + \frac{\eta}{4} \frac{H^2}{\dot{\bar \phi}^2} \expval{\delta \phi^2}|_{k_\sm} \nonumber \\ 
  &= -H\frac{\expval{\delta \phi}}{\dot {\bar\phi}} + \frac{\eta}{4} \expval{\zeta^2}|_{k_\sm},
  \label{eq:zeta_tad}
\end{align}
where $\svev{\delta \phi_L} = \svev{\delta \phi}$ by definition of $\delta \phi_L$, $\svev{\delta \phi^2}|_{k_\sm} = \int^{k_\sm}_{k_\ir} \dd \ln k (k^3/(2\pi^2)) |u_k|^2$, and $\svev{\zeta^2}|_{k_\sm} \equiv (H^2/\dot{\bar \phi}^2)\svev{\delta \phi^2}|_{k_\sm}$.
Note that, in our setup, $\expval{\zeta^2}|_{k_\sm}$ is constant in $\tau \geq \taui$. 
We can remove only the constant $\expval{\zeta}$ by the (time-independent) rescaling of the spatial coordinates. 
If $\expval{\zeta}$ depends on time, we need to absorb its time dependence into the scale factor. 
From Eq.~(\ref{eq:in_in_def}), the in-in formalism gives 
\begin{align}
  \expval{\delta \phi(\bfx,\tau)}
  &=  \int^\tau_{-\infty} \dd \tau' a^4(\tau') \Im[ u_0(\tau) u_0^*(\tau')] \left[ V_\tho(\tau') \int \frac{\dd^3 k}{(2\pi)^3} |u_{k}(\tau')|^2 + 2 V_{c,\fo}(\tau') \right],
  \label{eq:dt_one_pt}
\end{align}
where $\Im[ u_0(\tau) u_0^*(\tau')] \equiv \lim_{q \to 0} \Im[ u_q(\tau) u_q^*(\tau')]$. 
Note that, in $q \to 0$ limit, $\Im[ u_q(\tau) u_q^*(\tau')]$ becomes independent of $q$~\cite{Inomata:2025bqw}.
In the main text of this work, we make $\expval{\zeta}$ constant in $\tau \geq \tauf$ for simplicity by tuning the counter potential in $\tau \geq \taui$ as
\begin{align}
  V_{c,\fo}(\tau(\geq \taui)) = -\frac{V_\tho(\tau)}{2} \int^{k_\uv a(\tau)/a_i}_{k_\sm} \frac{\dd k}{k} \frac{k^3}{2\pi^2} |u_{k}(\tau)|^2, 
  \label{eq:v_c1_tad_cut}
\end{align}
where all the time dependencies here should be regarded as the function of $\bar \phi$ through $\tau = \bar\phi^{-1}(\bar \phi)$ with $\bar\phi^{-1}$ being the inverse function of $\bar \phi(\tau)$.
We stress that this counter potential is just a choice and is not a necessary condition for curvature conservation. 
Even if we adopt a different counter potential, which makes $\svev{\zeta}$ time-dependent, we can still obtain the same relation between $\zeta$ and $\delta \phi$ (Eq.~(\ref{eq:zeta_dN3})) by redefining the scale factor appropriately, as shown in Appendix~\ref{app:gen_c_pot}.
We can also take a similar counter potential in the dimensional regularization, as discussed in Appendix~\ref{app:dim_reg} (Eq.~(\ref{eq:v_c1_tad_cut_app})).
Note that we do not specify the form of $V_{c,\fo}$ in $\tau < \taui$.
Substituting Eq.~(\ref{eq:v_c1_tad_cut}) into Eq.~(\ref{eq:dt_one_pt}), we obtain 
\begin{align}
  \expval{\delta \phi(\tau(\geq \taui))}
  &=  \int^\tau_{\taui} \dd \tau' a^4(\tau') V_\tho(\tau') \Im[ u_0(\tau) u_0^*(\tau')] \int^{k_\sm}_{k_\ir} \frac{\dd k}{k} \frac{k^3}{2\pi^2} |u_{k}(\tau')|^2 + D(\tau),
  \label{eq:phi_tadpole}
\end{align}
where we have omitted the space argument on the left-hand side because the tadpole of $\delta \phi$ is independent of the space by definition. 
$D(\tau)$ is defined as
\begin{align}
    D(\tau) \equiv \int^{\taui}_{-\infty} \dd \tau' a^4(\tau') \Im[ u_0(\tau) u_0^*(\tau')] \left[ V_\tho(\tau') \int \frac{\dd^3 k}{(2\pi)^3} |u_{k}(\tau')|^2 + 2 V_{c,\fo}(\tau') \right].
\end{align}
We can see that, in $\tau \geq \taui$, the tadpole $\svev{\delta \phi}$ is newly induced only by the perturbations on $k < k_\sm$.
This physically means that the choice of the counter potential in Eq.~(\ref{eq:v_c1_tad_cut}) makes the background potential include the backreaction effects from the perturbations on $k > k_\sm$ in $\tau \geq \taui$.
Eq.~(\ref{eq:phi_tadpole}) comes from the backreaction effects that are not included in the background potential.
Note that fixing the counter potential determines the physical meaning of the background potential.

To explicitly see how Eq.~(\ref{eq:phi_tadpole}) leads to constant $\expval{\zeta}$, we transform Eq.~(\ref{eq:phi_tadpole}).
Using the expression in Eq.~(\ref{eq:u_k_wrons_theta}), we first reexpress $\Im[\cdots]$ as~\cite{Inomata:2022yte}: 
\begin{align}
  \Im[u_{q}(\tau)u^*_{q}(\tau')]|_{q \tau' \to 0} &= - |u_{q}(\tau)||u_{q}(\tau')| \int^\tau_{\tau'} \dd \tau'' \frac{1}{2 a^2(\tau'')|u_{q}(\tau'')|^2} \nonumber \\ 
  &= -\int^\tau_{\tau'} \dd \tau'' \frac{\dot{\bar \phi}(\tau) \dot{\bar \phi}(\tau')}{2 a^2(\tau'')\dot{\bar \phi}^2(\tau'')},
    \label{eq:im_uu0}
\end{align}
where we have used the fact that $|u_q(\tau)| \propto \dot{\bar \phi}(\tau)$ in the superhorizon limit, $q \tau \to 0$.
Using this, we can reexpress Eq.~(\ref{eq:phi_tadpole}) as
\begin{align}
  \expval{\delta \phi(\tau(\geq \taui))}
  &=  \int^{k_\sm}_{k_\ir} \frac{\dd k}{k} \frac{k^3}{2\pi^2} |u_{k}(\tau)|^2 \int^\tau_{\tau_i} \dd \tau' a^4(\tau') V_\tho(\tau') \Im[ u_0(\tau) u_0^*(\tau')] \frac{\dot{\bar \phi}^2(\tau')}{\dot{\bar \phi}^2(\tau)} + D(\tau) \nonumber \\
  &= -\int^{k_\sm}_{k_\ir} \frac{\dd k}{k} \frac{k^3}{2\pi^2} |u_{k}(\tau)|^2 \frac{1}{\dot{\bar \phi}(\tau)} \int^\tau_{\tau_i} \dd \tau' a^4(\tau') V_\tho(\tau') \dot{\bar \phi}^3(\tau') \int^\tau_{\tau'}\dd \tau'' \frac{1}{2 a^2(\tau'') \dot{\bar \phi}^2(\tau'')} + D(\tau).
  \label{eq:phi_tad}
\end{align}
To proceed, we use the following relation:
\begin{align}
  \frac{\dd \left(a^2 \dot {\phi}^2 \eta'\right)}{\dd \tau} &\simeq -a^4 \frac{2V_\tho \dot \phi^3}{H},
      \label{eq:v_3_dot_eta}
\end{align}
where we have neglected the terms additionally suppressed by $\epsilon$. See Appendix~\ref{app:rel} for the derivation of this relation.
Eq.~(\ref{eq:v_3_dot_eta}) enables us to further transform Eq.~(\ref{eq:phi_tad}) as 
\begin{align}
    \expval{\delta \phi(\tau(\geq \taui))}
  &= \frac{H}{\dot{\bar \phi}(\tau)} \left[\frac{\eta(\tau) - \eta(\tau_i)}{4} - \eta'(\taui) F(\tau,\taui) \right] \expval{\delta \phi^2(\tau)}|_{k_\sm} + D(\tau), 
  \label{eq:dphi_l_tad}
\end{align}
where we have integrated by parts and 
\begin{align}
  \label{eq:F_def0}
  F(\tau, \taui) = a^2(\taui) \dot{\bar \phi}^2(\taui) \int^\tau_{\taui}\dd \tau'' \frac{1}{4 a^2(\tau'') \dot{\bar \phi}^2(\tau'')}.
\end{align}
We here require the SR period ($\taui \leq \tau \leq \tauf$) to approximately lead to the following relation: 
\begin{align}
    \Im[u_{q}(\tau)u^*_{q}(\tau')] \propto \dbphi(\tau) \text{ for } \tau \geq \tauf \text{ and } \tau' \leq \taui.
    \label{eq:im_cond}
\end{align}
This is realized if $\eta$ is sufficiently greater than $-3$ during the SR period.
From this, we can see $D(\tau) \propto \dbphi(\tau)$ in $\tau \geq \tauf$.
By comparing Eqs.~(\ref{eq:im_uu0}) and (\ref{eq:F_def0}), we can also see that $F(\tau,\taui)$ approaches a constant after $\taui$ and finally becomes $F(\tau,\taui) = F(\tauf,\taui) (=\text{const.})$ in $\tau \geq \tauf$.
In general, the modes that exit the horizon during the period with $\eta < -3$, such as the USR period ($\eta = -6$), do not freeze until that period ends.
Note again that $V_\tho$ can be nonzero even during the SR period.
We also remark that Eq.~(\ref{eq:im_cond}) is one of the necessary conditions for the SR period. We will mention an additional condition for the SR period later in Eq.~(\ref{eq:im_cond_strict}).

Substituting Eq.~(\ref{eq:dphi_l_tad}) into Eq.~(\ref{eq:zeta_tad}), we can see the constancy of $\expval{\zeta}$:
\begin{align}
  \expval{\zeta} = \left[\frac{\eta(\tau_i)}{4} + \eta'(\taui)F(\tauf,\taui) \right]\expval{\zeta^2}|_{k_\sm} - H\frac{D(\tauf)}{\dbphi(\tauf)} \quad \text{in } \tau \geq \tauf.
\end{align}
This means that we can safely remove the constant $\expval{\zeta}$ by just rescaling the spatial coordinates when we consider the curvature power spectrum in $\tau \geq \tauf$. 
In the following analysis, we simply set $\expval{\zeta} = 0$ by implicitly rescaling the spatial coordinates.

Let us also consider the one-loop correction (backreaction) to $\dbphi$. 
By taking the physical time (not conformal time) derivative of Eq.~(\ref{eq:dphi_l_tad}), we obtain
\begin{align}
  \svev{\dot{\delta \phi}} = \frac{\eta(\tau) H}{2} \svev{\delta \phi} + \frac{H}{\dbphi(\tau)} \frac{\dot \eta(\tau)}{4} \expval{\delta \phi^2}|_{k_\sm},
  \label{eq:dphi_tadpole}
\end{align}
where we have used $\delta \phi_k \propto \dbphi$ on $k < k_\sm$ in $\tau \geq \tauf$.
By using this expression, we can rewrite Eq.~(\ref{eq:zeta_dN2}) as 
\begin{align}
  \zeta(\bfx) &= -\frac{H}{\svev{\dot \phi}}\delta \phi_L(\bfx) + \frac{\eta}{4} \frac{H^2}{\dot{\bar \phi}^2} \delta \hat \phi_L^2(\bfx) - \frac{\eta^2}{12} \frac{H^3}{\dot{\bar \phi}^3} \delta \phi_L^3(\bfx) + \frac{\dot \eta H^2}{4\dot{\bar \phi}^3} \left(\frac{1}{3} \delta \phi_L^3(\bfx) - \delta \phi_L(\bfx) \svev{\delta \phi^2}|_{k_\sm} \right),
  \label{eq:zeta_dN3}
\end{align}
where $\svev{\dot \phi} = \dbphi + \svev{\delta \dot \phi}$ and $\delta \hat \phi_L = \delta \phi_L - \svev{\delta \phi}$, and we have neglected the higher-order contributions.
Note that the last term, proportional to $\dot \eta$, vanishes when we consider the power spectrum of the curvature perturbations at one-loop level.

\subsection{One-loop power spectrum}

From Eq.~(\ref{eq:zeta_dN3}), we can express the curvature power spectrum up to one-loop order in $\tau \geq \tauf$ as
\begin{align}
  \expval{\zeta_{\bfq}(\tau) \zeta_{\bfq'}(\tau)} &= \frac{H^2}{\langle\dot \phi(\tau) \rangle^2} \expval{\delta \phi_{\bfq}(\tau) \delta \phi_{\bfq'}(\tau)} \nonumber \\ 
  & \qquad + \langle \zeta_{\bfq}(\tau) \zeta_{\bfq'}(\tau)\rangle_{12} + \langle \zeta_{\bfq}(\tau) \zeta_{\bfq'}(\tau)\rangle_{22} + \langle \zeta_{\bfq}(\tau) \zeta_{\bfq'}(\tau)\rangle_{13} + \langle \zeta_{\bfq}(\tau) \zeta_{\bfq'}(\tau)\rangle_B,
  \label{eq:zeta_one_loop}
\end{align} 
where $\delta \phi_{\bfq} = \int \dd^3 x \ee^{-i \bfq\cdot \bfx} \delta \phi(\bfx) = \int \dd^3 x \ee^{-i \bfq\cdot \bfx} \delta \phi_L(\bfx)$ because of $q < k_\sm$ and therefore the first line is independent of the smoothing procedure. 
The terms in the second line are
\begin{align}
  \label{eq:dn_loop_12}
  \langle \zeta_{\bfq}(\tau) \zeta_{\bfq'}(\tau)\rangle_{12} &= - \frac{\eta(\tau)}{4} \frac{H^3}{\dot {\bar \phi}^3(\tau)} \left( \expval{\delta \phi_{\bfq}(\tau) \int_S \frac{\dd^3 k}{(2\pi)^3} \delta \phi_{\bfq'-\bfk}(\tau) \delta \phi_{\bfk}(\tau)} +  \expval{\int_S \frac{\dd^3 k}{(2\pi)^3} \delta \phi_{\bfq-\bfk}(\tau) \delta \phi_{\bfk}(\tau) \delta \phi_{\bfq'}(\tau) } \right), \\ 
  \label{eq:dn_loop_22}  
  \langle \zeta_{\bfq}(\tau) \zeta_{\bfq'}(\tau)\rangle_{22} &= \frac{\eta^2(\tau)}{16} \frac{H^4}{\dot {\bar \phi}^4(\tau)} \vev{\int_S \frac{\dd^3 k}{(2\pi)^3} \delta \phi^I_{\bfq-\bfk}(\tau) \delta \phi^I_{\bfk}(\tau) \int_S \frac{\dd^3 k'}{(2\pi)^3} \delta \phi^I_{\bfq'-\bfk'}(\tau) \delta \phi^I_{\bfk'}(\tau)}, \\ 
  \langle \zeta_{\bfq}(\tau) \zeta_{\bfq'}(\tau)\rangle_{13} &= \frac{\eta^2(\tau)}{2} \frac{H^4}{\dot {\bar \phi}^4(\tau)} \int_S \frac{\dd^3 k}{(2\pi)^3} |u_k(\tau)|^2 \vev{\delta \phi^I_{\bfq}(\tau) \delta \phi^I_{\bfq'}(\tau)}.
  \label{eq:zeta_ps_loop}
\end{align}
We have added the boundary contribution $\langle\zeta_{\bfq} \zeta_{\bfq'}\rangle_B$ from the modification of the smoothing cutoff, which we will discuss in Sec.~\ref{subsec:bound_term}.
The wavenumber integral region $S$ only includes $k_\ir < k < k_\sm$ modes. 
For example, for $\int_S \frac{\dd^3 k}{(2\pi)^3} \delta \phi_{\bfq'-\bfk} \delta \phi_{\bfk}$, we set $S = \{ \mathbf{k} \in \mathbb{R}^3 \mid k_\ir < k < k_\sm, k_\ir < |\bfq' - \bfk| < k_\sm \}$.
Note that the unrestricted integral $\int \frac{\dd^3 k}{(2\pi)^3} \delta \phi_{\bfq'-\bfk} \delta \phi_{\bfk}$ effectively has the same integral region with the replacement of $k_\sm \to k_\uv a(\tau)/a_i$ because of the IR and UV cutoffs in the mode function (see Eq.~(\ref{eq:uk_cutoff})).
We call the first line in Eq.~(\ref{eq:zeta_one_loop}) the in-in loop contribution and the second line the $\delta N$ loop contributions, similarly to Ref.~\cite{Iacconi:2026uzo}.

\section{In-in loop contribution}
\label{sec:in_in_loop}

The main goal of this paper is to show the cancellation of the time dependence of $\svev{\zeta_{\bfq}(\tau) \zeta_{\bfq'}(\tau)}$ (Eq.~(\ref{eq:zeta_one_loop})) in $\tau \geq \tau_f$.
In this section, we extract the time-dependent part of the in-in loop contribution in $\tau \geq \tau_f$, which will cancel the time dependence of the $\delta N$ loop contributions.
Specifically, we disentangle the in-in loop contribution in Eq.~(\ref{eq:zeta_one_loop}):
\begin{align}
 H^2\frac{\expval{\delta \phi_{\bfq}(\tau) \delta \phi_{\bfq'}(\tau)}}{\langle \dot{\phi}(\tau) \rangle^2} = \frac{(2\pi)^3 \delta(\bfq + \bfq') H^2 \frac{2\pi^2}{q^3}\mathcal P_{\delta \phi}(q,\tau)}{\left(\dot{\bar \phi}(\tau) + \langle \delta \dot \phi(\tau) \rangle\right)^2},
 \label{eq:zeta_cons0}
\end{align}
where the power spectrum $\mathcal P_{\delta \phi}$ includes the tree and loop contributions.
The point is that some parts of $\mathcal P_{\delta \phi}$ and $\langle \delta \dot \phi \rangle$ cancel in some parameter regions. 
In the following, we see which contributions cancel and which survive.

\subsection{Expressions of one-loop power spectrum and backreaction}

To proceed, we first summarize the expressions of $\expval{\delta \phi_{\bfq}(\tau) \delta \phi_{\bfq'}(\tau)}$ and $\langle \delta \dot \phi(\tau) \rangle$.
Using the in-in formalism Eq.~(\ref{eq:in_in_def}), we can express the two-point function as~\cite{Inomata:2025bqw}
\begin{align}
  &\expval{\delta \phi_{\bfq}(\tau) \delta \phi_{\bfq'}(\tau)} = \vev{\delta \phi^I_{\bfq}(\tau) \delta \phi^I_{\bfq'}(\tau) } + 2\, \Im\left[\int^\tau_{-\infty} \dd \tau'\vev{ \delta \phi^I_{\bfq}(\tau) \delta \phi^I_{\bfq'}(\tau) (H_{\text{int},4}(\tau') + H_{\inte,2}(\tau'))}\right] \nonumber \\
  &+ 2\, \Re\left[ \int^\tau_{-\infty}\dd \tau' \int^{\tau'}_{-\infty}\dd \tau'' \vev{ \left(H_{\text{int},3}(\tau') \delta \phi^I_{\bfq}(\tau) \delta \phi^I_{\bfq'}(\tau) - \delta \phi^I_{\bfq}(\tau) \delta \phi^I_{\bfq'}(\tau) H_{\inte,3}(\tau') \right) (H_{\text{int},3}(\tau'') + H_{\text{int},1}(\tau''))}\right] \nonumber \\
  &= (2\pi)^3 \delta(\bfq + \bfq') \frac{2\pi^2}{q^3} \left[ \mathcal P_{\delta \phi,\tre}(q,\tau) + \mathcal P_{\delta \phi, 1\vx}(q,\tau) + \mathcal P_{\delta \phi,\tad}(q,\tau) + \mathcal P_{\delta \phi, 2\vx}(q,\tau) \right].
  \label{eq:two_vx_one_loop}
\end{align}
The tree-level power spectrum is given by 
\begin{align}
  \mathcal P_{\delta \phi,\tre}(q,\tau) = \frac{q^3}{2\pi^2}|u_q(\tau)|^2.
\end{align}
The one-vertex contribution comes from $H_{\inte,4}$ and $H_{\inte,2}$:
\begin{align}
  \mathcal P_{\delta \phi, 1\vx}(q,\tau) 
  &=  \frac{q^3}{\pi^2} \int^\tau_{-\infty} \dd \tau' a^4(\tau') \Im\left[ u_q(\tau) u^*_q(\tau') \right] \Re\left[ u_q(\tau) u^*_q(\tau') \right] \left(V_\foo(\tau')\int\frac{\dd^3 k}{(2\pi)^3} |u_k(\tau')|^2 + 2V_{c,\so}(\tau')\right).
  \label{eq:1vx}
\end{align}
Because of the cutoff of $u_k$ in Eq.~(\ref{eq:uk_cutoff}), we can express the wavenumber integral as
\begin{align}
  \int\frac{\dd^3 k}{(2\pi)^3} |u_k(\tau')|^2 = \int^{k_\uv \frac{a(\tau')}{a_i}}_{k_\ir} \frac{k^2 \dd k}{2\pi^2} |u_k(\tau')|^2.
\end{align}
In contrast to our previous works~\cite{Inomata:2024lud,Inomata:2025bqw,Inomata:2025pqa}, we do not impose $k_\ir \gg q$. Instead, we impose $k_\ir < q$ to take into account the contributions from $k \lesssim q$.
The second line of Eq.~(\ref{eq:two_vx_one_loop}) leads to $\mathcal P_{\tad}$ and $\mathcal P_{2\vx}$, where
\begin{align}
  \mathcal P_{\delta \phi, \tad}(q,\tau) 
  &= \frac{2q^3}{\pi^2} \int^\tau_{-\infty} \dd \tau' \int^{\tau'}_{-\infty} \dd \tau''a^4(\tau') a^4(\tau'') V_\tho(\tau') \Im\left[ u_q(\tau) u^*_q(\tau') \right] \Re\left[ u_q(\tau) u^*_q(\tau') \right] \nonumber \\
  &\qquad \times \Im[ u_{0}(\tau') u_{0}^*(\tau'')]  \left( V_\tho(\tau'') \int \frac{\dd^3 k}{(2\pi)^3} |u_{k}(\tau'')|^2 + 2 V_{c,\fo}(\tau'')\right),
  \label{eq:tad}
\end{align}
\begin{align}
  \mathcal P_{\delta \phi, 2\vx}(q,\tau) 
  &= \mathcal P^a_{\delta \phi, 2\vx}(q,\tau) + \mathcal P^b_{\delta \phi, 2\vx}(q,\tau),
  \label{eq:2vx}
\end{align}
\begin{align}
  \label{eq:p_a}
  \mathcal P^a_{\delta \phi, 2\vx}(q,\tau) &= \frac{q^3}{\pi^2}  \int^\tau_{-\infty} \dd \tau' \int^{\tau}_{-\infty} \dd \tau'' a^4(\tau')  a^4(\tau'') V_\tho(\tau') V_\tho(\tau'') \Im[u_q(\tau) u_q^*(\tau')] \Im[u_q(\tau) u_q^*(\tau'')] \nonumber \\
  &\qquad \qquad \qquad \times \int \frac{\dd^3 k}{(2\pi)^3} \Re[u_k(\tau')u^*_k(\tau'')u_{|\bfq - \bfk|}(\tau')u^*_{|\bfq - \bfk|}(\tau'')], \\
  \label{eq:p_b}  
  \mathcal P^b_{\delta \phi, 2\vx}(q,\tau) 
  &= \frac{4q^3}{\pi^2} \int^\tau_{-\infty} \dd \tau' \int^{\tau'}_{-\infty} \dd \tau'' a^4(\tau') 
  a^4(\tau'') V_\tho(\tau') V_\tho(\tau'') \Im[u_q(\tau) u_q^*(\tau')] \nonumber \\
  &\qquad \qquad \qquad  \times  \int \frac{\dd^3 k}{(2\pi)^3} \Im[u_{|\bfq-\bfk|}(\tau')u^*_{|\bfq-\bfk|}(\tau'')] \Re[u_q(\tau) u_q^*(\tau'')] \Re[u_{k}(\tau')u^*_{k}(\tau'')].
\end{align}

\begin{figure}[t]
\centering

\begin{minipage}{0.45\textwidth}
\centering

\begin{tikzpicture}
\begin{feynman}

\vertex (a) at (-3,0);
\vertex (b) at (-1.5,0);
\vertex (c) at (0,0);
\vertex (e) at (0.0,1.3);
\vertex (f) at (3,0);


\diagram*[large] {

(a) 
    -- [anti fermion] (c)
    -- [anti majorana] (f),

(c) -- [anti fermion, half left, looseness=1.7] (e)
    -- [fermion, half left, looseness=1.7] (c),
};
     \vertex [crossed dot, fill=white] (i) at ($(a)!0.5!(f) + (0, 1.3)$) {};      

      \vertex [crossed dot, fill=white] (i) at ($(a)!0.5!(f) - (-1.55, 0)$) {};
      \vertex [] (i) at ($(a)!0.5!(f) - (1.6, 0.3)$) {\(\mathbf{q}\)}; 
      
      \vertex [] (i) at ($(a)!0.5!(f) - (-1., 0.3)$) {\( \mathbf{q}\)};
      \vertex [] (i) at ($(a)!0.5!(f) - (-2, 0.3)$) {\( -\mathbf{q}\)};
      
      \vertex [] (i) at ($(a)!0.5!(f) - (-1., -0.7)$) {\(-\mathbf{k}\)};
      \vertex [] (i) at ($(a)!0.5!(f) - (1., -0.7)$) {\(\mathbf{k}\)};   

      \vertex [] (i) at ($(a)!0.5!(f) + (3, 0.2)$) {\(\mathbf{\tau}\)}; 
      \vertex [] (i) at ($(a)!0.5!(f) + (-3, 0.2)$) {\(\mathbf{\tau}\)}; 
      \vertex [] (i) at ($(a)!0.5!(f) + (0, 0.2)$) {\(\mathbf{\tau'}\)}; 

\end{feynman}
\end{tikzpicture}

\end{minipage}
\begin{minipage}{0.45\textwidth}
\centering

\begin{tikzpicture}
\begin{feynman}

\vertex (a) at (-3,0);
\vertex (b) at (-1.5,0);
\vertex (c) at (0,0);
\vertex (e) at (0.0,1.3);
\vertex (f) at (3,0);

\diagram*[large]{

(a) 
    -- [anti majorana] (c)
    -- [fermion] (f),

(c) -- [anti fermion, half left, looseness=1.7] (e)
    -- [fermion, half left, looseness=1.7] (c),
};
     \vertex [crossed dot, fill=white] (i) at ($(a)!0.5!(f) + (0, 1.3)$) {};      

      \vertex [crossed dot, fill=white] (i) at ($(a)!0.5!(f) + (-1.45, 0)$) {};
      \vertex [] (i) at ($(a)!0.5!(f) + (1.6, -0.3)$) {\(-\mathbf{q}\)}; 
      
      \vertex [] (i) at ($(a)!0.5!(f) + (-1., -0.3)$) {\(-\mathbf{q}\)};
      \vertex [] (i) at ($(a)!0.5!(f) + (-2, -0.3)$) {\( \mathbf{q}\)};
      
      \vertex [] (i) at ($(a)!0.5!(f) - (-1., -0.7)$) {\(-\mathbf{k}\)};
      \vertex [] (i) at ($(a)!0.5!(f) - (1., -0.7)$) {\(\mathbf{k}\)};   

      \vertex [] (i) at ($(a)!0.5!(f) + (3, 0.2)$) {\(\mathbf{\tau}\)}; 
      \vertex [] (i) at ($(a)!0.5!(f) + (-3, 0.2)$) {\(\mathbf{\tau}\)}; 
      \vertex [] (i) at ($(a)!0.5!(f) + (0, 0.2)$) {\(\mathbf{\tau'}\)}; 

\end{feynman}
  \end{tikzpicture}
\end{minipage}

\vspace{5pt}

\begin{minipage}{0.45\textwidth}
\centering

\begin{tikzpicture}
\begin{feynman}

\vertex (a) at (-3,0);
\vertex (b) at (-1.5,0);
\vertex (c) at (0,0);

\vertex (f) at (3,0);


\diagram*[large] {

(a) 
    -- [anti fermion] (c)
    -- [anti majorana] (f),
};
     \vertex [crossed dot, fill=black] (i) at ($(a)!0.5!(f) + (0, 0)$) {};           

      \vertex [crossed dot, fill=white] (i) at ($(a)!0.5!(f) - (-1.55, 0)$) {};
      \vertex [] (i) at ($(a)!0.5!(f) - (1.6, 0.3)$) {\(\mathbf{q}\)}; 
      
      \vertex [] (i) at ($(a)!0.5!(f) - (-1., 0.3)$) {\( \mathbf{q}\)};
      \vertex [] (i) at ($(a)!0.5!(f) - (-2, 0.3)$) {\( -\mathbf{q}\)};
      
      \vertex [] (i) at ($(a)!0.5!(f) + (3, 0.2)$) {\(\mathbf{\tau}\)}; 
      \vertex [] (i) at ($(a)!0.5!(f) + (-3, 0.2)$) {\(\mathbf{\tau}\)}; 
      \vertex [] (i) at ($(a)!0.5!(f) + (0, 0.4)$) {\(\mathbf{\tau'}\)}; 

\end{feynman}
\end{tikzpicture}

\end{minipage}
\begin{minipage}{0.45\textwidth}
\centering

\begin{tikzpicture}
\begin{feynman}

\vertex (a) at (-3,0);
\vertex (b) at (-1.5,0);
\vertex (c) at (0,0);

\vertex (f) at (3,0);

\diagram*[large]{

(a) 
    -- [anti majorana] (c)
    -- [fermion] (f),
};
     \vertex [crossed dot, fill=black] (i) at ($(a)!0.5!(f) + (0, 0)$) {};      

      \vertex [crossed dot, fill=white] (i) at ($(a)!0.5!(f) + (-1.45, 0)$) {};
      \vertex [] (i) at ($(a)!0.5!(f) + (1.6, -0.3)$) {\(-\mathbf{q}\)}; 
      
      \vertex [] (i) at ($(a)!0.5!(f) + (-1., -0.3)$) {\(-\mathbf{q}\)};
      \vertex [] (i) at ($(a)!0.5!(f) + (-2, -0.3)$) {\( \mathbf{q}\)};

      \vertex [] (i) at ($(a)!0.5!(f) + (3, 0.2)$) {\(\mathbf{\tau}\)}; 
      \vertex [] (i) at ($(a)!0.5!(f) + (-3, 0.2)$) {\(\mathbf{\tau}\)}; 
      \vertex [] (i) at ($(a)!0.5!(f) + (0, 0.4)$) {\(\mathbf{\tau'}\)}; 

\end{feynman}
  \end{tikzpicture}
\end{minipage}
\caption{Diagrams for the one-vertex contribution, $\mathcal P_{\delta \phi, 1\vx}$.}
\label{fig:1vx}

\end{figure}
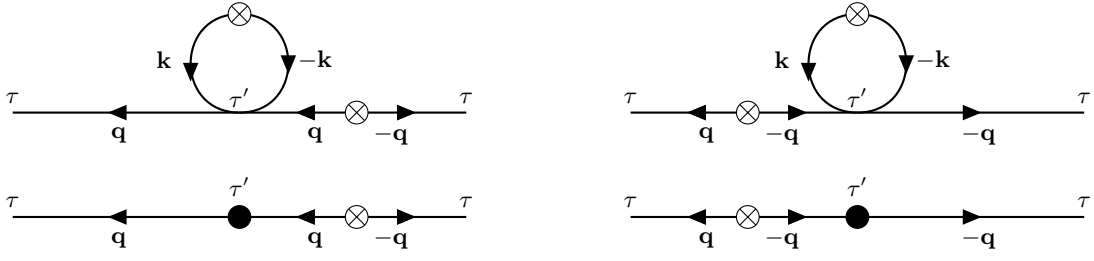

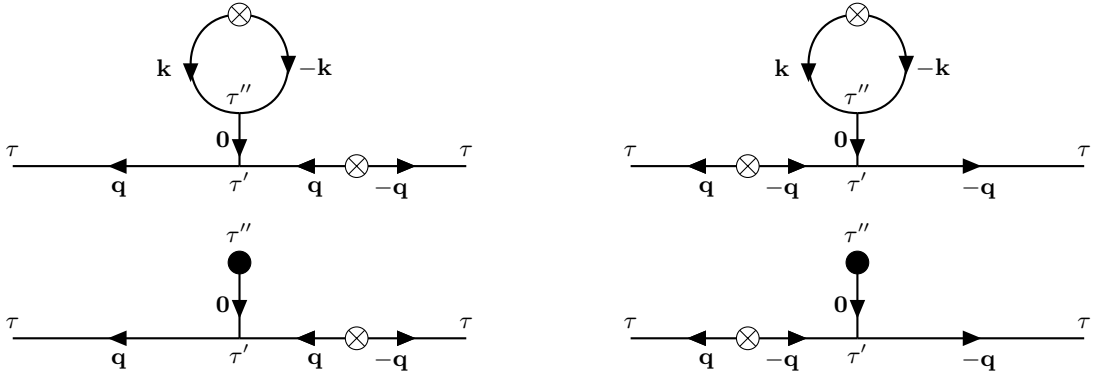
\begin{figure}[t]
\centering

\begin{minipage}{0.45\textwidth}
\centering

\begin{tikzpicture} 
\begin{feynman}

\vertex (a) at (-3,0);
\vertex (b) at (-1.5,0);
\vertex (c) at (0,0);
\vertex (d) at (0.0,0.7);
\vertex (e) at (0.0,2.);

\vertex (f) at (3,0);


\diagram*[large] {

(a) 
    -- [anti fermion] (c)
    -- [anti majorana] (f),

(c) -- [anti fermion] (d) -- [anti fermion, half left, looseness=1.7] (e)
    -- [fermion, half left, looseness=1.7] (d),
};
     \vertex [crossed dot, fill=white] (i) at ($(a)!0.5!(f) + (0, 2.)$) {};      

      \vertex [crossed dot, fill=white] (i) at ($(a)!0.5!(f) - (-1.55, 0)$) {};
      \vertex [] (i) at ($(a)!0.5!(f) - (1.6, 0.3)$) {\(\mathbf{q}\)}; 
      
      \vertex [] (i) at ($(a)!0.5!(f) - (-1., 0.3)$) {\( \mathbf{q}\)};
      \vertex [] (i) at ($(a)!0.5!(f) - (-2, 0.3)$) {\( -\mathbf{q}\)};
      
      \vertex [] (i) at ($(a)!0.5!(f) - (-1., -1.3)$) {\(-\mathbf{k}\)};
      \vertex [] (i) at ($(a)!0.5!(f) - (1., -1.3)$) {\(\mathbf{k}\)};   

      \vertex [] (i) at ($(a)!0.5!(f) + (-0.22, 0.35)$) {\(\mathbf{0}\)}; 

      \vertex [] (i) at ($(a)!0.5!(f) + (3, 0.2)$) {\(\mathbf{\tau}\)}; 
      \vertex [] (i) at ($(a)!0.5!(f) + (-3, 0.2)$) {\(\mathbf{\tau}\)}; 
      \vertex [] (i) at ($(a)!0.5!(f) + (0, -0.2)$) {\(\mathbf{\tau'}\)}; 
      \vertex [] (i) at ($(a)!0.5!(f) + (0, 0.95)$) {\(\mathbf{\tau''}\)}; 

\end{feynman}
\end{tikzpicture}

\end{minipage}
\begin{minipage}{0.45\textwidth}
\centering

\begin{tikzpicture} 
\begin{feynman}

\vertex (a) at (-3,0);
\vertex (b) at (-1.5,0);
\vertex (c) at (0,0);
\vertex (d) at (0.0,0.7);
\vertex (e) at (0.0,2.);

\vertex (f) at (3,0);

\diagram*[large]{

(a) 
    -- [anti majorana] (c)
    -- [fermion] (f),

(c) -- [anti fermion] (d) -- [anti fermion, half left, looseness=1.7] (e)
    -- [fermion, half left, looseness=1.7] (d),
};
     \vertex [crossed dot, fill=white] (i) at ($(a)!0.5!(f) + (0, 2.)$) {};      

      \vertex [crossed dot, fill=white] (i) at ($(a)!0.5!(f) + (-1.45, 0)$) {};
      \vertex [] (i) at ($(a)!0.5!(f) + (1.6, -0.3)$) {\(-\mathbf{q}\)}; 
      
      \vertex [] (i) at ($(a)!0.5!(f) + (-1., -0.3)$) {\(-\mathbf{q}\)};
      \vertex [] (i) at ($(a)!0.5!(f) + (-2, -0.3)$) {\( \mathbf{q}\)};
      
      \vertex [] (i) at ($(a)!0.5!(f) - (-1., -1.3)$) {\(-\mathbf{k}\)};
      \vertex [] (i) at ($(a)!0.5!(f) - (1., -1.3)$) {\(\mathbf{k}\)};   

      \vertex [] (i) at ($(a)!0.5!(f) + (-0.22, 0.35)$) {\(\mathbf{0}\)}; 

      \vertex [] (i) at ($(a)!0.5!(f) + (3, 0.2)$) {\(\mathbf{\tau}\)}; 
      \vertex [] (i) at ($(a)!0.5!(f) + (-3, 0.2)$) {\(\mathbf{\tau}\)}; 
      \vertex [] (i) at ($(a)!0.5!(f) + (0, -0.2)$) {\(\mathbf{\tau'}\)}; 
      \vertex [] (i) at ($(a)!0.5!(f) + (0, 0.95)$) {\(\mathbf{\tau''}\)};       

\end{feynman}
  \end{tikzpicture}
\end{minipage}

\vspace{2pt}

\begin{minipage}{0.45\textwidth}
\centering

\begin{tikzpicture} 
\begin{feynman}

\vertex (a) at (-3,0);
\vertex (b) at (-1.5,0);
\vertex (c) at (0,0);
\vertex (d) at (0.0,1.0);
\vertex (e) at (0.0,2.7);

\vertex (f) at (3,0);


\diagram*[large] {

(a) 
    -- [anti fermion] (c)
    -- [anti majorana] (f),

(c) -- [anti fermion] (d), 
};
     \vertex [crossed dot, fill=black] (i) at ($(a)!0.5!(f) + (0, 1.0)$) {};           

      \vertex [crossed dot, fill=white] (i) at ($(a)!0.5!(f) - (-1.55, 0)$) {};
      \vertex [] (i) at ($(a)!0.5!(f) - (1.6, 0.3)$) {\(\mathbf{q}\)}; 
      
      \vertex [] (i) at ($(a)!0.5!(f) - (-1., 0.3)$) {\( \mathbf{q}\)};
      \vertex [] (i) at ($(a)!0.5!(f) - (-2, 0.3)$) {\( -\mathbf{q}\)};

      \vertex [] (i) at ($(a)!0.5!(f) + (-0.22, 0.45)$) {\(\mathbf{0}\)}; 

      \vertex [] (i) at ($(a)!0.5!(f) + (3, 0.2)$) {\(\mathbf{\tau}\)}; 
      \vertex [] (i) at ($(a)!0.5!(f) + (-3, 0.2)$) {\(\mathbf{\tau}\)}; 
      \vertex [] (i) at ($(a)!0.5!(f) + (0, -0.2)$) {\(\mathbf{\tau'}\)}; 
      \vertex [] (i) at ($(a)!0.5!(f) + (0, 1.4)$) {\(\mathbf{\tau''}\)}; 

\end{feynman}
\end{tikzpicture}

\end{minipage}
\begin{minipage}{0.45\textwidth}
\centering

\begin{tikzpicture} 
\begin{feynman}

\vertex (a) at (-3,0);
\vertex (b) at (-1.5,0);
\vertex (c) at (0,0);
\vertex (d) at (0.0,1.0);

\vertex (f) at (3,0);

\diagram*[large]{

(a) 
    -- [anti majorana] (c)
    -- [fermion] (f),

(c) -- [anti fermion] (d), 
};
     \vertex [crossed dot, fill=black] (i) at ($(a)!0.5!(f) + (0, 1.0)$) {};      

      \vertex [crossed dot, fill=white] (i) at ($(a)!0.5!(f) + (-1.45, 0)$) {};
      \vertex [] (i) at ($(a)!0.5!(f) + (1.6, -0.3)$) {\(-\mathbf{q}\)}; 
      
      \vertex [] (i) at ($(a)!0.5!(f) + (-1., -0.3)$) {\(-\mathbf{q}\)};
      \vertex [] (i) at ($(a)!0.5!(f) + (-2, -0.3)$) {\( \mathbf{q}\)};
      
      \vertex [] (i) at ($(a)!0.5!(f) + (-0.22, 0.45)$) {\(\mathbf{0}\)}; 

      \vertex [] (i) at ($(a)!0.5!(f) + (3, 0.2)$) {\(\mathbf{\tau}\)}; 
      \vertex [] (i) at ($(a)!0.5!(f) + (-3, 0.2)$) {\(\mathbf{\tau}\)}; 
      \vertex [] (i) at ($(a)!0.5!(f) + (0, -0.2)$) {\(\mathbf{\tau'}\)}; 
      \vertex [] (i) at ($(a)!0.5!(f) + (0, 1.4)$) {\(\mathbf{\tau''}\)};       

\end{feynman}
  \end{tikzpicture}
\end{minipage}

\caption{Diagrams for the tadpole contribution, $\mathcal P_{\delta \phi, \tad}$.}
\label{fig:tad}
\end{figure}

\begin{figure}[t]
\centering

\begin{minipage}{0.45\textwidth}
\centering

\begin{tikzpicture}
\begin{feynman}

\vertex (a) at (-3,0);
\vertex (b) at (-0.7,0);
\vertex (c) at (0.7,0);

\vertex (f) at (3,0);


\diagram*[large] {

(a) 
    -- [anti fermion] (b),
(c) -- [anti majorana] (f),

(b) -- [anti majorana, half left, looseness=1.7] (c)
    -- [fermion, half left, looseness=1.7] (b),
};
      \vertex [crossed dot, fill=white] (i) at ($(a)!0.5!(f) + (0, 0.7)$) {};      
      \vertex [crossed dot, fill=white] (i) at ($(a)!0.5!(f) - (-1.85, 0)$) {};
      \vertex [] (i) at ($(a)!0.5!(f) - (1.9, 0.3)$) {\(\mathbf{q}\)};      
      \vertex [] (i) at ($(a)!0.5!(f) - (-2.3, 0.3)$) {\( -\mathbf{q}\)};
      \vertex [] (i) at ($(a)!0.5!(f) - (-1.3, 0.3)$) {\( \mathbf{q}\)};      
      \vertex [] (i) at ($(a)!0.5!(f) - (-0.7, -0.75)$) {\(-\mathbf{k}\)};
      \vertex [] (i) at ($(a)!0.5!(f) - (0.7, -0.75)$) {\(\mathbf{k}\)};   
      \vertex [] (i) at ($(a)!0.5!(f) - (-0.03, 0.9)$) {\(\mathbf{q}-\mathbf{k}\)}; 

      \vertex [] (i) at ($(a)!0.5!(f) + (3, 0.2)$) {\(\mathbf{\tau}\)}; 
      \vertex [] (i) at ($(a)!0.5!(f) + (-3, 0.2)$) {\(\mathbf{\tau}\)}; 
      \vertex [] (i) at ($(a)!0.5!(f) + (0.5, 0.)$) {\(\mathbf{\tau''}\)}; 
      \vertex [] (i) at ($(a)!0.5!(f) + (-0.45, 0.)$) {\(\mathbf{\tau'}\)}; 

\end{feynman}
\end{tikzpicture}

\end{minipage}
\begin{minipage}{0.45\textwidth}
\centering

\begin{tikzpicture}
\begin{feynman}

\vertex (a) at (-3,0);
\vertex (b) at (-0.7,0);
\vertex (c) at (0.7,0);

\vertex (f) at (3,0);


\diagram*[large] {

(a) 
    -- [anti majorana] (b),
(c) -- [fermion] (f),

(b) -- [anti majorana, half left, looseness=1.7] (c)
    -- [anti fermion, half left, looseness=1.7] (b),
};
      \vertex [crossed dot, fill=white] (i) at ($(a)!0.5!(f) + (0, 0.7)$) {};      
      \vertex [crossed dot, fill=white] (i) at ($(a)!0.5!(f) - (1.85, 0)$) {};

      \vertex [] (i) at ($(a)!0.5!(f) - (-1.9, 0.3)$) {\(-\mathbf{q}\)};      
      \vertex [] (i) at ($(a)!0.5!(f) - (2.3, 0.3)$) {\( \mathbf{q}\)};
      \vertex [] (i) at ($(a)!0.5!(f) - (1.3, 0.3)$) {\( -\mathbf{q}\)};  

      \vertex [] (i) at ($(a)!0.5!(f) - (-0.7, -0.75)$) {\(-\mathbf{k}\)};
      \vertex [] (i) at ($(a)!0.5!(f) - (0.7, -0.75)$) {\(\mathbf{k}\)};   
      \vertex [] (i) at ($(a)!0.5!(f) - (-0.03, 0.9)$) {\(\mathbf{k}-\mathbf{q}\)}; 

      \vertex [] (i) at ($(a)!0.5!(f) + (3, 0.2)$) {\(\mathbf{\tau}\)}; 
      \vertex [] (i) at ($(a)!0.5!(f) + (-3, 0.2)$) {\(\mathbf{\tau}\)}; 
      \vertex [] (i) at ($(a)!0.5!(f) + (0.5, 0.)$) {\(\mathbf{\tau'}\)}; 
      \vertex [] (i) at ($(a)!0.5!(f) + (-0.45, 0.)$) {\(\mathbf{\tau''}\)}; 

\end{feynman}
\end{tikzpicture}

\end{minipage}
\caption{Diagrams for the cut-in-the-side contribution, $\mathcal P_{\delta \phi, 2\vx}^b$.}
\label{fig:2vx_b}

\end{figure}
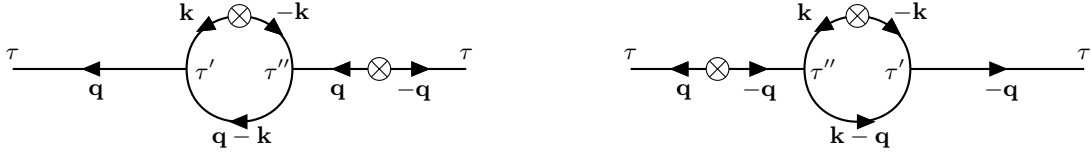

The Feynman diagrams for these one-loop power spectra are shown in Figs.~\ref{fig:1vx}-\ref{fig:2vx_b}.
The lines and vertices in the diagrams are\footnote{
  The simple arrow line actually corresponds to the retarded Green function for $u_q(\tau)$: $g_q(\tau;\tau') = -2\Theta(\tau-\tau') a^2(\tau') \Im[u_q(\tau) u_q^*(\tau')]$.
  See Appendix B of Ref.~\cite{Inomata:2022yte} for the expressions of $\mathcal P_{\delta \phi}$ with the Green function.
}
\begin{eqnarray}
   &\begin{tikzpicture} [baseline={(0,-0.1)}]
  \begin{feynman}

\vertex (a) at (-1.,0);
\vertex (c) at (1.,0);

\diagram*[large] {
(a) -- [fermion] (c)
};
      \vertex [] (i) at ($(a)!0.5!(c) - (0., 0.3)$) {\(\mathbf{q}\)};
      \vertex [] (i) at ($(a)!0.5!(c) + (-1., 0.2)$) {\(\mathbf{\tau'}\)}; 
      \vertex [] (i) at ($(a)!0.5!(c) + (1., 0.2)$) {\(\mathbf{\tau}\)}; 

      \end{feynman}
  \end{tikzpicture}
  = -2\Theta(\tau-\tau') a^2(\tau') \Im[u_q(\tau) u_q^*(\tau')], \qquad 
  \begin{tikzpicture} [baseline={(0,-0.1)}]
  \begin{feynman}

\vertex (a) at (-1.5,0);
\vertex (c) at (0,0);
\vertex (f) at (1.5,0);

\diagram*[large] {
(a) -- [anti fermion] (c)
    -- [fermion] (f),
};
      \vertex [crossed dot, fill=white] (i) at ($(a)!0.5!(f) + (0, 0)$) {};      
      \vertex [] (i) at ($(a)!0.5!(f) - (0.8, 0.3)$) {\(\mathbf{q}\)};
      \vertex [] (i) at ($(a)!0.5!(f) - (-0.8, 0.3)$) {\( -\mathbf{q}\)};            

      \vertex [] (i) at ($(a)!0.5!(f) + (-1.5, 0.2)$) {\(\mathbf{\tau}\)}; 
      \vertex [] (i) at ($(a)!0.5!(f) + (1.5, 0.2)$) {\(\mathbf{\tau'}\)}; 

      \end{feynman}
  \end{tikzpicture}
  = \Re[u_q(\tau) u^*_q(\tau')], \quad 
\\
&\begin{tikzpicture} [baseline={(0,-0.1)}]
\begin{feynman}

\vertex (a) at (-1.3,0);
\vertex (b) at (-.7,0);
\vertex (c) at (0.7,0);

\vertex (f) at (1.3,0);

\vertex (g) at (0,1);
\vertex (h) at (0,-1);


\diagram*[large] {

(a) 
    --  (b),
(c) --  (f),

(b) -- [anti majorana, half left, looseness=1.7] (c)
    -- [anti majorana, half left, looseness=1.7] (b),

(g) -- [dashed] (h),
};
      \vertex [crossed dot, fill=white] (i) at ($(a)!0.5!(f) + (0, 0.7)$) {};      
      \vertex [crossed dot, fill=white] (i) at ($(a)!0.5!(f) - (0, 0.7)$) {};
      \vertex [] (i) at ($(a)!0.5!(f) - (-0.7, -0.75)$) {\(-\mathbf{k}\)};
      \vertex [] (i) at ($(a)!0.5!(f) - (0.7, -0.75)$) {\(\mathbf{k}\)};
      \vertex [] (i) at ($(a)!0.5!(f) - (-0.8, 0.85)$) {\(\mathbf{k}-\mathbf q\)};
      \vertex [] (i) at ($(a)!0.5!(f) - (0.8, 0.85)$) {\(\mathbf{q}-\mathbf{k}\)};            

      \vertex [] (i) at ($(a)!0.5!(f) + (0.5, 0.)$) {\(\mathbf{\tau''}\)}; 
      \vertex [] (i) at ($(a)!0.5!(f) + (-0.45, 0.)$) {\(\mathbf{\tau'}\)}; 
\end{feynman}
\end{tikzpicture}
= \Re[u_k(\tau')u^*_k(\tau'') u_{|\bfq - \bfk|}(\tau')u^*_{|\bfq - \bfk|}(\tau'')], 
\\
&\begin{tikzpicture} [baseline={(0,-0.1)}]
  \begin{feynman}

\vertex (a) at (-0.5,0);
\vertex (d) at (0.,0);

\diagram*[large] {
(a)-- (d)
};
     \vertex [crossed dot, fill=black] (i) at ($(a)!0.5!(d) + (0.4, 0.)$) {};  
      \vertex [] (i) at ($(a)!0.5!(d) + (0.4, 0.3)$) {\(\mathbf{\tau}\)};
      \end{feynman}
  \end{tikzpicture}
  = -a^2(\tau) V_{c,\fo}(\tau), \quad     
  \begin{tikzpicture} [baseline={(0,-0.1)}]
  \begin{feynman}

\vertex (a) at (-0.5,0);
\vertex (d) at (0.5,0);

\diagram*[large] {
(a)-- (d)
};
     \vertex [crossed dot, fill=black] (i) at ($(a)!0.5!(d) + (0, 0)$) {};  
      \vertex [] (i) at ($(a)!0.5!(d) + (-0., 0.3)$) {\(\mathbf{\tau}\)};
      \end{feynman}
  \end{tikzpicture}
  = -a^2(\tau) V_{c,\so}(\tau), \quad   
    \begin{tikzpicture} [baseline={(0,0.2)}]
  \begin{feynman}

\vertex (a) at (-1*0.5,0);
\vertex (b) at (0,1.732*0.3333*0.5);
\vertex (c) at (0,1.732*0.5);
\vertex (d) at (1*0.5,0);

\diagram*[large] {
(a) -- (b) -- {(c),(d)}
};
      \vertex [] (i) at ($(a)!0.5!(d) + (-0.2, 0.4)$) {\(\mathbf{\tau}\)};
      \end{feynman}
  \end{tikzpicture}
  = -a^2(\tau) V_\tho(\tau), \quad  
  \begin{tikzpicture} [baseline={(0,-0.1)}]
  \begin{feynman}

\vertex (a) at (-0.4,-0.4);
\vertex (b) at (-0.4,0.4);
\vertex (c) at (0,0);
\vertex (d) at (0.4,0.4);
\vertex (e) at (0.4,-0.4);

\diagram*[large] {
{(a),(b)} -- (c) -- {(d),(e)}
};
      \vertex [] (i) at ($(a)!0.5!(d) + (-0., 0.3)$) {\(\mathbf{\tau}\)};
      \end{feynman}
  \end{tikzpicture}
  = -a^2(\tau) V_\foo(\tau),
\end{eqnarray}
where we have used markers similar to those employed in Ref.~\cite{Crocce:2005xy} (see also Ref.~\cite{Ema:2024hkj} for the cutting rule in the in-in formalism).
To obtain each $\mathcal P_{\delta \phi}$, we multiply the products of the lines and vertices in its diagram by $q^3/(2\pi^2)$, $\int \dd^3 k/(2\pi)^3$ (if a loop exists), $\int^\tau_{-\infty} \dd \tau' \int^\tau_{-\infty} \dd \tau''$ (if $\tau''$ does not exist in the diagram, omit $\int \dd \tau''$), and $1/S$ with $S$ the symmetry factor of the diagram.
From the shapes of the diagrams, we call $\mathcal P^a_{2\vx}$ and $\mathcal P^b_{2\vx}$ the cut-in-the-middle and the cut-in-the-side contributions, respectively.

Next, we summarize the expression of $\langle \delta \dot \phi \rangle$.
Taking the time derivative of Eq.~(\ref{eq:dt_one_pt}) and doing some transformations, we finally obtain~\cite{Inomata:2025bqw} 
\begin{align}
  \langle \delta \dot \phi(\tau) \rangle
   &= \langle \delta \dot \phi(\tau) \rangle_{1\vx} + \langle \delta \dot \phi(\tau) \rangle_\tad  + \langle \delta \dot \phi(\tau) \rangle_{\overline{2\vx}} + \langle \delta \dot \phi(\tau) \rangle_{\ir},
  \label{eq:dt_one_pt_g2}
\end{align}
where 
\begin{align}
&  \langle \delta \dot \phi(\tau) \rangle_{1\vx} 
  = \int^\tau_{-\infty} \dd \tau' a^4(\tau') \Im[u_0(\tau)u_0^*(\tau')]\dot{\bar \phi}(\tau') \left[ V_\foo(\tau') \int \frac{\dd^3 k}{(2\pi)^3} |u_{k}(\tau')|^2 + 2V_{c,\so}(\tau') \right],
  \label{eq:br_1vx} \\
 &\langle \delta \dot \phi(\tau) \rangle_\tad \nonumber \\ 
& = 2 \int^\tau_{-\infty} \dd \tau'\int^{\tau'}_{-\infty} \dd \tau''\, a^4(\tau') a^4(\tau'')\Im[u_0(\tau) u^*_0(\tau')] V_\tho(\tau') \dot {\bar\phi}(\tau') \Im[u_0(\tau')u^*_0(\tau'')] \nonumber \\ 
& \qquad \qquad \qquad \qquad \qquad \qquad \qquad \qquad \qquad 
\times \left[ V_\tho(\tau'') \int \frac{\dd^3 k}{(2\pi)^3} |u_{k}(\tau'')|^2 + 2 V_{c,\fo}(\tau'') \right],
\label{eq:tad_back} \\
  &\langle \delta \dot \phi(\tau) \rangle_{\overline{2\vx}}\nonumber \\ 
  &= 4 \int^\tau_{-\infty} \dd \tau' \int^{\tau'}_{-\infty} \dd \tau'' a^4(\tau') a^4(\tau'') V_\tho(\tau') V_\tho(\tau'') \dot{\bar \phi}(\tau'')\Im[u_0(\tau) u_0^*(\tau')] \int \frac{\dd^3 k}{(2\pi)^3} \Im[u_k(\tau') u^*_k(\tau'')] \Re[u_k(\tau') u^*_k(\tau'')], 
  \label{eq:ddot_2vx_bar} \\ 
  &\langle \delta \dot \phi(\tau) \rangle_{\ir} = \int^\tau_{-\infty} \dd \tau' a^4(\tau')\Im[u_0(\tau) u_0^*(\tau')] H  V_\tho(\tau') \mathcal P_{\delta \phi,\tre}\left( k_\ir,\tau' \right).
    \label{eq:ddot_ir}
\end{align}
Note that those expressions of $\svev{\delta \dot \phi}$ are always valid in contrast to the expression in Eq.~(\ref{eq:dphi_tadpole}), which is valid only in $\tau \geq \taui$ with the specific choice of counter potential, Eq.~(\ref{eq:v_c1_tad_cut}). 


\begin{figure}[t]
\centering

\begin{minipage}{0.45\textwidth}
\centering

\begin{tikzpicture}
\begin{feynman}

\vertex (a) at (-3,0);
\vertex (b) at (-0.7,0);
\vertex (c) at (0.7,0);

\vertex (f) at (3,0);

\vertex (g) at (0,1);
\vertex (h) at (0,-1);


\diagram*[large] {

(a) 
    -- [anti fermion] (b),
(c) -- [fermion] (f),

(b) -- [anti majorana, half left, looseness=1.7] (c)
    -- [anti majorana, half left, looseness=1.7] (b),

(g) -- [dashed] (h),
};
      \vertex [crossed dot, fill=white] (i) at ($(a)!0.5!(f) + (0, 0.7)$) {};      
      \vertex [crossed dot, fill=white] (i) at ($(a)!0.5!(f) - (0, 0.7)$) {};
      \vertex [] (i) at ($(a)!0.5!(f) - (1.9, 0.3)$) {\(\mathbf{q}\)};
      \vertex [] (i) at ($(a)!0.5!(f) - (-1.9, 0.3)$) {\( -\mathbf{q}\)};            
      \vertex [] (i) at ($(a)!0.5!(f) - (-0.7, -0.75)$) {\(-\mathbf{k}\)};
      \vertex [] (i) at ($(a)!0.5!(f) - (0.7, -0.75)$) {\(\mathbf{k}\)};
      \vertex [] (i) at ($(a)!0.5!(f) - (-0.8, 0.85)$) {\(\mathbf{k}-\mathbf q\)};
      \vertex [] (i) at ($(a)!0.5!(f) - (0.8, 0.85)$) {\(\mathbf{q}-\mathbf{k}\)};            

      \vertex [] (i) at ($(a)!0.5!(f) + (3, 0.2)$) {\(\mathbf{\tau}\)}; 
      \vertex [] (i) at ($(a)!0.5!(f) + (-3, 0.2)$) {\(\mathbf{\tau}\)}; 
      \vertex [] (i) at ($(a)!0.5!(f) + (0.5, 0.)$) {\(\mathbf{\tau''}\)}; 
      \vertex [] (i) at ($(a)!0.5!(f) + (-0.45, 0.)$) {\(\mathbf{\tau'}\)}; 

\end{feynman}
\end{tikzpicture}

\end{minipage}
\caption{Diagram for the cut-in-the-middle contribution, $\mathcal P_{\delta \phi, 2\vx}^a$.}
\label{fig:2vx_a}

\end{figure}
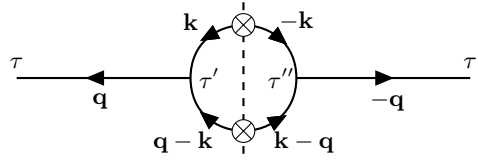


\subsection{Disentangling in-in loops}

Using the above expressions, we disentangle the in-in loop contribution, which can be expressed as
\begin{align}
 H^2\frac{\expval{\delta \phi_{\bfq}(\tau) \delta \phi_{\bfq'}(\tau)}}{\langle \dot{\phi}(\tau) \rangle^2} = \frac{(2\pi)^3 \delta(\bfq + \bfq') H^2 \frac{2\pi^2}{q^3}\mathcal P_{\delta \phi}(q,\tau)}{\left(\dot{\bar \phi}(\tau) + \langle \delta \dot \phi(\tau) \rangle\right)^2} \simeq \frac{(2\pi)^3 \delta(\bfq + \bfq') H^2 \frac{2\pi^2}{q^3}\mathcal P_{\delta \phi,\tre}(q,\tau) \left(1 + \frac{\mathcal P_{\delta \phi}(q,\tau) - \mathcal P_{\delta \phi,\tre}(q,\tau)}{\mathcal P_{\delta \phi,\tre}(q,\tau)} \right)}{ \dot{\bar \phi}^2(\tau) \left(1 + 2\frac{\langle \delta \dot \phi(\tau) \rangle}{\dot {\bar \phi}(\tau)}\right)},
 \label{eq:zeta_cons}
\end{align}
where we have neglected $\svev{\delta \dot \phi}^2$ of the denominator in the second equality because it contributes as a two-loop contribution, and
\begin{align}
  \frac{\mathcal P_{\delta \phi}(q,\tau) - \mathcal P_{\delta \phi, \tre}(q,\tau)}{\mathcal P_{\delta \phi,\tre}(q,\tau)} &= \frac{\mathcal P_{\delta \phi, 1\vx}(q,\tau) + \mathcal P_{\delta \phi,\tad}(q,\tau) + \mathcal P^a_{\delta \phi, 2\vx}(q,\tau) + \mathcal P^b_{\delta \phi, 2\vx}(q,\tau)}{\mathcal P_{\delta \phi,\tre}(q,\tau)}, \\
  \frac{\langle \delta \dot \phi(\tau) \rangle}{\dot {\bar \phi}(\tau)} &= \frac{\langle \delta \dot \phi(\tau) \rangle_{1\vx} + \langle \delta \dot \phi(\tau) \rangle_\tad + \langle \delta \dot \phi(\tau) \rangle_{\overline{2\vx}} + \langle \delta \dot \phi(\tau) \rangle_\text{IR}}{\dot {\bar \phi}(\tau)}.
  \label{eq:cons_cond}
\end{align}
Comparing Eqs.~(\ref{eq:1vx}) and (\ref{eq:br_1vx}), we can see 
\begin{align}
  \frac{\mathcal P_{\delta \phi, 1\vx}(q,\tau)}{\mathcal P_{\delta \phi,\tre}(q,\tau)} -2 \frac{\langle \delta \dot \phi(\tau) \rangle_{1\vx}}{\dot {\bar \phi}(\tau)} &= 2 \int^\tau_{-\infty} \dd \tau' a^4(\tau') \left[V_\foo(\tau')\int\frac{\dd^3 k}{(2\pi)^3} |u_k(\tau')|^2 + 2V_{c,\so}(\tau')\right] \nonumber \\ 
  &\qquad \times \left(\frac{\Im\left[ u_q(\tau) u^*_q(\tau') \right] \Re\left[ u_q(\tau) u^*_q(\tau') \right]}{|u_q(\tau)|^2} -  \frac{\dot{\bar \phi}(\tau')}{\dot{\bar \phi}(\tau)}\Im[u_0(\tau)u_0^*(\tau')] \right) \nonumber \\ 
  &\equiv \Delta_{1\vx}(q,\tau).
  \label{eq:1vx_cond}
\end{align}
The time-integral contributions from $\tau' \geq \taui$ do not contribute to $\Delta_{1\vx}(q,\tau)$ because $u_q(\tau') \propto \dbphi(\tau')$ and $\Im[u_q(\tau) u^*_q(\tau')] = \Im[u_0(\tau) u^*_0(\tau')]$. 
On the other hand, the time-integral contributions from $\tau' < \taui$ can yield a nonzero contribution to $\Delta_{1\vx}(q,\tau)$.
From this observation and $\Im[u_q(\tau) u^*_q(\tau')] \propto \dbphi(\tau)$ when $\tau \geq \tauf$ and $\tau' < \taui$ (Eq.~(\ref{eq:im_cond})), we can see that $\Delta_{1\vx}(q,\tau)$ is constant in $\tau \geq \tauf$.

Similarly, comparing Eqs.~(\ref{eq:tad}) and (\ref{eq:tad_back}), we obtain
\begin{align}
  &\frac{\mathcal P_{\delta \phi, \tad}(q,\tau)}{\mathcal P_{\delta \phi,\tre}(q,\tau)} -2 \frac{\langle \delta \dot \phi(\tau) \rangle_{\tad}}{\dot {\bar \phi}(\tau)} \nonumber \\ 
  &= 4 \int^\tau_{-\infty} \dd \tau' \int^{\tau'}_{-\infty} \dd \tau''a^4(\tau') a^4(\tau'') V_\tho(\tau') \Im[ u_{0}(\tau') u_{0}^*(\tau'')]  \left( V_\tho(\tau'') \int \frac{\dd^3 k}{(2\pi)^3} |u_{k}(\tau'')|^2 + 2 V_{c,\fo}(\tau'')\right) \nonumber \\
  &\qquad \times \left( \frac{\Im\left[ u_q(\tau) u^*_q(\tau') \right] \Re\left[ u_q(\tau) u^*_q(\tau') \right]}{|u_q(\tau)|^2}- \frac{\dot {\bar\phi}(\tau')}{\dbphi(\tau)} \Im[u_0(\tau)u^*_0(\tau')] \right) \nonumber \\ 
  &\equiv \Delta_{\tad}(q,\tau).
  \label{eq:tad_cond}
\end{align}
Similarly to $\Delta_{1\vx}(q,\tau)$, we can easily see that $\Delta_\tad(q,\tau)$ is constant in $\tau \geq \tauf$.

For the two-vertex contributions, we need to be careful because they do not become constant for $\tau \geq \tauf$ in general.
$\mathcal P^a_{\delta \phi, 2\vx}$ is not canceled by any other $\svev{\delta \dot \phi}$ terms, while $\mathcal P^b_{\delta \phi, 2\vx}$ is canceled by $\langle \delta \dot \phi(\tau) \rangle_{\overline{2\vx}}$ only under some conditions.
Let us see the conditional cancellation in the following. 
From Eqs.~(\ref{eq:p_b}) and (\ref{eq:ddot_2vx_bar}), we obtain 
\begin{align}
  &\frac{\mathcal P^b_{\delta \phi, 2\vx}(q,\tau)}{\mathcal P_{\delta \phi,\tre}(q,\tau)} - 2 \frac{\langle \delta \dot \phi(\tau) \rangle_{\overline{2\vx}}}{\dot {\bar \phi}(\tau)} 
  = 8 \int^\tau_{-\infty} \dd \tau' \int^{\tau'}_{-\infty} \dd \tau'' a^4(\tau') a^4(\tau'') V_\tho(\tau') V_\tho(\tau'') \nonumber \\ 
  &\qquad \times \int \frac{\dd^3 k}{(2\pi)^3} \left( \Im[u_q(\tau) u_q^*(\tau')] \Im[u_{|\bfk-\bfq|}(\tau') u^*_{|\bfk-\bfq|}(\tau'')]\frac{\Re[u_q(\tau) u_q^*(\tau'')]}{|u_q(\tau)|^2} \right. \nonumber \\ 
  &\qquad\qquad\qquad\qquad \left. - \frac{\dot{\bar \phi}(\tau'')}{\dot{\bar \phi}(\tau)}\Im[u_0(\tau) u_0^*(\tau')] \Im[u_k(\tau') u^*_k(\tau'')] \right) \Re[u_k(\tau') u^*_k(\tau'')] \nonumber \\   
  &\equiv \Delta_{2\vx(b)}(q,\tau).
  \label{eq:2vx_cond}
\end{align}
If $V_\tho(\tau'') = 0$ during the initial period ($\tau'' < \taui$), we find $\Im[u_q(\tau)u_q^*(\tau')] = \Im[u_0(\tau)u_0^*(\tau')]$ when $V_\tho(\tau') \neq 0$ ($\tau' \geq \tau_i$), where we have neglected the terms suppressed by $q\tau'$ (recall $|q\tau_i| \ll 1$). 
In that case, we further find $\Im[u_{|\bfk - \bfq|}(\tau')u_{|\bfk - \bfq|}^*(\tau'')] = \Im[u_k(\tau')u_k^*(\tau'')]$ when $V_\tho(\tau'') \neq 0$ once we neglect the terms suppressed by $q/k$ or $k\tau''$.
Using these, we can see $\Delta_{2\vx(b)} =0$ if $V_\tho = 0$ during the initial period.
On the other hand, if $V_\tho \neq 0$ during the initial period, we generally find $\Delta_{2\vx(b)} \neq 0$. 
This means that the nonzero $\Delta_{2\vx(b)}$ comes from the time integral over $\tau'' < \taui$.
Based on these discussions, we can reexpress $\Delta_{2\vx(b)}(q,\tau)$ as 
\begin{align}
  &\Delta_{2\vx(b)}(q,\tau) = 8 \int^\tau_{-\infty} \dd \tau' \int^{\min[\tau',\taui]}_{-\infty} \dd \tau'' a^4(\tau') a^4(\tau'') V_\tho(\tau') V_\tho(\tau'') \nonumber \\ 
  &\qquad\qquad \qquad \times \int \frac{\dd^3 k}{(2\pi)^3} \left( \Im[u_q(\tau) u_q^*(\tau')] \Im[u_{|\bfk-\bfq|}(\tau') u^*_{|\bfk-\bfq|}(\tau'')]\frac{\Re[u_q(\tau) u_q^*(\tau'')]}{|u_q(\tau)|^2} \right. \nonumber \\ 
  &\qquad\qquad\qquad\qquad\qquad \qquad  \left. - \frac{\dot{\bar \phi}(\tau'')}{\dot{\bar \phi}(\tau)}\Im[u_0(\tau) u_0^*(\tau')] \Im[u_k(\tau') u^*_k(\tau'')] \right)\Re[u_k(\tau') u^*_k(\tau'')] \nonumber \\   
  &= 8 \Theta(\tau-\tau_i)\int^\tau_{\tau_i} \dd \tau' a^4(\tau')  V_\tho(\tau') \Im[u_0(\tau) u_0^*(\tau')]\frac{1}{\dot{\bar \phi}(\tau)} \mathcal T(q,\tau,\tau',\taui) + C(q,\tau,\taui),
  \label{eq:delta_2vx}
\end{align}
where 
\begin{align}
  \label{eq:cal_T}
    \mathcal T(q,\tau, \tau',\taui) &\equiv \int^{\taui}_{-\infty} \dd \tau'' a^4(\tau'') V_\tho(\tau'') \nonumber \\ 
     &\quad \times \int \frac{\dd^3 k}{(2\pi)^3} \left( \dot{\bar \phi}(\tau)\Im[u_{|\bfk-\bfq|}(\tau') u^*_{|\bfk-\bfq|}(\tau'')]\frac{\Re[u_q(\tau) u_q^*(\tau'')]}{|u_q(\tau)|^2} - \dot{\bar \phi}(\tau'')\Im[u_k(\tau') u^*_k(\tau'')] \right)\Re[u_k(\tau') u^*_k(\tau'')], \\
  \label{eq:C_const}    
    C(q,\tau,\taui) &\equiv 8 \int^{\taui}_{-\infty} \dd \tau' \int^{\tau'}_{-\infty} \dd \tau'' a^4(\tau') a^4(\tau'') V_\tho(\tau') V_\tho(\tau'') \nonumber \\ 
  &\qquad \times \int \frac{\dd^3 k}{(2\pi)^3} \left( \Im[u_q(\tau) u_q^*(\tau')] \Im[u_{|\bfk-\bfq|}(\tau') u^*_{|\bfk-\bfq|}(\tau'')]\frac{\Re[u_q(\tau) u_q^*(\tau'')]}{|u_q(\tau)|^2} \right. \nonumber \\ 
  &\qquad\qquad\qquad\qquad \qquad\qquad\qquad \left. - \frac{\dot{\bar \phi}(\tau'')}{\dot{\bar \phi}(\tau)}\Im[u_0(\tau) u_0^*(\tau')] \Im[u_k(\tau') u^*_k(\tau'')] \right)\Re[u_k(\tau') u^*_k(\tau'')].
\end{align}
For $\tau \geq \tauf$, $\mathcal T$ and $C$ are $\tau$-independent, while $\Delta_{2\vx(b)}$ is $\tau$-dependent.
Note that $\Im[u_0(\tau)u^*_0(\tau')] \propto u_0(\tau) \propto \dot{\bar \phi}(\tau)$ in $\tau \geq \tauf$ and $\tau' \leq \taui$. 
The nonzero $\mathcal T$ is from the contribution in $k \lesssim q$ because the contribution in $k \gg q$ and $|q \tau''| \gtrsim \mathcal O(1)$ is suppressed by $q/k$ due to the fast oscillations of $u_k(\tau')u^*_k(\tau'')$ with $k(\tau'-\tau'') \gg 1$\footnote{See also Appendix D of Ref.~\cite{Inomata:2025bqw} for the discussion on the UV contributions in $k(\tau'-\tau'') \gg 1$.} and, from Eq.~(\ref{eq:cal_T}), we can easily see that the contribution in $k \gg q$ and $|q \tau''| \ll 1$ is suppressed by $q/k$.

With these expressions, we can express the in-in loop contribution in $\tau \geq \taui$ as
\begin{align}
  \frac{H^2}{\langle\dot \phi(\tau) \rangle^2} \expval{\delta \phi_{\bfq}(\tau) \delta \phi_{\bfq'}(\tau)} &= \langle \zeta_{\bfq}(\tau) \zeta_{\bfq'}(\tau)\rangle_{\tre} + \langle \zeta_{\bfq}(\tau) \zeta_{\bfq'}(\tau)\rangle_{1\vx} + \langle \zeta_{\bfq}(\tau) \zeta_{\bfq'}(\tau)\rangle_{\tad}  \nonumber \\ 
  &\quad + \langle \zeta_{\bfq}(\tau) \zeta_{\bfq'}(\tau)\rangle_{2\vx(a)} + \langle \zeta_{\bfq}(\tau) \zeta_{\bfq'}(\tau)\rangle_{2\vx(b)} + \langle \zeta_{\bfq}(\tau) \zeta_{\bfq'}(\tau)\rangle_\ir,
  \label{eq:in_in_loop_exp}
\end{align}
where 
\begin{align}
  \llangle \zeta_{\bfq}(\tau) \zeta_{-\bfq}(\tau)\rrangle_{\tre} &=  \frac{2\pi^2}{q^3} \mathcal P_{\zeta,\tre}(q), \\
  \llangle \zeta_{\bfq}(\tau) \zeta_{-\bfq}(\tau)\rrangle_{1\vx} &=  \Delta_{1\vx}(q,\tau)\frac{2\pi^2}{q^3} \mathcal P_{\zeta,\tre}(q),\\
  \llangle \zeta_{\bfq}(\tau) \zeta_{-\bfq}(\tau)\rrangle_{\tad} &=  \Delta_{\tad}(q,\tau)\frac{2\pi^2}{q^3} \mathcal P_{\zeta,\tre}(q),\\    
  \label{eq:2vx_a}  
  \llangle \zeta_{\bfq}(\tau) \zeta_{-\bfq}(\tau)\rrangle_{2\vx(a)} &= \frac{H^2}{\dot {\bar\phi}^2(\tau)}\frac{2\pi^2}{q^3} \mathcal P^a_{\delta \phi, 2\vx}(q,\tau), \\
  \label{eq:2vx_b}
  \llangle \zeta_{\bfq}(\tau) \zeta_{-\bfq}(\tau)\rrangle_{2\vx(b)} &=  \Delta_{2\vx(b)}(q,\tau)\frac{2\pi^2}{q^3} \mathcal P_{\zeta,\tre}(q),\\
  \label{eq:ir}
  \llangle \zeta_{\bfq}(\tau) \zeta_{-\bfq}(\tau)\rrangle_{\ir} &=  -2 \frac{\langle \delta \dot\phi (\tau)\rangle_\ir}{\dot {\bar \phi}(\tau)} \frac{2\pi^2}{q^3} \mathcal P_{\zeta,\tre}(q).
\end{align}
Note $\langle \zeta_\bfq \zeta_{\bfq'} \rangle  = (2\pi)^3 \delta(\bfq + \bfq') \llangle \zeta_\bfq \zeta_{-\bfq} \rrangle$ and $\mathcal P_{\zeta,\tre}(q) = (H^2/\dot{\bar \phi}^2(\tau)) \mathcal P_{\delta \phi,\tre}(q,\tau)$, which is constant in $\tau \geq \taui$.

For clarity, we reexpress the total curvature power spectrum Eq.~(\ref{eq:zeta_one_loop}) in $\tau \geq \tauf$ as 
\begin{align}
  \expval{\zeta_{\bfq}(\tau) \zeta_{\bfq'}(\tau)} &= (2\pi)^3 \delta(\bfq + \bfq') \nonumber \\ 
  &\quad \times \left(\llangle \zeta_{\bfq}(\tauf) \zeta_{-\bfq}(\tauf)\rrangle_{\tre}  + \llangle \zeta_{\bfq}(\tauf) \zeta_{-\bfq}(\tauf)\rrangle_{1\vx} + \llangle \zeta_{\bfq}(\tauf) \zeta_{-\bfq}(\tauf)\rrangle_{\tad}  \right. \nonumber \\ 
  &\qquad \ + \llangle \zeta_{\bfq}(\tau) \zeta_{-\bfq}(\tau)\rrangle_{2\vx(a)} + \llangle \zeta_{\bfq}(\tau) \zeta_{-\bfq}(\tau)\rrangle_{2\vx(b)} + \llangle \zeta_{\bfq}(\tau) \zeta_{-\bfq}(\tau)\rrangle_\ir \nonumber \\ 
  &\qquad \  \left.  + \llangle \zeta_{\bfq}(\tau) \zeta_{-\bfq}(\tau)\rrangle_{12} + \llangle \zeta_{\bfq}(\tau) \zeta_{-\bfq}(\tau)\rrangle_{22} + \llangle \zeta_{\bfq}(\tau) \zeta_{-\bfq}(\tau)\rrangle_{13} + \llangle \zeta_{\bfq}(\tau) \zeta_{-\bfq}(\tau)\rrangle_B \right),
  \label{eq:zeta_ps_loop2}
\end{align}
where we have used the fact that $\llangle \zeta_\bfq \zeta_{-\bfq} \rrangle_{\tre/1\vx/\tad}$ are constant in $\tau \geq \tauf$.

\section{Cancellation of the time dependence in one-loop curvature power spectrum}
\label{sec:cancellation}

In this section, we see the cancellation of the time dependence of the one-loop curvature power spectrum. 
Specifically, we will see that the $\tau$ dependencies in Eq.~(\ref{eq:zeta_ps_loop2}) cancel in $\tau \geq \tauf$. 
Unless otherwise noted, we assume $\tau \geq \tauf$ for $\tau$ that appear in this section, especially when we discuss $\ssvev{\zeta_\bfq(\tau) \zeta_{-\bfq}(\tau)}$.

We have already seen that $\llangle \zeta_{\bfq}(\tau) \zeta_{-\bfq}(\tau) \rrangle_{\tre/1\vx/\tad}$ are constant. 
We separate the remaining loop contributions into three sectors depending on the structure of the loop integrals: 1) the cut-in-the-middle contributions, 2) the cut-in-the-side contributions, and 3) the boundary contributions.
In the following, we show that the loop contributions cancel within each sector.

\subsection{Cut-in-the-middle contributions}

We begin with the cut-in-the-middle contributions. 
These contributions get $q^3$ suppression if there is a large hierarchy $k \gg q$~\cite{Inomata:2025bqw}.
Given this, we implicitly assume $k \lesssim q$ throughout this subsection. 
We first transform the $\mathcal P^a_{\delta \phi, 2\vx}$ contribution, Eqs.~(\ref{eq:p_a}) and (\ref{eq:2vx_a}), as 
\begin{align}
  \label{eq:p_a_2vx_2}
   \llangle \zeta_{\bfq}(\tau) \zeta_{-\bfq}(\tau)\rrangle _{2\vx(a)} &= \frac{2H^2}{\dot {\bar\phi}^2(\tau)} \int^\tau_{-\infty} \dd \tau' \int^{\tau}_{-\infty} \dd \tau'' a^4(\tau')  a^4(\tau'') V_\tho(\tau') V_\tho(\tau'') \Im[u_q(\tau) u_q^*(\tau')] \Im[u_q(\tau) u_q^*(\tau'')] \nonumber \\
  &\qquad \qquad \qquad \times \int \frac{\dd^3 k}{(2\pi)^3} \Re[u_k(\tau')u^*_k(\tau'')u_{|\bfq - \bfk|}(\tau')u^*_{|\bfq - \bfk|}(\tau'')] \nonumber \\ 
  &= \frac{2H^2}{\dot {\bar\phi}^2(\tau)} \int \frac{\dd^3 k}{(2\pi)^3} |I(\bfq, \bfk,\tau)|^2,
\end{align}
where 
\begin{align}
  \label{eq:I_def}
  I(\bfq,\bfk,\tau) &\equiv \int^\tau_{-\infty} \dd \tau' a^4(\tau') V_\tho(\tau') \Im[u_q(\tau) u_q^*(\tau')] u_k(\tau') u_{|\bfq - \bfk|}(\tau').
\end{align}
For convenience, we express this as 
\begin{align}
 I(\bfq, \bfk, \tau )&= \int^\tau_{\tau_i} \dd \tau' a^4(\tau') V_\tho(\tau') \Im[u_q(\tau) u_q^*(\tau')] u_k(\tau') u_{|\bfq - \bfk|}(\tau') +  J(\bfq, \bfk, \tau,\taui),
\end{align}
where 
\begin{align}
  J(\bfq, \bfk, \tau,\taui) = \int^{\taui}_{-\infty} \dd \tau' a^4(\tau') V_\tho(\tau') \Im[u_q(\tau) u_q^*(\tau')] u_k(\tau') u_{|\bfq - \bfk|}(\tau').
\end{align}
Note that $J(\bfq, \bfk, \tau,\taui) \propto |u_q(\tau)| \propto \dot{\bar \phi}(\tau)$ in $\tau \geq \tauf$.

Using Eqs.~(\ref{eq:im_uu0}) and (\ref{eq:v_3_dot_eta}), we can reexpress $I$ as 
\begin{align}
  I(\bfq,\bfk,\tau) &= - u_k(\tau) u_{|\bfq-\bfk|}(\tau) \frac{1}{\dot{\bar \phi}(\tau)} \int^{\tau}_{\tau_i} \dd \tau' a^4(\tau') V_\tho(\tau') \dot{\bar \phi}^3(\tau') \int^{\tau}_{\tau'} \dd \tau'' \frac{1}{2a^2(\tau'')\dot{\bar \phi}^2(\tau'')} + J(\bfq, \bfk, \tau, \taui) \nonumber \\ 
  &= u_k(\tau) u_{|\bfq-\bfk|}(\tau) \frac{H}{\dot{\bar \phi}(\tau)} \frac{\eta(\tau) - \eta(\tau_i)}{4}  + \tilde J(\bfq, \bfk, \tau, \taui),
  \label{eq:I_rel}
\end{align}
where we have used the implicit assumption in this subsection, $k \lesssim q$, and 
\begin{align}
  \tilde J(\bfq, \bfk, \tau, \taui) = J(\bfq, \bfk, \tau, \taui) - u_k(\tau) u_{|\bfq-\bfk|}(\tau) \frac{H}{\dot{\bar \phi}(\tau)} \eta'(\taui)F(\tau,\taui).
\end{align}
From Eq.~(\ref{eq:im_cond}), we can see that $\tilde J(\bfq,\bfk,\tau,\taui) \propto \dbphi(\tau)$ in $\tau \geq \tauf$ and $k \lesssim q$.
We have separated the contribution $\tilde J$ because the contribution from $|q\tau'| \gtrsim 1$ prevents us from using Eq.~(\ref{eq:im_uu0}).
For later convenience, we here generalize this relation as
\begin{align}
   \label{eq:ab_rel}
   &\int^{\tau}_{-\infty} \dd \tau' a^4(\tau') V_\tho(\tau') \Im[u_q(\tau) u_q^*(\tau')] X(\tau') Y(\tau')= X(\tau) Y(\tau) \frac{H}{\dot{\bar \phi}(\tau)} \frac{\eta(\tau) - \eta(\tau_i)}{4} + \tilde J_{XY}(\tau,\taui), \\
   &\tilde J_{XY}(\tau,\taui) \equiv \int^{\taui}_{-\infty} \dd \tau' a^4(\tau') V_\tho(\tau') \Im[u_q(\tau) u_q^*(\tau')] X(\tau') Y(\tau') - X(\tau) Y(\tau) \frac{H}{\dot{\bar \phi}(\tau)} \eta'(\taui) F(\tau,\taui),
\end{align}
where $X(\tau)$ and $Y(\tau)$ are any functions whose $\tau$-dependence is proportional to $\dot{\bar \phi}(\tau)$ in $\tau \geq \taui$. 
Similarly, we can see $\tilde J_{XY}(\tau,\taui) \propto \dot{\bar \phi}(\tau)$ in $\tau \geq \tauf$.
Substituting Eq.~(\ref{eq:I_rel}) into Eq.~(\ref{eq:p_a_2vx_2}), we obtain 
\begin{align}
  \llangle \zeta_{\bfq}(\tau) \zeta_{-\bfq}(\tau)\rrangle_{2\vx(a)} &= \frac{H^4}{8 \dot{\bar \phi}^4(\tauf)} (\eta(\tau) - \eta(\tau_i))^2 \int \frac{\dd^3 k}{(2\pi)^3} |u_k(\tauf)|^2 |u_{|\bfq-\bfk|}(\tauf)|^2  \nonumber \\ 
  & \quad  + \frac{H^3}{\dot{\bar \phi}^3(\tauf)} (\eta(\tau) - \eta(\tau_i)) \int \frac{\dd^3 k}{(2\pi)^3} \Re[u^*_k(\tauf) u^*_{|\bfq-\bfk|}(\tauf) \tilde J(\bfq, \bfk, \tauf, \taui)] \nonumber \\ 
  &\quad + 2 \frac{H^2}{\dot{\bar \phi}^2(\tauf)} \int \frac{\dd^3 k}{(2\pi)^3} |\tilde J(\bfq, \bfk, \tauf, \taui)|^2,
  \label{eq:2a}
\end{align}
where we have used $\tilde J(\bfq,\bfk,\tau,\taui) \propto \dbphi(\tau)$ in $\tau \geq \tauf$.
The $\tau$ dependence only appears through $\eta(\tau)$.

Next, we see the bispectrum contribution $\llangle\zeta_{\bfq}(\tau) \zeta_{-\bfq}(\tau)\rrangle_{12}$ in Eq.~(\ref{eq:dn_loop_12}). 
We can express the bispectrum itself as
\begin{align}
  \expval{\delta \phi_{\bfq}(\tau) \delta \phi_{\bfq'-\bfk}(\tau) \delta \phi_{\bfk}(\tau)} &= 2\, \Im\left[ \int^{\tau}_{-\infty} \dd \tau' \vev{\delta \phi_{\bfq}(\tau)  \delta \phi_{\bfq'-\bfk}(\tau) \delta \phi_{\bfk}(\tau) H_{\inte,3}(\tau')} \right] \nonumber \\ 
  &= (2\pi)^3 \delta(\bfq + \bfq') ({\mathcal B}_a(\bfq', \bfk, \tau) + {\mathcal B}_b(\bfq', \bfk, \tau)), 
  \label{eq:bi}
\end{align} 
where 
\begin{align}
  {\mathcal B}_a(\bfq, \bfk, \tau)&\equiv 2 \Re[u^*_k(\tau) u^*_{|\bfk - \bfq|}(\tau) I(\bfq,\bfk,\tau)], \\
  {\mathcal B}_b(\bfq, \bfk, \tau)&\equiv 4 \int^\tau_{-\infty} \dd \tau' a^4(\tau') V_\tho(\tau') \Im[u_{|\bfk-\bfq|}(\tau) u^*_{|\bfk-\bfq|}(\tau')] \Re[u_q(\tau) u^*_q(\tau')] \Re[u_k(\tau) u_k^*(\tau')].
  \label{eq:D_b}
\end{align}
${\mathcal B}_a$ is related to the cut-in-the-middle contribution, while ${\mathcal B}_b$ is related to both the cut-in-the-side contribution and the boundary contribution, as we will see in the next subsections.
For convenience, we express the bispectrum contribution in Eq.~(\ref{eq:dn_loop_12}) as 
\begin{align}
  \llangle\zeta_{\bfq}(\tau) \zeta_{-\bfq}(\tau)\rrangle_{12} = \llangle\zeta_{\bfq}(\tau) \zeta_{-\bfq}(\tau)\rrangle_{12(a)} + \llangle\zeta_{\bfq}(\tau) \zeta_{-\bfq}(\tau)\rrangle_{12(b)},
  \label{eq:zeta_bi_12}
\end{align}
where 
\begin{align}
  \llangle\zeta_{\bfq}(\tau) \zeta_{-\bfq}(\tau)\rrangle_{12(\alpha)} = - \frac{\eta(\tau)}{2} \frac{H^3}{\dot {\bar \phi}^3(\tau)}\int_S \frac{\dd^3 k}{(2\pi)^3} {\mathcal B}_{\alpha}(\bfq,\bfk, \tau), \quad \alpha \in \{a, b\}.
  \label{eq:bi_ab}
\end{align}
We have used the fact that $\int_S \dd k^3 \mathcal B_\alpha(\bfq,\bfk,\tau)$ does not depend on the direction of $\bfq$, which leads to $\int_S \dd k^3 \mathcal B_\alpha(\bfq',\bfk,\tau) = \int_S \dd k^3 \mathcal B_\alpha(-\bfq,\bfk,\tau) = \int_S \dd k^3 \mathcal B_\alpha(\bfq,\bfk,\tau)$.
Then, we obtain 
\begin{align}
  \llangle\zeta_{\bfq}(\tau) \zeta_{-\bfq}(\tau)\rrangle_{12(a)} &= - \frac{\eta(\tau)}{2} \frac{H^3}{\dot {\bar \phi}^3(\tauf)}\int_S \frac{\dd^3 k}{(2\pi)^3} \left[\frac{1}{2} \frac{H}{\dot{\bar \phi}(\tau_f)} (\eta(\tau)- \eta(\tau_i)) |u_k(\tau_f)|^2 |u_{|\bfk - \bfq|}(\tau_f)|^2  \right. \nonumber \\ 
  &\qquad \qquad \qquad \qquad \qquad \qquad 
  \left. 
  + 2  \Re[u^*_k(\tau_f) u^*_{|\bfk - \bfq|}(\tau_f) \tilde J(\bfq, \bfk, \tau_f, \taui)]\right].
  \label{eq:12a}
\end{align}

We can also express $\llangle\zeta_{\bfq}(\tau) \zeta_{-\bfq}(\tau)\rrangle_{22}$ in Eq.~(\ref{eq:dn_loop_22}) as 
\begin{align}
  \llangle\zeta_{\bfq}(\tau) \zeta_{-\bfq}(\tau)\rrangle_{22} &= \frac{H^4}{8 \dot{\bar \phi}^4(\tau_f)} \eta^2(\tau) \int_S \frac{\dd^3 k}{(2\pi)^3} |u_k(\tau_f)|^2 |u_{|\bfq-\bfk|}(\tau_f)|^2.
  \label{eq:diff_pure}
\end{align}

Combining Eqs.~(\ref{eq:2a}), (\ref{eq:12a}), and (\ref{eq:diff_pure}), we can see the cancellation of the $\tau$ dependence in the cut-in-the-middle contributions:
\begin{align}
  \label{eq:cancel_sum_x}
  \sum_{X \in \{ 2\vx(a), 12(a), 22 \}} \llangle\zeta_{\bfq}(\tau) \zeta_{-\bfq}(\tau)\rrangle_X &= \frac{H^4}{8 \dot{\bar \phi}^4(\tauf)} \eta^2(\tau_i) \int \frac{\dd^3 k}{(2\pi)^3} |u_k(\tauf)|^2 |u_{|\bfq-\bfk|}(\tauf)|^2  \nonumber \\ 
  & \quad  - \frac{H^3}{\dot{\bar \phi}^3(\tauf)} \eta(\tau_i) \int \frac{\dd^3 k}{(2\pi)^3} \Re[u^*_k(\tauf) u^*_{|\bfq-\bfk|}(\tauf) \tilde J(\bfq, \bfk, \tauf, \taui)] \nonumber \\ 
  & \quad 
  + 2 \frac{H^2}{\dot{\bar \phi}^2(\tauf)} \int \frac{\dd^3 k}{(2\pi)^3} |\tilde J(\bfq, \bfk, \tauf, \taui)|^2. 
\end{align}
Recall that the contributions from the loop wavenumber $k \gg q$ are negligible in the cut-in-the-middle contributions, which allows us to approximate $\int_S \dd^3 k \simeq \int \dd^3 k$ to obtain the above equation.
Note again that the difference of $\int_S \dd^3 k$ and $\int \dd^3 k$ only comes from the UV cutoff difference, $k_\sm \text{ vs. } k_\uv a(\tau)/a_i$ (see below Eq.~(\ref{eq:zeta_ps_loop})).

\subsection{Cut-in-the-side contributions}

Next, we discuss the cut-in-the-side contributions. 
We begin with the in-in loop contribution $\llangle \zeta_{\bfq}(\tau) \zeta_{-\bfq}(\tau)\rrangle_{2\vx(b)}$, Eqs.~(\ref{eq:2vx_cond}) and (\ref{eq:2vx_b}):
\begin{align}
  \label{eq:2vxb}
  \llangle \zeta_{\bfq}(\tau) \zeta_{-\bfq}(\tau)\rrangle_{2\vx(b)} &= \frac{2\pi^2}{q^3} \mathcal P_{\zeta,\tre}(q) \left[ 8 \int^{\tau}_{\tau_i} \dd \tau' a^4(\tau') V_\tho(\tau') 
  \Im[u_0(\tau) u_0^*(\tau')] \frac{1}{\dot{\bar \phi}(\tau)} \mathcal T(q,\tau_f,\tau',\taui) + C(q,\tauf,\taui) \right] \nonumber \\ 
  &= \frac{2\pi^2}{q^3} \mathcal P_{\zeta,\tre}(q) \left[ 8 \int^{\tau}_{\tauis} \dd \tau' a^4(\tau') V_\tho(\tau') \Im[u_0(\tau) u_0^*(\tau')] \dot{\bar \phi}^2(\tau')\frac{1}{\dot{\bar \phi}(\tau)} \frac{\mathcal T(q,\tau_f,\tau',\taui)}{\dot{\bar \phi}^2(\tau')} \right.  \nonumber \\ 
  &\qquad\qquad\qquad\quad \left. + 8 \int^{\tauis}_{\taui} \dd \tau' a^4(\tau') V_\tho(\tau') \Im[u_0(\tau) u_0^*(\tau')] \dot{\bar \phi}^2(\tau')\frac{1}{\dot{\bar \phi}(\tau)} \frac{\mathcal T(q,\tau_f,\tau',\taui)}{\dot{\bar \phi}^2(\tau')} + C(q,\tauf,\taui) \right]
   \nonumber \\ 
  &= \frac{2\pi^2}{q^3} \mathcal P_{\zeta,\tre}(q) \left[2 H (\eta(\tau)- \eta(\tauis)) \frac{\mathcal T(q,\tau_f,\tau_f,\taui)}{\dot{\bar \phi}^2(\tau_f)} + \tilde C(q,\tauf,\tauis,\taui) \right],
\end{align}
where we have used $\mathcal T(q,\tau,\tau',\taui)= \mathcal T(q,\tauf,\tau',\taui)$ and $C(q,\tau,\taui) = C(q,\tauf,\taui)$ in $\tau \geq \tauf$, and 
\begin{align}
  \tilde C(q,\tauf,\tauis, \taui) &= C(q,\tauf,\taui) + 8 \int^{\tauis}_{\taui} \dd \tau' a^4(\tau') V_\tho(\tau') \Im[u_0(\tauf) u_0^*(\tau')] \dot{\bar \phi}^2(\tau')\frac{1}{\dot{\bar \phi}(\tauf)} \frac{\mathcal T(q,\tau_f,\tau',\taui)}{\dot{\bar \phi}^2(\tau')} \nonumber \\ 
  &\qquad - 8 H \eta'(\tauis) F(\tauf,\tauis) \frac{\mathcal T(q,\tau_f,\tau_f,\taui)}{\dot{\bar \phi}^2(\tau_f)}.
\end{align}
We define $\tauis$ as the time in the interval $\taui < \tauis < \tauf$ such that the following relation holds in two cases:
\begin{align}
 \Im[u_0(\tau)u^*_0(\tau')] \propto \dbphi(\tau) \text{ in 1) } \tau \geq \tauis \text{ and } \tau' \leq \taui, \text{ and 2) } \tau \geq \tauf \text{ and } \tau' \leq \tauis.
 \label{eq:im_cond_strict}
\end{align}
We require the SR period to last long enough that an appropriate $\tauis$ exists.
Note that, $\mathcal T(q,\tau_f,\tau',\taui)/\dot{\bar \phi}^2(\tau')$ in the second line of Eq.~(\ref{eq:2vxb}) is independent of $\tau'$ in $\tau' \geq \tauis$, which enables us to perform the $\tau'$ integral by using Eq.~(\ref{eq:ab_rel}).

We now turn to the contribution from ${\mathcal B}_b$, Eq.~(\ref{eq:D_b}). 
To proceed, we decompose ${\mathcal B}_b$ as 
\begin{align}
  {\mathcal B}_b(\bfq,\bfk,\tau) = {\mathcal B}_{\Delta b}(\bfq,\bfk,\tau) + {\mathcal B}_{\tilde b}(q,k,\tau),
\end{align}
where
\begin{align}
  {\mathcal B}_{\tilde b}(q,k, \tau) &\equiv 4 \frac{|u_q(\tau)|^2}{\dbphi(\tau)} \int^\tau_{-\infty} \dd \tau' a^4(\tau') V_\tho(\tau') \dbphi(\tau') \Im[u_{k}(\tau) u^*_{k}(\tau')] \Re[u_k(\tau) u_k^*(\tau')].
  \label{eq:d_til_b}
\end{align}
We can see ${\mathcal B}_b(\bfq,\bfk,\tau) = {\mathcal B}_{\tilde b}(q,k,\tau)$ in the limit of $k \gg q$ and $|k\tau| \ll 1$ because the time integral contribution from $|\tau'| \gtrsim 1/q$ in Eq.~(\ref{eq:D_b}) is negligible due to the fast oscillation of $\Im[u_k(\tau) u^*_k(\tau')] \Re[u_k(\tau) u_k^*(\tau')]$, which enables us to transform $\Re[u_q(\tau)u^*_q(\tau')] = |u_q(\tau)|^2 \dbphi(\tau')/\dbphi(\tau)$ in ${\mathcal B}_b$, Eq.~(\ref{eq:D_b}), in that limit.
Similar to Eq.~(\ref{eq:bi_ab}), we define 
\begin{align}
  \label{eq:zeta_db_tb}
  \llangle\zeta_{\bfq}(\tau) \zeta_{-\bfq}(\tau)\rrangle_{12(\beta)} = - \frac{\eta(\tau)}{2} \frac{H^3}{\dot {\bar \phi}^3(\tau)} \int_S \frac{\dd^3 k}{(2\pi)^3}  {\mathcal B}_{\beta}(\bfq,\bfk,\tau), \quad \beta \in \{\Delta b, \tilde b\},
\end{align}
where, for $\beta = \tilde b$, the wavenumber arguments of $\mathcal B_\beta$ are just $q$ and $k$.
Note that $\llangle\zeta_{\bfq}(\tau) \zeta_{-\bfq}(\tau)\rrangle_{12(b)} = \llangle\zeta_{\bfq}(\tau) \zeta_{-\bfq}(\tau)\rrangle_{12(\Delta b)} + \llangle\zeta_{\bfq}(\tau) \zeta_{-\bfq}(\tau)\rrangle_{12(\tilde b)}$.
In this subsection, we focus on $\llangle\zeta_{\bfq} \zeta_{-\bfq}\rrangle_{12(\Delta b)}$, relevant to the cut-in-the-side contribution.
We can express it as 
\begin{align}
  \label{eq:zeta_12_db}
  &\llangle\zeta_{\bfq}(\tau) \zeta_{-\bfq}(\tau)\rrangle_{12(\Delta b)} = - 2\eta(\tau) \frac{H^3|u_q(\tau)|^2}{\dot {\bar \phi}^4(\tau)} \int^{\tau}_{-\infty} \dd \tau' a^4(\tau') V_\tho(\tau') \nonumber \\ 
  &\times \int_S \frac{\dd^3 k}{(2\pi)^3} \left( \dot{\bar \phi}(\tau)\Im[u_{|\bfk-\bfq|}(\tau) u^*_{|\bfk-\bfq|}(\tau')]\frac{\Re[u_q(\tau) u_q^*(\tau')]}{|u_q(\tau)|^2} 
    - \dot{\bar \phi}(\tau')\Im[u_k(\tau) u^*_k(\tau')] \right)\Re[u_k(\tau) u^*_k(\tau')].
\end{align}
The contribution from the integral over $\tau' > \taui$ is negligible because $u_q(\tau) \propto \dot{\bar \phi}(\tau)$, $u_q(\tau') \propto \dot{\bar \phi}(\tau')$, and $\Im[u_{|\bfk-\bfq|}(\tau) u^*_{|\bfk-\bfq|}(\tau')] = \Im[u_k(\tau) u^*_k(\tau')]$ in $\tau' > \taui$ once we neglect the terms suppressed by $q/k$ or $q\tau'$, similarly to Eq.~(\ref{eq:2vx_cond}).
Furthermore, the wavenumber integral contribution from $k \gg q$ in $\tau' \leq \taui$ is negligible following the same logic presented below Eq.~(\ref{eq:C_const}) and therefore we can approximate $\int_S \dd^3 k \simeq \int \dd^3 k$, similarly to Eq.~(\ref{eq:cancel_sum_x}).
Taking these into account, we can reexpress Eq.~(\ref{eq:zeta_12_db}) as
\begin{align}
  &\llangle\zeta_{\bfq}(\tau) \zeta_{-\bfq}(\tau)\rrangle_{12(\Delta b)} = - 2\eta(\tau)H^3 \frac{|u_q(\tau)|^2}{\dot{\bar \phi}^4(\tau)}\mathcal T(q,\tau,\tau,\taui),
  \label{eq:12db}
\end{align}
where $\mathcal T$ is defined in Eq.~(\ref{eq:cal_T}).
The point is that $\frac{|u_q(\tau)|^2}{\dot{\bar \phi}^4(\tau)}\mathcal T(q,\tau,\tau,\taui)$ is $\tau$-independent in $\tau \geq \tauf$ and the remaining $\tau$-dependence is only from $\eta(\tau)$.
Then, we can reexpress it as 
\begin{align}
  &\llangle\zeta_{\bfq}(\tau) \zeta_{-\bfq}(\tau)\rrangle_{12(\Delta b)} = - \frac{4\pi^2}{q^3} \mathcal P_{\zeta,\tre}(q) H \eta(\tau) \frac{\mathcal T(q,\tau_f,\tau_f,\taui)}{\dot{\bar \phi}^2(\tau_f)}.
  \label{eq:del_b_dif}
\end{align}
Combining this and Eq.~(\ref{eq:2vxb}), we can see the cancellation of the $\tau$ dependence in the cut-in-the-side contributions:
\begin{align}
  \label{eq:cancel_sum_y}  
  \sum_{Y \in \{ 2\vx(b), 12(\Delta b) \}} \llangle\zeta_{\bfq}(\tau) \zeta_{-\bfq}(\tau)\rrangle_Y = \frac{2\pi^2}{q^3} \mathcal P_{\zeta,\tre}(q) \left[-2 H \eta(\tauis) \frac{\mathcal T(q,\tau_f,\tau_f,\taui)}{\dot{\bar \phi}^2(\tau_f)} + \tilde C(q,\tauf,\tauis,\taui) \right].
\end{align}

\subsection{Boundary contributions}
\label{subsec:bound_term}

Finally, we see the cancellation of the $\tau$ dependence of the boundary contributions.
Let us begin with the IR boundary contribution from the in-in loop, Eqs.~(\ref{eq:ddot_ir}) and (\ref{eq:ir}):
\begin{align}
  &\llangle \zeta_\bfq(\tau) \zeta_{-\bfq}(\tau)\rrangle_\ir = -2\frac{\langle \delta \dot \phi(\tau) \rangle_{\ir}}{\dot{\bar \phi}(\tau)} \frac{2\pi^2}{q^3} \mathcal P_{\zeta,\tre}(q) \nonumber \\ 
  &= -\frac{2}{\dot{\bar \phi}(\tau)} \frac{2\pi^2}{q^3} \mathcal P_{\zeta,\tre}(q) \int^\tau_{-\infty} \dd \tau' a^4(\tau')\Im[u_0(\tau) u_0^*(\tau')] H V_\tho(\tau') \mathcal P_{\delta \phi,\tre}(k_\ir,\tau') \nonumber \\ 
  &=  - \left[ \frac{\eta(\tau) - \eta(\tau_i)}{2} - 2\eta'(\taui) F(\tauf,\taui) \right] \frac{2\pi^2}{q^3} \mathcal P_{\zeta,\tre}(q) \mathcal P_{\zeta,\tre}(k_\ir) \nonumber \\ 
  &\qquad - \frac{2}{\dot{\bar \phi}(\tau)} \frac{2\pi^2}{q^3} \mathcal P_{\zeta,\tre}(q) \int^{\taui}_{-\infty} \dd \tau' a^4(\tau')\Im[u_0(\tau) u_0^*(\tau')] H V_\tho(\tau') \mathcal P_{\delta \phi,\tre}(k_\ir,\tau'), 
    \label{eq:ddot_ir2}
\end{align}
where we have used Eq.~(\ref{eq:ab_rel}).
The term in the second to last line depends on $\tau$, while the term in the last line is time-independent in $\tau \geq \tauf$ because $1/\dot{\bar \phi}(\tau)$ and $\Im[u_0(\tau) u_0^*(\tau')]$ cancel their $\tau$ dependencies.

Next, we see the remaining bispectrum contribution $\llangle \zeta_{\bfq} \zeta_{-\bfq} \rrangle_{12(\tilde b)}$, defined in Eqs.~(\ref{eq:d_til_b}) and (\ref{eq:zeta_db_tb}).
To proceed, we first show that Maldacena's consistency relation~\cite{Maldacena:2002vr} requires the following relation: 
\begin{align}
    -\frac{2\pi^2}{k^3} \frac{\dd \mathcal  P_{\zeta,\tre}(k)}{\dd \ln k} = - \left(\frac{H}{\dot {\bar \phi}(\tau)}\right)^3 \mathcal A(k,\tau)
  + \eta(\tau) \left(\frac{H}{\dot {\bar \phi}(\tau)}\right)^4 |u_k(\tau)|^2,
  \label{eq:mal_cons}
\end{align}
where $\mathcal A(k,\tau) \equiv \mathcal B_{\tilde b}(q,k,\tau)/|u_q(\tau)|^2$ and we have assumed $\zeta_{\tre,\bfk} = \text{const.}$ on superhorizon.
Let us begin with Maldacena's consistency relation in the squeezed configuration $r \ll k$~\cite{Maldacena:2002vr} 
\begin{align}
  \expval{\zeta_\bfr \zeta_{\bfr'-\bfk} \zeta_{\bfk}} = -(2\pi)^3 \delta(\bfr + \bfr') \frac{2\pi^2}{r^3} \mathcal P_{\zeta,\tre}(r) \frac{2\pi^2}{k^3} \frac{\dd \mathcal P_{\zeta,\tre}(k)}{\dd \ln k} + \mathcal O((r/k)^2),
  \label{eq:zeta_bi}
\end{align}
where we have assumed that all the three modes ($\zeta_\bfr, \zeta_\bfk, \zeta_{\bfr'-\bfk}$) are constant on superhorizon scales.
We intentionally use $\bfr$ instead of $\bfq$ for the soft modes because we assume $r \ll k_\ir$.
Note that this relation is always satisfied for the bispectrum of the superhorizon curvature perturbations in single-field inflation models. 
Using the nonlinear relation between $\zeta$ and $\delta \phi$, Eq.~(\ref{eq:zeta_dN3}), we can reexpress the curvature bispectrum with $\delta \phi$ as
\begin{align}
  \expval{\zeta_\bfr \zeta_{\bfr'-\bfk} \zeta_{\bfk}} &= - \left(\frac{H}{\dot {\bar \phi}(\tau)}\right)^3 \expval{\delta \phi_\bfr(\tau) \delta \phi_{\bfr'-\bfk}(\tau) \delta \phi_{\bfk}(\tau)} \nonumber \\ 
  &\quad + \frac{\eta(\tau)}{4} \left(\frac{H}{\dot {\bar \phi}(\tau)}\right)^4 \int \frac{\dd^3 p}{(2\pi)^3} \left( \expval{\delta \phi_\bfr(\tau) \delta \phi_{\bfr'-\bfk}(\tau) \delta \phi_{\bfp}(\tau) \delta \phi_{\bfk-\bfp}(\tau)} + \expval{\delta \phi_\bfr(\tau) \delta \phi_{\bfp}(\tau) \delta \phi_{\bfr'-\bfk-\bfp}(\tau) \delta \phi_{\bfk}(\tau)} \right. \nonumber \\ 
  &\qquad \qquad \qquad \qquad \qquad \qquad \quad
  \left.
  + \expval{\delta \phi_{\bfp}(\tau) \delta \phi_{\bfr-\bfp}(\tau) \delta \phi_{\bfr'-\bfk}(\tau) \delta \phi_{\bfk}(\tau)} \right) \nonumber \\ 
  &= (2\pi)^3 \delta(\bfr + \bfr') \left[- \left(\frac{H}{\dot {\bar \phi}(\tau)}\right)^3 {\mathcal B}_{\tilde b}(r,k,\tau) + \eta(\tau) \left(\frac{H}{\dot {\bar \phi}(\tau)}\right)^4 |u_r(\tau)|^2 |u_k(\tau)|^2 + \mathcal O((r/k)^2) \right],
  \label{eq:zeta_bi2}
\end{align}
where we have used ${\mathcal B}_b(\bfr,\bfk,\tau) = {\mathcal B}_{\tilde b}(r,k,\tau)$, valid in $r \ll k$ and $|r\tau| \ll 1$ (see below Eq.~(\ref{eq:d_til_b})).
Equating this with Eq.~(\ref{eq:zeta_bi}), we can see that Eq.~(\ref{eq:mal_cons}) is satisfied in $k > k_\ir$.
Then, using Eq.~(\ref{eq:mal_cons}), we can see
\begin{align}
  \llangle\zeta_{\bfq}(\tau) \zeta_{-\bfq}(\tau)\rrangle_{12(\tilde b)} + \llangle\zeta_{\bfq}(\tau) \zeta_{-\bfq}(\tau)\rrangle_{13} &= -\frac{\eta(\tau)}{2} \frac{2\pi^2}{q^3}\mathcal P_{\zeta,\tre}(q) \int^{k_\sm}_{k_\ir} \dd \ln k  \frac{\dd \mathcal P_{\zeta,\tre}(k)}{\dd \ln k} \nonumber \\ 
  &= -\frac{\eta(\tau)}{2} \frac{2\pi^2}{q^3}\mathcal P_{\zeta,\tre}(q) (\mathcal P_{\zeta,\tre}(k_\sm) - \mathcal P_{\zeta,\tre}(k_\ir)),
  \label{eq:del_n_bound}
\end{align}
where we have used the expressions of Eqs.~(\ref{eq:zeta_ps_loop}) and (\ref{eq:zeta_db_tb}).
This relation can also be seen in Ref.~\cite{Iacconi:2026uzo}.
Combining this and Eq.~(\ref{eq:ddot_ir2}), we find 
\begin{align}
  \sum_{Z\in \{\ir, 12(\tilde b), 13 \}} \llangle\zeta_{\bfq}(\tau) \zeta_{-\bfq}(\tau)\rrangle_Z &= -\frac{\eta(\tau)}{2} \frac{2\pi^2}{q^3} \mathcal P_{\zeta,\tre}(q) \mathcal P_{\zeta,\tre}(k_\sm) + \left[ \frac{\eta(\tau_i)}{2} + 2\eta'(\taui) F(\tauf,\taui) \right] \frac{2\pi^2}{q^3} \mathcal P_{\zeta,\tre}(q) \mathcal P_{\zeta,\tre}(k_\ir) \nonumber \\ 
  &\qquad - \frac{2}{\dot{\bar \phi}(\tauf)} \frac{2\pi^2}{q^3} \mathcal P_{\zeta,\tre}(q) \int^{\taui}_{-\infty} \dd \tau' a^4(\tau')\Im[u_0(\tauf) u_0^*(\tau')] H V_\tho(\tau') \mathcal P_{\delta \phi,\tre}(k_\ir,\tau').
  \label{eq:boundary_0}
\end{align}
We can see that the $\tau$ dependence remains in the first term. 

So far, we have discussed all the contributions except for $\llangle \zeta_{\bfq} \zeta_{\bfq'} \rrangle_B$ in Eq.~(\ref{eq:zeta_ps_loop2}). 
We have seen the cancellation of the $\tau$ dependence in the cut-in-the-middle and the cut-in-the-side one-loop contributions, Eqs.~(\ref{eq:cancel_sum_x}) and (\ref{eq:cancel_sum_y}), while we have seen the non-cancellation of the $\tau$ dependence among the boundary contributions except for $\llangle \zeta_{\bfq} \zeta_{\bfq'} \rrangle_B$, Eq.~(\ref{eq:boundary_0}).
In the following, we explain the remaining contribution, $\llangle \zeta_{\bfq} \zeta_{\bfq'} \rrangle_B$.

We first explain why we introduce that contribution.
The remaining boundary term at $k_\sm$ in Eq.~(\ref{eq:boundary_0}) can actually be interpreted as due to the violation of the separate universe picture by the introduction of the smoothing cutoff in our $\delta N$ formalism setup.
We explain this by using the so-called large-gauge transformations, which do not vanish at spatial infinity~\cite{Weinberg:2003sw,Hinterbichler:2012nm}.
In comoving gauge, the diffeomorphism invariance allows us to perform the large-gauge transformation for the superhorizon limit curvature $\zeta$: $\bfx \to \tilde \bfx = \ee^{\lambda} \bfx$ and $\zeta \to \tilde \zeta - \lambda$ with constant $\lambda$.
By using this, we can always locally remove the superhorizon curvature perturbation for each Hubble patch if the separate universe picture is valid. 
In spatially-flat gauge, this gauge transformation corresponds to $\bfx \to \tilde \bfx = \ee^{\lambda} \bfx$ and $t \to \tilde t = t - \lambda/H$ with constant $\lambda$, which leads to $\delta \phi \to \tilde{\delta \phi} = \delta \phi + \lambda\dot{\bar \phi}/H + \mathcal O(\lambda^2)$ for the superhorizon-limit $\delta \phi$ while keeping $\zeta = 0$.
For example, if we take $\lambda = -(H/\dbphi)\delta \phi$, we get $\tilde{\delta \phi} = 0$.
This physically means that, if the separate universe picture is valid, we can locally remove the superhorizon $\delta \phi$ by shifting the inflaton background value.
This indicates that the evolution stage of each Hubble patch is fully characterized by the local inflaton background ($=$global background + superhorizon $\delta \phi$).
Under the transformation $\bfx \to \tilde \bfx = \ee^{\lambda} \bfx$ and $t \to \tilde t = t - \lambda/H$, the comoving wavenumber changes as $k \to \tilde k = \ee^{-\lambda} k$.
This leads to $\int^{k_\sm} \dd k f(k,\tau) \to \int^{\ee^{-\lambda}k_\sm} \dd \tilde k f(\tilde k,\tilde \tau)$ for an arbitrary function $f$ if the smoothing cutoff scale is homogeneous (space-independent).
If the cutoff is not invariant under the large-gauge transformation, we cannot fully remove the effect of the superhorizon $\delta \phi$ by the large-gauge transformation.
This physically means that a local observer inside a Hubble patch can measure the superhorizon-limit $\delta \phi$, which is inconsistent with the separate universe picture.

We here show how the inhomogeneous smoothing cutoff restores the separate universe picture in the $\delta N$ formalism.
The following argument is based on Appendix C of our previous work~\cite{Inomata:2025bqw}.\footnote{
  In our previous work~\cite{Inomata:2025bqw}, we have focused on the IR boundary term from the in-in loop with the assumption of $k_\ir \gg q$. 
}
We modify the cutoff $k_\sm$ to $k_\sm (1- H \delta \phi(\bfx,\tau)/\dot{\bar \phi}(\tau))$, where we have assumed that $\delta \phi(\bfx, \tau)$ is the superhorizon perturbation whose scale is given by $1/q$ with $q \ll k_\sm$.
From the wavenumber transformation $k \to \tilde k = \ee^{-\lambda} k$ under the large-gauge transformation, the wavenumber integral transforms as 
\begin{align}
\int^{k_\sm \left(1- H \frac{\delta \phi(\bfx,\tau)}{\dot{\bar \phi}(\tau)} \right)} \dd k f(k,\tau) \to \int^{\ee^{-\lambda} k_\sm \left(1- H \frac{\delta \phi(\bfx,\tau)}{\dot{\bar \phi}(\tau)} \right)} \dd \tilde k f(\tilde k,\tilde \tau)  = \int^{k_\sm \left(1- H \frac{\tilde{\delta \phi}(\bfx,\tilde \tau)}{\dot{\bar \phi}(\tilde\tau)} \right)} \dd \tilde k f(\tilde k,\tilde \tau),
\end{align}
where $f$ is an arbitrary function again and we have neglected $\mathcal O(\lambda^2)$ contributions and used the time independence of $\delta \phi/\dbphi$ for the superhorizon $\delta \phi$ and $\delta \phi \to \tilde{\delta \phi} = \delta \phi + \lambda\dot{\bar \phi}/H$. 
We can see that the expression of the upper bound is invariant under the large-gauge transformation.
On the other hand, for the scale of $k_\ir \ll q$, we cannot regard $\delta \phi(\bfx)$ (whose scale is $\sim 1/q$) as the background.
Because of this, there is no reason to modify the IR cutoff at $k_\ir$. 
Indeed, we have already seen the cancellation of the time-dependence of the IR boundary terms in the in-in loop (Eq.~(\ref{eq:ddot_ir2})) and the $\delta N$ loop (Eq.~(\ref{eq:del_n_bound})).

Let us see that this modified cutoff leads to the cancellation of the time-dependence of the remaining boundary contribution in Eq.~(\ref{eq:boundary_0}). 
We here rewrite this modification as $k_\sm \to k_\sm (1 +\zeta(\bfx))$ by using the relation $\zeta = -(H/\dbphi)\delta \phi$.
To proceed, let us first recall the tadpole of the curvature, Eq.~(\ref{eq:zeta_tad}): 
\begin{align}
  \langle \zeta \rangle = - H \frac{\expval{\delta \phi}}{\dbphi} + \frac{\eta}{4} \expval{\zeta^2}|_{k_\sm},
  \tag{\ref{eq:zeta_tad}}
\end{align}
where $\svev{\zeta^2}|_{k_\sm} = (H^2/\dot{\bar \phi}^2)\int^{k_\sm}_{k_\ir} \dd \ln k (k^3/2\pi^2) |u_k|^2$ again.
With $k_\sm \to k_\sm (1 + \zeta(\bfx))$, the tadpole becomes space-dependent as
\begin{align}
    \label{eq:svev_zeta_inh}
  \langle \zeta \rangle_\inh(\bfx,\tau) &= - H \frac{\expval{\delta \phi}}{\dbphi} + \frac{\eta(\tau)}{4} \expval{\zeta^2}|_{k_\sm\left(1 + \zeta(\bfx)\right)} \nonumber \\ 
  &= \expval{\zeta}_\homo + \frac{\eta(\tau)}{4} \mathcal P_{\zeta,\tre}(k_\sm) \zeta(\bfx),
\end{align}
where $\expval{\zeta}_\homo$ is the homogeneous tadpole, identical to Eq.~(\ref{eq:zeta_tad}), and we have expanded the second term with respect to $\zeta(\bfx)$ and neglected the higher-order contributions.
Note that, even after the change of $k_\sm \to k_\sm (1 + \zeta(\bfx))$, $\svev{\delta \phi}$ does not contribute to $\svev{\zeta}_\inh$.
The perturbation term that newly arises after substituting $k_\sm \to k_\sm (1 + \zeta(\bfx))$ into Eq.~(\ref{eq:v_c1_tad_cut}) contributes as a part of $V_{c,\so} \delta \phi(\bfx)$, which is taken into account as the tadpole loop contribution and does not change the discussion in the previous (sub)sections.
We define the Fourier mode of this inhomogeneous tadpole as 
\begin{align}
  \zeta_{\bfq,\text{inh}} &= \int \dd^3 x\, \ee^{-i\bfq \cdot \bfx }\expval{\zeta}_{\text{inh}}(\bfx) \nonumber \\ 
  &= \frac{\eta(\tau)}{4} \mathcal P_{\zeta,\tre}(k_\sm) \zeta_{\bfq}.
\end{align}
$\ssvev{\zeta_{\bfq} \zeta_{-\bfq}}_B$ corresponds to the contribution from this $\zeta_{\bfq,\text{inh}}$:
\begin{align}
 \ssvev{\zeta_{\bfq}(\tau) \zeta_{-\bfq}(\tau)}_B &\equiv  \ssvev{\zeta_{\bfq}(\tau) \zeta_{-\bfq,\text{inh}}(\tau) + \zeta_{\bfq,\text{inh}}(\tau) \zeta_{-\bfq}(\tau)} \nonumber \\ 
 &=  \frac{\eta(\tau)}{2} \frac{2\pi^2}{q^3} \mathcal P_{\zeta,\tre}(q) \mathcal P_{\zeta,\tre}(k_\sm).
 \label{eq:inh_b}
\end{align}
This is the only contribution that arises from the inhomogeneous cutoff at the one-loop level.
We can also obtain $\ssvev{\zeta_\bfq \zeta_{-\bfq}}_B$ by directly expanding the coarse-grained field $\delta \phi_L$ with respect to $\zeta(\bfx)$ in the inhomogeneous cutoff, shown in Appendix~\ref{app:another_deri}.
$\ssvev{\zeta_\bfq \zeta_{-\bfq}}_B$ cancels the $\tau$ dependence that remains in Eq.~(\ref{eq:boundary_0}).
We stress that, since the $\delta N$ formalism is based on the separate universe picture, it is self-consistent to impose that the way of introducing the smoothing cutoff respects the separate universe picture.
Combining Eqs.~(\ref{eq:cancel_sum_x}), (\ref{eq:cancel_sum_y}), (\ref{eq:boundary_0}), and (\ref{eq:inh_b}), we can finally see that $\svev{\zeta_{\bfq}(\tau) \zeta_{\bfq'}(\tau)}$ is constant in $\tau \geq \tauf$, which shows the conservation of the superhorizon curvature perturbations at one-loop level. 
This is the main result of this paper.

Before closing this section, we mention that the cutoff scales fixed in physical scales are consistent with the separate universe picture. 
Apart from the smoothing cutoff, we have also employed the UV cutoff regularization $\int^{k_\uv a(t)/a_*} \dd k$ for the in-in loops throughout this work. 
This UV cutoff is indeed invariant under the large-gauge transformation: $\bfx \to \tilde \bfx = \ee^{\lambda} \bfx$, $t \to \tilde t = t - \lambda/H$, and $k \to \tilde k = \ee^{-\lambda} k$. 
Using these relations, we can see the invariance as $\int^{k_\uv a(t)/a_*} \dd k f(k,t) \to \int^{\ee^{-\lambda} k_\uv a(t)/a_*} \dd \tilde k f(\tilde k,\tilde t) =  \int^{k_\uv a(\tilde t)/a_*} \dd \tilde k f(\tilde k,\tilde t)$.
In Appendix~\ref{app:dim_reg}, we reproduce the in-in loop expression with dimensional regularization, which also respects the large-gauge transformation invariance because there is no UV cutoff in that regularization scheme. 
Also, if we fix the smoothing cutoff in physical scales as $k_\sm a(t)/a_\sm$ with $a_\sm$ being some reference scale factor, we do not need to make the smoothing cutoff inhomogeneous because the cutoff $k_\sm a(t)/a_\sm$ is consistent with the separate universe picture as is. 
See Appendix~\ref{app:smoothing} for details.

\section{Conclusion}
\label{sec:conclusion}

In this paper, we have shown the conservation of superhorizon curvature perturbations at the one-loop level in single-field inflation. 
We have taken into account loop wavenumbers (momenta) from all scales, in particular the regime $k \lesssim q$ with $k$ and $q$ being the loop and the power spectrum wavenumber, respectively.
This regime $k \lesssim q$ was neglected in our previous works that assumed a large hierarchy $k \gg q$~\cite{Inomata:2024lud,Inomata:2025bqw,Inomata:2025pqa}.
Working in spatially-flat gauge, we have computed the one-loop corrections to the two-point function of the inflaton fluctuations and the background evolution, which we call the in-in loop contributions, and then have further computed loop contributions that arise from the non-linear relation between the inflaton fluctuations and the curvature perturbations by using the $\delta N$ formalism, which we call the $\delta N$ loop contributions.

We here clarify the assumptions we made in our analysis.
We have considered the transition: the initial period $\to$ SR period $\to$ the late period, where the linear (tree-level) curvature perturbations on $k \lesssim q$ exit the horizon and become constant during the initial period, and remain constant after that.
This requires $|q \tau_i| \ll 1$.
We have also assumed the decoupling limit with $\epsilon \ll 1$ throughout this paper.
We have not specified the inflaton potential for the initial and the late period, which can include transient non-SR periods, such as the USR period and/or a parametric resonance period with oscillatory feature, as long as the tree-level curvature perturbations on $k \lesssim q$ are constant after the initial period.
This requires the SR period to be long enough if we consider a transient non-SR period during the late period.
We have assumed that $\eta$ is always sufficiently greater than $-3$ during the SR period.
In contrast to our previous works~\cite{Inomata:2024lud,Inomata:2025bqw,Inomata:2025pqa}, we have allowed $V_\tho \neq 0$ during any period.
Apart from the inflation setup, we have taken the smoothing cutoff satisfying $q/k_\sm \ll 1$ so that we can take into account the contributions on $k \lesssim q$ in the $\delta N$ formalism.

We have used the $\delta N$ formalism after the initial period and shown that the time dependence of the one-loop power spectrum at $q$ cancels during the late period.
Specifically, we have shown that the cancellation of the time dependence in the one-loop contributions is nontrivial for $k \lesssim q$: the in-in and the $\delta N$ loop contributions do not separately vanish, and the curvature conservation emerges only after all contributions are combined consistently.

Throughout this work, we have restricted ourselves to canonical single-field inflation models as the simplest case. 
It would be worthwhile to generalize our calculation to non-canonical models in the context of effective field theory.
Also, given that we have focused on the one-loop level, the extension of our analysis to higher-order loops without the hierarchy $q \ll k$ is an interesting future direction (see Ref.~\cite{Fang:2025kgf} for the recent analysis with $q \ll k$).

\acknowledgements
The author thanks Jacopo Fumagalli for helpful comments on a draft of this paper. 

\appendix

\section{Smoothing scale fixed in physical scales}
\label{app:smoothing}

In the main text, we have introduced the smoothing cutoff fixed in comoving scales and have smoothed out the modes on $k > k_\sm$ in the $\delta N$ formalism. 
In this Appendix, we see how the $\delta N$ expression changes if we instead introduce the smoothing cutoff fixed in physical scales and smooth out the modes on $k > k_\sm a(\tau)/a_\sm$, where $a_\sm$ is a reference value of the scale factor.
For most part of the calculation, we can use the same expressions as in the main text by replacing $k_\sm$ with $k_\sm a(\tau)/a_\sm$ as long as the tree-level curvature $\zeta_{\tre,\bfk}$ on $k < k_\sm a(\tau)/a_\sm$ is always constant in $\tau \geq \tau_i$.
The major difference from the main text is that we do not need to modify the cutoff to be inhomogeneous because the cutoff fixed in physical scales is consistent with the separate universe picture, as mentioned in Sec.~\ref{subsec:bound_term}. 
In the following, we will see how the term corresponding to Eq.~(\ref{eq:inh_b}) appears with the smoothing cutoff fixed in physical scales.

In the case of the smoothing cutoff fixed in physical scales, the background evolution is continuously affected by the modes that cross the smoothing scale.
This effect appears as the stochastic noise term in the $\delta N$ formalism.
We have the coarse-grained field:
\begin{align}
  \delta \phi_L = \int_{k<k_\sm a(\tau)/a_\sm} \frac{\dd^3 k}{(2\pi)^3} \delta \phi_\bfk \ee^{i \bfk \cdot \bfx}.
\end{align}
By differentiating this, we obtain 
\begin{align}
  \delta \dot \phi_L = \int_{k<k_\sm a(\tau)/a_\sm} \frac{\dd^3 k}{(2\pi)^3} \delta \dot \phi_\bfk \ee^{i \bfk \cdot \bfx} + \xi(t,\bfx),
\end{align}
where 
\begin{align}
  \xi(t,\bfx) = \frac{a(t) H k_\sm}{a_\sm} \int \frac{\dd^3 k}{(2\pi)^3} \delta(k - k_\sm a(t)/a_\sm) \delta \phi_\bfk(t) \ee^{i \bfk \cdot \bfx}.
\end{align}
This $\xi$ term represents the stochastic noise effect, which does not appear in the case of the smoothing cutoff fixed in comoving scales. 
We here consider the e-folds for the inflaton to move from $\phi_a$ to $\phi_b$, denoted by $N(\phi_a,\phi_b)$.
The noise term gives an additional contribution to the e-folds as
\begin{align}
  N(\phi_a,\phi_b) = \bar N(\phi_a,\phi_b) + N_\xi(\phi_a,\phi_b),
\end{align}
where $\bar N(\phi_a,\phi_b)$ is the contribution independent of $\xi$ and $N_\xi$ is a functional of $\xi$. 
Specifically, it can be expanded as a functional Taylor series:
\begin{align}
  N_\xi(\phi_a,\phi_b) = \int^{t_b}_{t_a} \dd t' \frac{\delta N}{\delta \xi(t')} \xi(t') + \frac{1}{2} \int^{t_b}_{t_a} \dd t_1 \int^{t_b}_{t_a} \dd t_2 \frac{\delta^2 N}{\delta \xi(t_1) \delta \xi(t_2)} \xi(t_1) \xi(t_2) + \mathcal O(\xi^3),
  \label{eq:n_xi}
\end{align}
where $t_a(t_b)$ is the time when $\bar \phi = \phi_a(\phi_b)$ with $\xi(t) = 0$.
We are interested in $\delta N$ induced by $\delta \phi_L(t_a)$ and the two-point correlation function of $\zeta (= \delta N)$.
Since $\xi(t)$ in $t>t_a$ does not correlate with $\delta \phi_L(t_a)$ (this is why we treat them separately), the part of $\delta N_\xi$ relevant to $\ssvev{\zeta_{\bfq} \zeta_{-\bfq}}$ at one-loop level only comes from the second term in Eq.~(\ref{eq:n_xi}).
Note that the correlation of the first terms is suppressed by the volume factor, $q^3$.
In other words, if we focus on the contributions relevant to $\ssvev{\zeta_{\bfq} \zeta_{-\bfq}}$, we must take the ensemble average for the $\xi^2$ in the second term.
Specifically, we first consider 
\begin{align}
  \svev{N_\xi(\phi_a,\phi_b)} &= \frac{1}{2} \int^{t_b}_{t_a} \dd t_1 \int^{t_b}_{t_a} \dd t_2 \frac{\delta^2 N}{\delta \xi(t_1) \delta \xi(t_2)} \svev{\xi(t_1) \xi(t_2)} \nonumber \\ 
  &= \frac{1}{2} \int^{t_b}_{t_a} \dd t_1 \int^{t_b}_{t_a} \dd t_2 \frac{\delta^2 N}{\delta \xi(t_1) \delta \xi(t_2)} \delta(t_1 - t_2)  H \mathcal P_{\delta \phi,\tre}(k_\sm a(t_1)/a_\sm) \nonumber \\ 
  &= \frac{1}{2} \int^{t_b}_{t_a} \dd t_1 \frac{\delta^2 N}{\delta \xi(t_1) \delta \xi(t_1)}  H \mathcal P_{\delta \phi,\tre}(k_\sm a(t_1)/a_\sm) \nonumber \\   
  &= \frac{1}{2} \int^{t_b}_{t_a} \dd t_1 N_{\phi\phi}(t_1)  H \mathcal P_{\delta \phi,\tre}(k_\sm a(t_1)/a_\sm) \nonumber \\ 
  &= \frac{1}{4} \int^{t_b}_{t_a} \dd t_1 \frac{H^2}{\dbphi^2(t_1)} \eta(t_1)  H \mathcal P_{\delta \phi,\tre}(k_\sm a(t_1)/a_\sm),  
  \label{eq:n_xi_vev}
\end{align}
where $N_{\phi\phi}(t_1)$ here denotes $\dd^2 N(\phi_1,\phi_b)/\dd \phi_1^2$ with $\phi_1 = \bar \phi(t_1)$.
We now consider $\delta N_\xi$ induced by $\delta \phi_L(t_a)$, which causes the change $\phi_a \to \phi_a + \delta \phi_L(t_a)$ and therefore $t_a \to t_a + \delta \phi_L/\dbphi$:
\begin{align}
  \delta N_\xi \simeq \frac{\dd \svev{N_\xi(\phi_a,\phi_b)}}{\dd \phi_a} \delta \phi_L = -\frac{1}{4} \frac{H^3}{\dot{\bar \phi}^3(t_a)} \eta(t_a)  \mathcal P_{\delta \phi,\tre}(k_\sm a(t_a)/a_\sm) \delta \phi_L(t_a),
  \label{eq:d_n_xi}
\end{align}
where we have neglected the contributions that are suppressed by $q^3$ in $\ssvev{\zeta_\bfq \zeta_{-\bfq}}$.
With the Fourier modes of $\delta N_\xi$, we can express the $\delta N_\xi$ contribution to $\ssvev{\zeta_\bfq \zeta_{-\bfq}}$ as:
\begin{align}
  \ssvev{\zeta_\bfq(\tau) \zeta_{-\bfq}(\tau)}_\xi &= \ssvev{\delta N_\bfq(\tau) \delta N_{\xi,-\bfq}(\tau) + \delta N_{\xi,\bfq}(\tau) \delta N_{-\bfq}(\tau)} \nonumber \\ 
  &= \frac{\eta(\tau)}{2} \frac{2\pi^2}{q^3} \mathcal P_{\zeta,\tre}(q) \mathcal P_{\zeta,\tre}(k_\sm a(\tau)/a_\sm),
  \label{eq:zeta_xi}
\end{align}
where we have replaced $t_a$ by the time corresponding to $\tau$ from Eq.~(\ref{eq:d_n_xi}) so that we can focus on the curvature perturbations at each time.
Eq.~(\ref{eq:zeta_xi}) is the same as Eq.~(\ref{eq:inh_b}) and cancels the $\tau$ dependence of the boundary term at $k_\sm a(\tau)/a_\sm$ in Eq.~(\ref{eq:boundary_0}).

As a cross-check, let us obtain Eq.~(\ref{eq:zeta_xi}) by focusing on the time dependence of the curvature tadpole.
By replacing $k_\sm \to k_\sm a(\tau)/a_\sm$, we obtain the backreaction as 
\begin{align}
    \expval{\delta \phi(\tau(\geq \taui))}
  &= \frac{H}{\dot{\bar \phi}(\tau)} \left[\frac{\eta(\tau) - \eta(\tau_i)}{4} - \eta'(\taui) F(\tau,\taui) \right] \expval{\delta \phi^2(\tau)}|_{k_\sm a(\tau)/a_\sm} + D(\tau),
  \label{eq:dphi_l_tad_app}
\end{align}
Let us recall the curvature tadpole expression, Eq.~(\ref{eq:zeta_tad}), with the replacement $k_\sm \to k_\sm a(\tau)/a_\sm$:
\begin{align}
  \langle \zeta \rangle = - H \frac{\expval{\delta \phi}}{\dbphi} + \frac{\eta}{4} \expval{\zeta^2}|_{k_\sm a(\tau)/a_\sm}.
\end{align}
Substituting Eq.~(\ref{eq:dphi_l_tad_app}) into this, we obtain 
\begin{align}
  \expval{\zeta} = \left[\frac{\eta(\tau_i)}{4} + \eta'(\taui)F(\tauf,\taui) \right]\expval{\zeta^2}|_{k_\sm a(\tau)/a_\sm } - H\frac{D(\tauf)}{\dbphi(\tauf)} \quad \text{in } \tau \geq \tauf.
  \label{eq:zeta_vev}
\end{align}
This is time-dependent because of the time dependence of the smoothing cutoff.
We absorb this time dependence of $\svev{\zeta}$ into the scale factor. 
Specifically, we redefine the scale factor as $\tilde a(\tau) = \ee^{\svev{\zeta(\tau)}} a(\tau)$.
With the modified scale factor, we redefine the curvature as
\begin{align}
  \tilde \zeta(\bfx) =  \delta \tilde N(\bfx) &= -\frac{\tilde H}{\dbphi}\delta \phi_L(\bfx) + \frac{\eta}{4} \frac{H^2}{\dot{\bar \phi}^2} \delta \phi_L^2(\bfx) - \frac{\eta^2-\dot\eta/H}{12} \frac{H^3}{\dot{\bar \phi}^3} \delta \phi_L^3(\bfx) - \svev{\zeta} \nonumber \\ 
  &= -\frac{H}{\dbphi}\delta \phi_L(\bfx) + \frac{\eta}{4} \frac{H^2}{\dot{\bar \phi}^2} \delta \phi_L^2(\bfx) - \frac{\eta^2-\dot\eta/H}{12} \frac{H^3}{\dot{\bar \phi}^3} \delta \phi_L^3(\bfx) - \svev{\zeta} -  \frac{\svev{\dot\zeta}}{\dbphi} \delta \phi_L(\bfx),
  \label{eq:til_zeta}
\end{align}
where $\tilde H \equiv \dot{\tilde a}/\tilde a$ and we have subtracted $\svev{\zeta}$ so that $\svev{\tilde \zeta} = 0$ and have taken into account $\tilde H$ only for the first term because the others only give higher-loop contributions from the difference between $H$ and $\tilde H$.

The time derivative of the curvature tadpole is explicitly given by 
\begin{align}
  \svev{\dot \zeta(\tau)} = \left[\frac{\eta(\tau_i)}{4} + \eta'(\taui)F(\tauf,\taui) \right] H \mathcal P_{\zeta,\tre}(k_\sm a(\tau)/a_\sm) \quad \text{in } \tau \geq \tauf.
\end{align}
Similarly, the time derivative of the inflaton backreaction is given by 
\begin{align}
  \svev{\dot{\delta \phi}(\tau)} = \frac{\eta(\tau) H}{2} \svev{\delta \phi} + \frac{H}{\dbphi(\tau)} \frac{\dot \eta(\tau)}{4} \expval{\delta \phi^2}|_{k_\sm a(\tau)/a_\sm} + \frac{H}{\dot{\bar \phi}(\tau)} \left[\frac{\eta(\tau) - \eta(\tau_i)}{4} - \eta'(\taui) F(\tauf,\taui) \right] H &\mathcal P_{\delta \phi,\tre}(k_\sm a(\tau)/a_\sm) \nonumber \\
  &\qquad \text{in } \tau \geq \tauf.
\end{align}
Using these expressions, we can reexpress Eq.~(\ref{eq:til_zeta}) as 
\begin{align}
  \tilde \zeta(\bfx) &= -\frac{H}{\svev{\dot \phi}}\delta \phi_L(\bfx) + \frac{\eta}{4} \frac{H^2}{\dot{\bar \phi}^2} \delta \hat \phi_L^2(\bfx) - \frac{\eta^2}{12} \frac{H^3}{\dot{\bar \phi}^3} \delta \phi_L^3(\bfx) + \frac{\dot \eta H^2}{4\dot{\bar \phi}^3} \left(\frac{1}{3} \delta \phi_L^3(\bfx) - \delta \phi_L(\bfx) \svev{\delta \phi^2}|_{k_\sm a(\tau)/a_\sm} \right) \nonumber \\ 
  &\qquad - \svev{\zeta} -\frac{1}{4} \frac{H^3}{\dot{\bar \phi}^3} \eta  \mathcal P_{\delta \phi,\tre}(k_\sm a(\tau)/a_\sm) \delta \phi_L(\bfx),
  \label{eq:zeta_dN3_app}
\end{align}
where note again $\svev{\dot \phi} = \dbphi + \svev{\delta \dot \phi}$ and $\delta \hat \phi_L = \delta \phi_L - \svev{\delta \phi}$.
This is the same as Eq.~(\ref{eq:zeta_dN3}) except for the last two terms.
In particular, the last term gives the same contribution as Eq.~(\ref{eq:zeta_xi}) in $\ssvev{\zeta_\bfq \zeta_{-\bfq}}$.
Note also that, in the main text, we implicitly subtracted $\svev{\zeta}$ with the time-independent rescaling of the spatial coordinates.

\section{General counter potential}
\label{app:gen_c_pot}

In the main text, we took the counter potential given by Eq.~(\ref{eq:v_c1_tad_cut}) and found the curvature tadpole time-independent in $\tau \geq \tauf$.
In this Appendix, we reproduce Eq.~(\ref{eq:zeta_dN3}), which is the key relation for the one-loop calculation, without specifying the form of the counter potential. 

By taking the time derivative of the curvature tadpole (Eq.~(\ref{eq:zeta_tad})), 
\begin{align}
  \expval{\zeta} = -\frac{H}{\dbphi}\svev{\delta\phi} + \frac{\eta}{4} \frac{H^2}{\dot {\bar \phi}^2} \svev{\delta \phi^2}|_{k_\sm}, 
  \tag{\ref{eq:zeta_tad}}
\end{align}
we obtain
\begin{align}
  \svev{\dot \zeta} = - \frac{H}{\dbphi} \svev{\delta \dot \phi} + \frac{\eta}{2} \frac{H^2}{\dbphi} \svev{\delta \phi} + \frac{\dot \eta}{4} \frac{H^2}{\dot{\bar \phi}^2} \svev{\delta \phi^2}|_{k_\sm}.
  \label{eq:dot_zeta_app}
\end{align}
We do not specify how $\svev{\delta \phi}$ depends on time and therefore $\svev{\dot \zeta} \neq 0$ in general.
As in Appendix~\ref{app:smoothing}, we absorb the time-dependence of the curvature tadpole into the scale factor, $a \to \tilde a = \ee^{\svev{\zeta}}a$. 
Then, similarly to Eqs.~(\ref{eq:til_zeta}) and (\ref{eq:zeta_dN3_app}), we can express the redefined curvature as 
\begin{align}
  \tilde \zeta(\bfx) &= -\frac{H}{\dbphi}\delta \phi_L(\bfx) + \frac{\eta}{4} \frac{H^2}{\dot{\bar \phi}^2} \delta \phi_L^2(\bfx) - \frac{\eta^2-\dot\eta/H}{12} \frac{H^3}{\dot{\bar \phi}^3} \delta \phi_L^3(\bfx) - \svev{\zeta} -  \frac{\svev{\dot\zeta}}{\dbphi} \delta \phi_L(\bfx) \nonumber \\ 
  &= -\frac{H}{\svev{\dot \phi}}\delta \phi_L(\bfx) + \frac{\eta}{4} \frac{H^2}{\dot{\bar \phi}^2} \delta \hat \phi_L^2(\bfx) - \frac{\eta^2}{12} \frac{H^3}{\dot{\bar \phi}^3} \delta \phi_L^3(\bfx) + \frac{\dot \eta H^2}{4\dot{\bar \phi}^3} \left(\frac{1}{3} \delta \phi_L^3(\bfx) - \delta \phi_L(\bfx) \svev{\delta \phi^2}|_{k_\sm} \right) - \svev{\zeta}.
\end{align}
This is the same as Eq.~(\ref{eq:zeta_dN3}) except for the explicit subtraction of $\svev{\zeta}$, which is implicitly subtracted in the main text.

Similarly, even if we fix the smoothing cutoff in physical scales and take a general counter potential, we can easily see that Eq.~(\ref{eq:zeta_dN3_app}) is realized by just replacing $k_\sm \to k_\sm a(\tau)/a_\sm$ and additionally taking into account the terms from the time derivative of the smoothing cutoff in Eq.~(\ref{eq:dot_zeta_app}). 

From these arguments, we can see that the curvature conservation does not depend on the choice of the counter potential.

\section{Relation between $\eta'$ and $V_\tho$}
\label{app:rel}

In this Appendix, we derive Eq.~(\ref{eq:v_3_dot_eta}), the relation between $\eta'$ (or $\dot \eta$) and $V_\tho$:
\begin{align}
  \frac{\dd \left(a^2 \dot {\phi}^2 \eta'\right)}{\dd \tau} &\simeq -a^4 \frac{2V_\tho \dot \phi^3}{H}.
      \tag{\ref{eq:v_3_dot_eta}}
\end{align}
To this end, we begin with the following equation:
\begin{align}
   \frac{\dd \left(a^3 \epsilon \dot \eta\right)}{\dd t} = 3 H a^3 \epsilon \dot \eta + a^3 (\epsilon \dot \eta)\dot\,. 
   \label{eq:dot_eta_0}
\end{align}
We assume $V_\so \propto \epsilon^0$ and $V_\tho \propto \epsilon^{-1/2}$ when $V_\tho \neq 0$, which is the case in the decoupling limit (see the discussion above Eq.~(\ref{eq:hamil})). 
Then, we can obtain the following relations:
\begin{align}
  \ddot {\phi} &= -3 H \dot {\phi} - V_\fo, \\ 
  \dot H &= -\frac{\dot \phi^2}{2 M_\Pl^2} = -\epsilon H^2, \\
  \ddot H & = -6H \dot H + \frac{1}{M_\Pl^2} \dot \phi V_\fo, \\
  \dddot H & = -6(\dot H^2 + H \ddot H) + \frac{1}{M_\Pl^2} \left( -3 H \dot \phi V_\fo - V_\fo^2 + \dot \phi^2 V_\so \right), \\
  \eta &\equiv \frac{\dot \epsilon}{H \epsilon} = \frac{\ddot H}{H \dot H} + 2 \epsilon = -6 + \frac{1}{H \dot H M_\Pl^2} \dot \phi V_\fo + 2 \epsilon, \label{eq:eta_def_exp} \\
  \dot \eta & = \frac{\dddot H}{H \dot H} - \frac{\ddot H (\dot H^2 + H \ddot H)}{(H \dot H)^2} + 2 \epsilon \left( \frac{\ddot H}{\dot H} + 2 \epsilon H \right) \nonumber \\
  &= -6 \left( \frac{\dot H}{H} + \frac{\ddot H}{\dot H} \right) - \frac{\ddot H}{H^2} - \frac{\ddot H^2}{H \dot H^2} + 2 \epsilon \left( \frac{\ddot H}{\dot H} + 2 \epsilon H \right) + \frac{1}{M_\Pl^2} \left( -\frac{3 \dot \phi}{\dot H} V_\fo - \frac{V_\fo^2}{H \dot H} + \frac{\dot \phi^2}{ H \dot H} V_\so \right)\nonumber \\
  & \simeq -6 \left( \frac{\dot H}{H} + \frac{\ddot H}{\dot H} \right) - \frac{\ddot H^2}{H \dot H^2} + \frac{1}{M_\Pl^2} \left( -\frac{3 \dot \phi}{\dot H} V_\fo - \frac{V_\fo^2}{H \dot H} + \frac{\dot \phi^2}{ H \dot H} V_\so \right) \nonumber \\
  &= -6 \left[ \frac{\dot H}{H} + \left( - 6H + \frac{\dot \phi V_\fo}{M_\Pl^2 \dot H} \right) \right] - \frac{1}{H} \left( - 6H + \frac{\dot \phi V_\fo}{M_\Pl^2 \dot H} \right)^2   + \frac{1}{M_\Pl^2} \left( -\frac{3 \dot \phi}{\dot H} V_\fo - \frac{V_\fo^2}{H \dot H} + \frac{\dot \phi^2}{ H \dot H} V_\so \right) \nonumber \\
  &= -6 \frac{\dot H}{H} + 6\frac{\dot \phi V_\fo}{M_\Pl^2 \dot H} - \frac{\dot \phi^2 V_\fo^2}{M_\Pl^4 H \dot H^2} + \frac{1}{M_\Pl^2} \left( -\frac{3 \dot \phi}{\dot H} V_\fo - \frac{V_\fo^2}{H \dot H} + \frac{\dot \phi^2}{ H \dot H} V_\so \right) \nonumber \\
  &= -6 \frac{\dot H}{H} + 3 \frac{\dot \phi V_\fo}{M_\Pl^2 \dot H} + \frac{V_\fo^2}{M_\Pl^2 H \dot H} + \frac{\dot \phi^2}{M_\Pl^2 H \dot H} V_\so \nonumber \\ 
  &\simeq 3 \frac{\dot \phi V_\fo}{M_\Pl^2 \dot H} + \frac{V_\fo^2}{M_\Pl^2 H \dot H} - \frac{2}{H} V_\so,  
  \label{eq:eta_dot}
\end{align}
where we take a shorthand notation $\phi = \bar \phi$ throughout this Appendix and have neglected the higher-order terms in $\epsilon$ at the approximate equalities, $\simeq$.
Using these, we find 
\begin{align}
  \epsilon \dot \eta 
  &\simeq - \frac{3 \dot \phi V_\fo}{M_\Pl^2 H^2} - \frac{V_\fo^2}{M_\Pl^2 H^3} - \frac{2 \epsilon }{H} V_\so.
\end{align}
Substituting the above expressions into the right-hand side of Eq.~(\ref{eq:dot_eta_0}), we obtain
\begin{align}
  \frac{\dd \left(a^3 \epsilon \dot \eta\right)}{\dd t} \simeq& 3 H a^3 \left[-\frac{3 \dot \phi V_\fo}{M_\Pl^2 H^2} - \frac{V_\fo^2}{M_\Pl^2 H^3} - \frac{2 \epsilon }{H} V_\so \right] + a^3 \left[ -\frac{3 \ddot \phi V_\fo}{M_\Pl^2 H^2} - \frac{3 \dot \phi^2 V_\so}{M_\Pl^2 H^2} - \frac{2V_\fo V_\so \dot \phi}{M_\Pl^2 H^3} - 2 \epsilon \eta V_\so - \frac{2\epsilon}{H} V_\tho \dot \phi \right] \nonumber \\
  = & 3 H a^3 \left[- \frac{3 \dot \phi V_\fo}{M_\Pl^2 H^2} - \frac{V_\fo^2}{M_\Pl^2 H^3} - \frac{2 \epsilon }{H} V_\so \right] \nonumber \\
  &+ a^3 \left[ -\frac{3 V_\fo (-3 H \dot \phi - V_\fo)}{M_\Pl^2 H^2} - \frac{3 \dot \phi^2 V_\so}{M_\Pl^2 H^2} - \frac{2V_\fo V_\so \dot \phi}{M_\Pl^2 H^3} - 2 \epsilon \eta V_\so - \frac{2\epsilon}{H} V_\tho \dot \phi \right] \nonumber \\
  = & a^3 \left[ -6 \epsilon V_\so - \frac{3 \dot \phi^2 V_\so}{M_\Pl^2 H^2} - \frac{2V_\fo V_\so \dot \phi}{M_\Pl^2 H^3} - 2 \epsilon \eta V_\so - \frac{2\epsilon}{H} V_\tho \dot \phi \right]\nonumber \\
  = & a^3 \left[ -2(\eta + 6) \epsilon V_\so - \frac{2V_\fo V_\so \dot \phi}{M_\Pl^2 H^3} - \frac{2\epsilon}{H} V_\tho \dot \phi \right].
  \label{eq:d_a3_ep_dot_eta}
\end{align}
Here, we use the following relation: (from Eq.~(\ref{eq:eta_def_exp}))
\begin{align}
  \frac{\dot \phi V_\fo}{H^3 M_\Pl^2} &= \frac{\dot H}{H^2} \frac{\dot \phi V_\fo}{H \dot H M_\Pl^2} \nonumber \\
  &= -\epsilon (\eta + 6 - 2\epsilon).
\end{align}
Using this, we can rewrite Eq.~(\ref{eq:d_a3_ep_dot_eta}) as 
\begin{align}
  \frac{\dd \left(a^3 \epsilon \dot \eta\right)}{\dd t} &\simeq -a^3 \frac{2\epsilon}{H} V_\tho \dot \phi \nonumber \\ 
  &= -a^3 \frac{V_\tho \dot \phi^3}{H^3 M_\Pl^2},
\end{align}
where we have neglected the higher-order terms in $\epsilon$ by using the relations $V_\so \propto \epsilon^0$ and $V_\tho \propto \epsilon^{-1/2}$ when $V_\tho \neq 0$.
With the conformal time, we can reexpress this relation as 
\begin{align}
  \frac{\dd \left(a^2 \dot {\phi}^2 \eta'\right)}{\dd \tau} &\simeq -a^4 \frac{2V_\tho \dot \phi^3}{H},
\end{align}
where we have neglected the $\epsilon$-suppressed term again.
This is Eq.~(\ref{eq:v_3_dot_eta}).

\section{Another derivation of $\ssvev{\zeta_\bfq \zeta_{-\bfq}}_B$}
\label{app:another_deri}

In this Appendix, we derive $\ssvev{\zeta_\bfq \zeta_{-\bfq}}_B$, Eq.~(\ref{eq:inh_b}), based on the delta function term that appears in the coarse-grained field after directly expanding the inhomogeneous cutoff. 
After the cutoff modification ($k_\sm \to k_\sm(1+\zeta(\bfx))$),  the coarse-grained field is given by 
\begin{align}
    \delta \phi_L(\bfx) &= \int \frac{\dd^3 k}{(2\pi)^3} \ee^{i \bfk \cdot \bfx} \Theta(k_\sm(1+\zeta(\bfx))- k) \delta \phi_\bfk \nonumber \\ 
    &=  \int \frac{\dd^3 k}{(2\pi)^3} \ee^{i \bfk \cdot \bfx} \Theta(k_\sm- k) \delta \phi_\bfk + \Delta \delta \phi_L(\bfx),
\end{align}
where 
\begin{align}
    \label{eq:Delta_dphi_L}
    \Delta \delta \phi_L(\bfx) \equiv \int \frac{\dd^3 k}{(2\pi)^3} \ee^{i \bfk \cdot \bfx} \delta(k_\sm- k) \delta \phi_\bfk k_\sm \zeta(\bfx).
\end{align}
We have neglected $\mathcal O(\delta \phi_\bfk \zeta(\bfx)^2)$ because it does not contribute to the one-loop power spectrum.
Recall the nonlinear $\delta N$ relation, Eq.~(\ref{eq:zeta_dN3}):
\begin{align}
    \zeta =-\frac{H}{\svev{\dot \phi}}\delta \phi_L(\bfx) + \frac{\eta}{4} \frac{H^2}{\dot{\bar \phi}^2} \delta \hat \phi_L^2(\bfx) + \mathcal O(\delta \phi_L^3),
\end{align}
where $\delta \hat \phi_L = \delta \phi_L - \svev{\delta \phi}$ and we have not explicitly written the $\mathcal O(\delta \phi_L^3)$ terms because the $\Delta \delta \phi_L$ modification to $\mathcal O(\delta \phi_L^3)$ only contributes to two or higher-order loop contributions.
In Fourier space, we can rewrite this as 
\begin{align}
    \zeta_\bfq =-\frac{H}{\svev{\dot \phi}}\delta \phi_\bfq + \frac{\eta}{4} \frac{H^2}{\dot{\bar \phi}^2} \delta \hat \phi_{L}^2|_{\bfq} + \mathcal O(\delta \phi_L^3),
\end{align}
where 
\begin{align}
    \delta \phi_\bfq &=  \int \dd^3 x \ee^{-i\bfq \cdot \bfx} \delta \phi_L(\bfx) =  \int \dd^3 x \ee^{-i\bfq \cdot \bfx} \delta \phi(\bfx), \\
    \delta \hat \phi_{L}^2|_{\bfq} &= \int \dd^3 x \ee^{-i\bfq \cdot \bfx} (\delta \phi^2_L(\bfx) + 2\delta \phi_L(\bfx)  \Delta \delta \phi_L(\bfx)). 
\end{align}
Note that the first term in $\zeta_\bfq$, proportional to $\delta \phi_\bfq$, does not include the cutoff modification contribution, $\Delta \delta \phi_L$.
This is why the expression of the in-in loop does not change even if we introduce the inhomogeneous cutoff, as mentioned below Eq.~(\ref{eq:svev_zeta_inh}).
On the other hand, the expressions of the $\delta N$ loop contributions change.
For convenience, we express the modification as 
\begin{align}
    \Delta\delta \hat \phi_{L}^2|_{\bfq} \equiv 2\int \dd^3 x \ee^{-i\bfq \cdot \bfx} \delta \phi_L(\bfx)  \Delta \delta \phi_L(\bfx).
\end{align}
This modification changes the two-point function, $\ssvev{\zeta_\bfq \zeta_{-\bfq}}$, by 
\begin{align}
    \Delta \ssvev{\zeta_\bfq \zeta_{-\bfq}} &= \frac{\eta}{2} \frac{H^2}{\dot{\bar \phi}^2} (\ssvev{\zeta_\bfq \Delta\delta \hat \phi_{L}^2|_{\bfq'}} + \ssvev{\Delta\delta \hat \phi_{L}^2|_{\bfq} \zeta_{\bfq'}}) \nonumber \\ 
    &= \frac{\eta}{2} \frac{H^2}{\dot{\bar \phi}^2} \mathcal P_{\delta \phi}(k_\sm) \frac{2\pi^2}{q^3} \mathcal P_\zeta(q) \nonumber\\ 
    &= \frac{\eta}{2} \frac{2\pi^2}{q^3} \mathcal P_\zeta(q) \mathcal P_{\zeta}(k_\sm),
    \label{eq:Delta_zeta}
\end{align}
where we have used $\int^\varepsilon_{-\varepsilon} \dd x \Theta(x) \delta(x) = [\Theta^2(x)/2]^\varepsilon_{-\varepsilon} = 1/2$.
Eq.~(\ref{eq:Delta_zeta}) is exactly the same as $\ssvev{\zeta_\bfq \zeta_{-\bfq}}_B$.

\section{Dimensional regularization}
\label{app:dim_reg}

In the main text, based on our previous papers~\cite{Inomata:2024lud,Inomata:2025bqw,Inomata:2025pqa}, we have adopted the cutoff UV regularization and used the following relation:
\begin{align}
  H^2\frac{\langle \delta \phi_{\bfq}(\tau) \delta \phi_{\bfq'}(\tau) \rangle}{\langle \dot{\phi}(\tau) \rangle^2} &= \langle \zeta_{\bfq}(\tau) \zeta_{\bfq'}(\tau)\rangle_{\tre} + \langle \zeta_{\bfq}(\tau) \zeta_{\bfq'}(\tau)\rangle_{1\vx} + \langle \zeta_{\bfq}(\tau) \zeta_{\bfq'}(\tau)\rangle_{\tad}  \nonumber \\ 
  &\quad + \langle \zeta_{\bfq}(\tau) \zeta_{\bfq'}(\tau)\rangle_{2\vx(a)} + \langle \zeta_{\bfq}(\tau) \zeta_{\bfq'}(\tau)\rangle_{2\vx(b)} + \langle \zeta_{\bfq}(\tau) \zeta_{\bfq'}(\tau)\rangle_\ir.
  \tag{\ref{eq:in_in_loop_exp}}
\end{align}
The main goal of this Appendix is to reproduce this equation with dimensional regularization.

We begin with the action in $D+1$ dimensions:
\begin{align}
  S = \mu^{D-3} \int \dd \tau \dd^D x\, a^{D+1} \mathscr{L}, \  \mathscr{L} = -\frac{1}{2} \partial^\mu \phi \partial_\mu \phi - V_b(\phi), 
  \label{eq:action_d}
\end{align}
where we have introduced the parameter $\mu$ with mass dimension so that we can keep the mass dimensions of the other variables the same as in $D=3$. 
From this, the momentum conjugate of $\delta \phi$ becomes
\begin{align}
  \Pi_{\delta \phi} = \frac{\partial (\mu^{D-3} a^{D+1} \mathscr{L})}{\partial \delta \phi'} = \mu^{D-3}a^{D-1} \delta \phi'.
\end{align}
From the commutation relations $[\delta \phi(\bfx),\Pi_{\delta \phi}(\bfy)] = \mu^{D-3} a^{D-1}[\delta \phi(\bfx),\delta \phi'(\bfy)] = i \delta^D(\bfx-\bfy)$, we can obtain the following Wronskian relation:
\begin{align}
  \Im[u_k(\tau) {u^*}'_k(\tau)] = \frac{1}{2 \mu^{D-3} a^{D-1}(\tau)}.
\end{align}
The Hamiltonian density is given by
\begin{align}
  \mu^{D-3} a^{D+1} \mathscr H = \Pi_{\delta \phi}\delta \phi' - \mu^{D-3} a^{D+1} \mathscr{L}.
\end{align}
The Hamiltonian density $\mathscr H$ is the same as Eq.~(\ref{eq:hamil}) and therefore $\mathscr H_{\inte,n}$ is also the same as Eq.~(\ref{eq:inte_hamil}), while the interaction Hamiltonian itself is given by 
\begin{align}
  H_{\inte,n} = \mu^{D-3}\int \dd^{D} x \, a^{D+1} \mathscr H_{\inte,n}.
\end{align}
In $D+1$ dimensions, the one-loop expressions of $\mathcal P_{\delta \phi}$, Eqs.~(\ref{eq:1vx}) and (\ref{eq:tad})-(\ref{eq:p_b}), are modified through $a^4 \to \mu^{D-3} a^{D+1}$ and $\int \dd^3 k \to \int \dd^D k$, in addition to the modification of the function form of $u_k$:
\begin{align}
  \label{eq:1vx_app}  
  \mathcal P_{\delta \phi, 1\vx}(q,\tau) &=  2 \frac{q^D \Omega_D }{(2\pi)^D} \mu^{D-3}\int^\tau_{-\infty} \dd \tau' a^{D+1}(\tau') \Im\left[ u_q(\tau) u^*_q(\tau') \right] \Re\left[ u_q(\tau) u^*_q(\tau') \right] \nonumber \\ 
  &\qquad \qquad \qquad \qquad \qquad \qquad \qquad \qquad 
  \times \left(V_\foo(\tau')\int\frac{\dd^D k}{(2\pi)^D} |u_k(\tau')|^2 + 2V_{c,\so}(\tau')\right), \\
  \label{eq:tad_app}  
  \mathcal P_{\delta \phi, \tad}(q,\tau) 
  &= 4 \frac{q^D \Omega_D }{(2\pi)^D} \mu^{2(D-3)}\int^\tau_{-\infty} \dd \tau' \int^{\tau'}_{-\infty} \dd \tau''a^{D+1}(\tau') a^{D+1}(\tau'') V_\tho(\tau') \Im\left[ u_q(\tau) u^*_q(\tau') \right] \Re\left[ u_q(\tau) u^*_q(\tau') \right] \nonumber \\
  &\qquad \times \Im[ u_{0}(\tau') u_{0}^*(\tau'')]  \left( V_\tho(\tau'') \int \frac{\dd^D k}{(2\pi)^D} |u_{k}(\tau'')|^2 + 2 V_{c,\fo}(\tau'')\right), \\
  \label{eq:p_a_app}
  \mathcal P^a_{\delta \phi, 2\vx}(q,\tau) &= 2 \frac{q^D \Omega_D }{(2\pi)^D} \mu^{2(D-3)} \int^\tau_{-\infty} \dd \tau' \int^{\tau}_{-\infty} \dd \tau'' a^{D+1}(\tau')  a^{D+1}(\tau'') V_\tho(\tau') V_\tho(\tau'') \Im[u_q(\tau) u_q^*(\tau')] \Im[u_q(\tau) u_q^*(\tau'')] \nonumber \\
  &\qquad \qquad \qquad \times \int \frac{\dd^D k}{(2\pi)^D} \Re[u_k(\tau')u^*_k(\tau'')u_{|\bfq - \bfk|}(\tau')u^*_{|\bfq - \bfk|}(\tau'')], \\
  \label{eq:p_b_app}  
  \mathcal P^b_{\delta \phi, 2\vx}(q,\tau)   &= 8 \frac{q^D \Omega_D }{(2\pi)^D} \mu^{2(D-3)} \int^\tau_{-\infty} \dd \tau' \int^{\tau'}_{-\infty} \dd \tau'' a^{D+1}(\tau') 
  a^{D+1}(\tau'') V_\tho(\tau') V_\tho(\tau'') \Im[u_q(\tau) u_q^*(\tau')] \nonumber \\
  &\qquad \times  \int \frac{\dd^D k}{(2\pi)^D} \Im[u_{|\bfq - \bfk|}(\tau')u^*_{|\bfq - \bfk|}(\tau'')] \Re[u_q(\tau) u_q^*(\tau'')]\Re[u_k(\tau')u^*_k(\tau'')],
\end{align}
where $\Omega_D = 2\pi^{D/2}/\Gamma(D/2)$ and the power spectrum is given by $\langle \delta \phi_\bfq \delta \phi_{\bfq'} \rangle = (2\pi)^D \delta^D(\bfq + \bfq') \frac{(2\pi)^D}{\Omega_D q^D} \mathcal P_{\delta \phi}(q)$.
Note that the mass dimension of $u_k$ is $[u_k] = -(D-2)/2$, while $[\delta \phi(\bfx)] = 1$.

Next, let us see how the form of $u_k$ is given in $D+1$ dimensions. 
From Eq.~(\ref{eq:action_d}), the equation of motion of $u_k$ is given by (see Eq.~(\ref{eq:u_eom}) for $3+1$ dimension case)
\begin{align}
  \left[\frac{\dd^2}{\dd \tau^2} + (D-1) \mathcal H \frac{\dd}{\dd \tau} + k^2 + a^2(\tau) V_\so(\tau) \right] u_k(\tau) = 0.
  \label{eq:u_eom_app}
\end{align}
This can be reexpressed with the physical time as 
\begin{align}
  \left[\frac{\dd^2}{\dd t^2} + D H \frac{\dd}{\dd t} + \frac{k^2}{a^2(t)} + V_\so(t) \right] u_k(t) = 0,
  \label{eq:u_eom_app_pt}
\end{align}
where $t (= \int^\tau \dd \tau' a(\tau'))$ and $u_k(t) = u_k(\tau(t))$.
The retarded Green function for this equation is given by 
\begin{align}
  \label{eq:g_tt}
  g_k(t;t') &= \Theta(t-t')\frac{u_k(t) u_k^*(t') - u^*_k(t) u_k(t')}{\dot u_k(t') u^{*}_k(t') - u_k(t') \dot u^*_k(t')} \nonumber \\
  &= -2\Theta(t-t')\mu^{D-3}a^D(t') \Im[u_k(t) u_k^*(t')].
\end{align}
This $g_k(t;t')$ satisfies
\begin{align}
  \left[ \frac{\dd^2}{\dd t^2} + DH \frac{\dd}{\dd t} + \frac{k^2}{a^2(t)} + V_\so(t) \right] g_k(t;t') = \delta(t - t').
  \label{eq:green_eom}
\end{align}
This Green function in $D+1$ dimensions is used to obtain the expressions of $\langle \delta \dot \phi \rangle$ in $D+1$ dimensions.

For the time being, let us focus on the case where $V_\so(\tau)$ can be approximated as a constant ($V_\so =m^2$) compared to the oscillation timescale $1/k$.
This case is important because we are interested in the UV behavior of $u_k$, which is relevant to the UV regularization.
For convenience, we normalize the variable as 
\begin{align}
  v_k \equiv a^{\frac{D-1}{2}} u_k.
\end{align}
With this, we can rewrite Eq.~(\ref{eq:u_eom_app}) as 
\begin{align}
  \left[\frac{\dd^2}{\dd \tau^2} + k^2 - \frac{\nu^2 - \frac{1}{4}}{\tau^2} \right] v_k(\tau) = 0,
\end{align}
where $\nu = \sqrt{D^2/4 - m^2/H^2}$. 
Solving this equation, we obtain 
\begin{align}
  v_k(\tau) = \mu^{-(D-3)/2} \frac{\sqrt{\pi}}{2} \ee^{i \frac{\pi}{2}(\nu + 1/2)}\sqrt{-\tau} H_\nu^\fo(-k\tau) \quad (V_\so = \text{const.}),
\end{align}
where we have imposed the Bunch-Davies condition, $v_k \simeq \mu^{-(D-3)/2}\ee^{-ik\tau}/\sqrt{2k}$ in $-k\tau \to \infty$, and put $\mu^{-(D-3)/2}$ to take into account the mass dimension of $u_k$.
From this, we obtain 
\begin{align}
  u_k(\tau) = \mu^{-(D-3)/2} H^{\frac{D-1}{2}} \left(-\tau\right)^{D/2} \frac{\sqrt{\pi}}{2} \ee^{i \frac{\pi}{2}(\nu + 1/2)} H_\nu^\fo(-k\tau) \quad (V_\so = \text{const.}).
\end{align}
Note that, in the dimensional regularization scheme, we do not need to introduce the UV cutoff $\Theta(k_\uv a(\tau)/a_* - k)$ in $u_k(\tau)$, given by Eq.~(\ref{eq:uk_cutoff}).
Taking the same procedure in Ref.~\cite{Inomata:2025bqw}, we can express the time derivative of $u_k(t)$ in $D+1$ dimensions with a time-dependent $V_\so$ as 
\begin{align}
  \frac{\dd u_k(\tau(t))}{\dd t} &= 2\mu^{D-3}\int^\tau_{-\infty} \dd \tau' a^{D+1}(\tau') V_\tho(\tau') \dot {\bar \phi}(\tau') \Im[u_k(\tau) u^*_k(\tau')] u_k(\tau') - H k^{-D/2}\frac{\dd (k^{D/2} u_k(\tau))}{\dd \ln k},
  \label{eq:du_dt_f}
\end{align}
where we have taken the Bunch-Davies solution as the initial condition of $u_k(\tau)$ in the far past ($\tau \to -\infty$).

We here discuss $\langle \delta \dot \phi \rangle$.
We begin with the tadpole $\expval{\delta \phi}$ in $D+1$ dimensions:
\begin{align}
  \expval{\delta \phi(\tau(t))}
  &=  \mu^{D-3}\int^t_{0} \dd t' a^D(t') \Im[ u_0(t) u_0^*(t')] \left[ V_\tho(t') \int \frac{\dd^D k}{(2\pi)^D} |u_{k}(t')|^2 + 2 V_{c,\fo}(t') \right] \nonumber \\ 
  &= -\frac{1}{2} \int^t_{0} \dd t' g_0(t;t') \left[ V_\tho(t') \int \frac{\dd^D k}{(2\pi)^D} |u_{k}(t')|^2 + 2 V_{c,\fo}(t') \right].
  \label{eq:dt_one_pt_app}
\end{align}
Similarly to Ref.~\cite{Inomata:2025bqw}, we obtain 
\begin{align}
  \langle \delta \dot \phi(\tau) \rangle
   &= \langle \delta \dot \phi(\tau) \rangle_{1\vx} + \langle \delta \dot \phi(\tau) \rangle_\tad + \langle \delta \dot \phi(\tau) \rangle_{2\vx},
  \label{eq:dt_one_pt_g2_app}
\end{align}
where 
\begin{align}
\langle \delta \dot \phi(\tau) \rangle_{1\vx} &= \mu^{D-3}\int^\tau_{-\infty} \dd \tau' a^{D+1}(\tau') \Im[u_0(\tau)u_0^*(\tau')]\dot{\bar \phi}(\tau') \left[ V_\foo(\tau') \int \frac{\dd^D k}{(2\pi)^D} |u_{k}(\tau')|^2 + 2V_{c,\so}(\tau') \right],
  \label{eq:br_1vx_app} \\
 \langle \delta \dot \phi(\tau) \rangle_\tad & = 2 \mu^{2(D-3)}\int^\tau_{-\infty} \dd \tau'\int^{\tau'}_{-\infty} \dd \tau''\, a^{D+1}(\tau') a^{D+1}(\tau'')\Im[u_0(\tau) u^*_0(\tau')] V_\tho(\tau') \dot {\bar\phi}(\tau') \Im[u_0(\tau')u^*_0(\tau'')] \nonumber \\ 
& \qquad \qquad 
\times \left[ V_\tho(\tau'') \int \frac{\dd^D k}{(2\pi)^D} |u_{k}(\tau'')|^2 + 2 V_{c,\fo}(\tau'') \right],
\label{eq:tad_back_app} \\  
\langle \delta \dot \phi(\tau) \rangle_{2\vx} &= \mu^{D-3}\int^\tau_{-\infty} \dd \tau' a^{D+1}(\tau') V_\tho(\tau') \Im[u_0(\tau)u^*_0(\tau')]  \frac{\dd}{\dd t'} \int \frac{\dd^D k}{(2\pi)^D} |u_{k}(t')|^2.
    \label{eq:2vx_tad_app}
\end{align}
In the case of the UV cutoff scheme, $\langle \delta \dot \phi(\tau) \rangle_{2\vx}$ includes the term from the time derivative acting on the UV boundary, which does not exist in dimensional regularization. 
Because of this, the calculation of $\langle \delta \dot \phi(t) \rangle_{2\vx}$ becomes slightly different in dimensional regularization. 
Using Eq.~(\ref{eq:du_dt_f}), we obtain 
\begin{align}
  \langle \delta \dot \phi(\tau) \rangle_{2\vx} &= \langle \delta \dot \phi(\tau) \rangle_{\overline{2\vx}} - \mu^{D-3}\int^\tau_{-\infty} \dd \tau' a^{D+1}(\tau') V_\tho(\tau') \Im[u_0(\tau)u^*_0(\tau')]  H\frac{\Omega_D}{(2\pi)^D} \int^\infty_{k_\ir} \dd \ln k \frac{\dd (k^{D} |u_k(\tau')|^2)}{\dd \ln k},
  \label{eq:2vx_app}
\end{align}
where 
\begin{align}
    \langle \delta \dot \phi(\tau) \rangle_{\overline{2\vx}}&= 4 \mu^{2(D-3)}\int^\tau_{-\infty} \dd \tau' \int^{\tau'}_{-\infty} \dd \tau'' a^{D+1}(\tau') a^{D+1}(\tau'') V_\tho(\tau') V_\tho(\tau'') \dot{\bar \phi}(\tau'')\Im[u_0(\tau) u_0^*(\tau')]\nonumber \\
  &\qquad \times \int \frac{\dd^D k}{(2\pi)^D} \Im[u_k(\tau') u^*_k(\tau'')] \Re[u_k(\tau') u^*_k(\tau'')].
  \label{eq:ddot_2vx_bar_app}
\end{align}
We here use the dimensional regularization trick. 
Recalling $u_k \propto 1/\sqrt{k}$ in the UV limit, we first take $D < 1$ and obtain 
\begin{align}
  \langle \delta \dot \phi(\tau) \rangle_{2\vx} &= \langle \delta \dot \phi(\tau) \rangle_{\overline{2\vx}} + \langle \delta \dot \phi(\tau) \rangle_\ir,
\end{align}
where 
\begin{align}
    &\langle \delta \dot \phi(\tau) \rangle_{\ir} = \mu^{D-3}\int^\tau_{-\infty} \dd \tau' a^{D+1}(\tau')\Im[u_0(\tau) u_0^*(\tau')] H  V_\tho(\tau') \mathcal P_{\delta \phi,\tre}\left( k_\ir,\tau' \right).
    \label{eq:ddot_ir_app}
\end{align}
The UV boundary term in Eq.~(\ref{eq:2vx_app}) disappears in $D < 1$, similarly to the cutoff scheme.
In addition, the other contributions also converge in $D < 1$. 
This means that we have obtained Eq.~(\ref{eq:in_in_loop_exp}) in $D < 1$ dimensions.
Then, we analytically continue the expressions from $D < 1$ to $D=3$ and reproduce Eq.~(\ref{eq:in_in_loop_exp}).
This continuation does not revive the UV boundary term, while the continuation gives rise to a divergence proportional to $1/(D-3)$ in $\int \frac{\dd^D k}{(2\pi)^D} |u_{k}(\tau'')|^2$, included in the contributions in Eq.~(\ref{eq:in_in_loop_exp}).
This kind of divergence is removed by the counter potential $V_{c,\fo}$.
For example, similarly to Eq.~(\ref{eq:v_c1_tad_cut}), we can take 
\begin{align}
  V_{c,\fo}(\tau) = \lim_{D \to 3} \left[ -\frac{V_\tho(\tau)}{2} \int^{\infty}_{k_\sm} \frac{\dd k}{k} \frac{k^D \Omega_D }{(2\pi)^D} |u_{k}(\tau)|^2 \right], 
  \label{eq:v_c1_tad_cut_app}
\end{align}
which cancels the divergence proportional to $1/(D-3)$.

\small
\bibliographystyle{apsrev4-1}
\bibliography{draft_one_loop_all_scales}

\end{document}